\documentclass[apj, numberedappendix]{emulateapj}

\usepackage{graphicx} 
\usepackage{amsmath} 
\usepackage{amssymb}
\usepackage{exscale, relsize}
\usepackage{amscd}
\usepackage{natbib}
\usepackage{lscape}

\newcommand{\eg}{\textit{e.g.}}
\newcommand{\ie}{\textit{ie.}}
\newcommand{\viz}{\textit{viz.}}
\newcommand{\cf}{\textit{cf.}}

\newcommand{\zphot}{z_{\mathrm{phot}}}
\newcommand{\zspec}{z_{\mathrm{spec}}}
\newcommand{\zform}{z_{\mathrm{form}}}

\newcommand{\masslimit}{10^{11}}
\renewcommand{\sun}{$_{\odot}$}

\newcommand{\ptot}{\gamma_{\mathrm{tot}}}
\newcommand{\pred}{\gamma_{\mathrm{red}}}
\newcommand{\pfrac}{\gamma_{\mathrm{frac}}}

\usepackage{morefloats}

\begin{document}

\title{The Rise of Massive Red Galaxies: \\ the Color--Magnitude and
  Color--Stellar Mass Diagrams for $\zphot \lesssim 2$ \\ from the
  MUltiwavelength Survey by Yale--Chile (MUSYC)}

\shorttitle{The Rise of Massive Red Galaxies}
\shortauthors{E N Taylor et al.}


\begin{abstract}
  We present the color--magnitude and color--stellar mass diagrams for
  galaxies with $\zphot \lesssim 2$, based on a $K^{\mathrm{(AB)}} <
  22$ catalog of the $\frac{1}{2} \times \frac{1}{2}~\square^{\circ}$
  Extended Chandra Deep Field South (ECDFS) from the MUltiwavelength
  Survey by Yale--Chile (MUSYC).  Our main sample of 7840 galaxies
  contains 1297 $M_* > \masslimit$ M\sun\ galaxies in the range $0.2 <
  \zphot < 1.8$.  We show empirically that this catalog is
  approximately complete for $M_* > \masslimit$ M\sun\ galaxies for
  $\zphot < 1.8$.  For this mass--limited sample, we show that the
  locus of the red sequence color--stellar mass relation evolves as
  $\Delta(u-r) \propto (-0.44 \pm 0.02) ~ \zphot$ for $\zphot \lesssim
  1.2$.  For $\zphot \gtrsim 1.3$, however, we are no longer able to
  reliably distinguish red and blue subpopulations from the observed
  color distribution; we show that this would require much deeper near
  infrared (NIR) data.  At $1.5 < \zphot < 1.8$, the comoving number
  density of $M_* > 10^{11}$ M\sun\ galaxies is $\approx 50 \%$ of the
  local value, with a red fraction of $\approx 33$ \%.  Making a
  parametric fit to the observed evolution, we find $n_\mathrm{tot}
  (z) \propto (1+\zphot)^{-0.52 \pm 0.12 (\pm 0.20)}$.  We find
  stronger evolution in the red fraction: $f_\mathrm{red}(z) \propto
  (1+\zphot)^{-1.17 \pm 0.18 (\pm 0.21)}$.  Through a series of
  sensitivity analyses, we show that the most important sources of
  systematic error are: 1.\ systematic differences in the analysis of
  the $z \approx 0$ and $z \gg 0$ samples; 2.\ systematic effects
  associated with details of the photometric redshift calculation; and
  3.\ uncertainties in the photometric calibration.  With this in
  mind, we show that our results based on photometric redshifts are
  consistent with a completely independent analysis which does not
  require redshift information for individual galaxies.  Our results
  suggest that, at most, 1/5 of local red sequence galaxies with $M_*
  > 10^{11}$ M\sun\ were already in place at $z \sim 2$.
\end{abstract}

\author{Edward N Taylor$^1$, 
  Marijn Franx$^1$, 
  Pieter G van Dokkum$^2$, 
  Eric F Bell$^3$, 
  Gabriel B Brammer$^2$, \\
  Gregory Rudnick$^4$, 
  Stijn Wuyts$^5$, 
  Eric Gawiser$^6$, 
  Paulina Lira$^7$, 
  C Megan Urry$^2$, 
  Hans-Walter Rix$^3$, 
}


\affil{$^1$ Sterrewacht Leiden, Leiden University, NL-2300 RA Leiden, Netherlands; ent@strw.leidenuniv.nl, \\
$^2$ Department of Astronomy, Yale University, New Haven, CT 06520-8101, 
$^3$ Max-Planck-Institut f\"ur Astronomie, \\ D-69117 Heidelberg, Germany, 
$^4$ Goldberg Fellow, National Optical Astronomy Observatories,
Tucson, AZ 85721; currently \\at Department of Physics and
Astronomy, University of Kansas, Lawrence, KS, 66045
$^5$ W. M. Keck Postdoctoral Fellow, \\Harvard-Smithsonian Center for
Astrophysics, Cambridge, MA 02138,
$^6$ Department of Physics and Astronomy, \\Rutgers University,
Piscataway, NJ, 08854
$^7$ Departamento de Astronom\'ia, Universidad de Chile, Santiago, Chile}



\section{Introduction} \label{ch:intro}

Observing the evolution of the massive galaxy population provides
basic constraints on cosmological models of structure formation, and
so helps to identify the physical processes that govern the formation
and evolution of massive galaxies.  In this context, the
color--magnitude diagram (CMD)---astronomy's most basic diagnostic
plot---has been particularly important and useful over the past five
years.  Physically, a galaxy's restframe color is determined by the
(luminosity weighted) mean stellar age, modulo the mean stellar
metallicity and extinction from dust in the ISM.  The restframe
optical brightness acts as a proxy for the total stellar mass,
although the connection between the two has a similarly complicated
dependence on star formation history, metallicity, and dust.  The CMD
thus offers two complementary means of characterizing the star
formation history of individual galaxies, in terms of the amount and
character of their starlight.

In the local universe, galaxies can be separated into two distinct but
overlapping populations in color--magnitude space \citep{Baldry2004}:
a relatively narrow and well--defined `red sequence', as distinct from
the more diffuse `blue cloud', with each following its own
color--magnitude relation (CMR).  Red sequence galaxies dominate the
bright galaxy population, and tend to have the more concentrated light
distributions typical of morphologically early type galaxies
\citep{Strateva2001, Blanton2003, Driver2006, vdWel2008}.  They
typically have stellar masses greater than $10^{10.5}$ M\sun\ and are
dominated by old stars, whereas blue cloud galaxies are typically less
massive and continue to be actively star forming \citep{Kauffmann2003,
BrinchmannEtAl, WyderEtAl}.  Further, red sequence galaxies lie
preferentially in higher density environments \citep{Hogg2003,
Blanton2005Env, Baldry2006}.  The emergent picture is of a population
of massive, quiescent, concentrated, and strongly clustered red
sequence galaxies, as distinct from the typically less massive,
disk--dominated, and star forming blue cloud population
\citep{Ellis2005}.  This paper focuses on the redshift evolution of
the red sequence galaxy population.

Using high quality photometric redshifts from the COMBO-17 survey,
\citet{Bell2004} showed that a red galaxy sequence is already in place
at $\zphot \sim 1$ \citep[see also, \eg,][]{ImEtAl, WeinerEtAl,
WillmerEtAl}.  Further, as in the local universe, the $\zphot \approx
0.7$ red sequence is dominated by passive, morphologically early type
galaxies \citep{Bell2004-Morph}.  The combined mass of red sequence
galaxies at $z \sim 1$ is at least half of the present day value
\citep{Bell2004, Faber2007, BrownEtAl2008}.  By contrast, the stellar
mass density of actively star forming blue cloud galaxies remains more
or less constant for $z \lesssim 1$ \citep{Arnouts2007, Bell2007},
even as the combined star formation rate drops by an order of
magnitude over the same interval \citep{Lilly1996, Madau1996,
Hopkins2004}.  These results---a steadily growing number of passively
evolving red galaxies, and a relatively constant number of actively
star forming blue galaxies---have led to the idea of a quenching
mechanism for star formation, operating to incite a transformation
that moves active galaxies from the blue cloud onto the passive red
sequence \citep{Menci2005, Croton2006, Cattaneo2006, DekelBirnboim,
DeLucia2007}.

Our specific goal in this paper is to quantify the evolution of
massive galaxies in general, and of red sequence galaxies in
particular, in the color--magnitude and color--stellar mass planes for
$\zphot \lesssim 2$.  The $1 \lesssim z \lesssim 2$ interval is
particularly interesting: whereas the $z \sim 1$ galaxy population
appears qualitatively similar to the local universe, at least in terms
of the existence and properties of red sequence galaxies, the
situation at $z \gtrsim 2$ may be quite different.  While massive,
passive galaxies have been confirmed at $z \gtrsim 1.5$
\citep{Daddi2005, McGrath2007} and even $z \gtrsim 2$
\citep{Kriek2006}, these galaxies do not appear to dominate the
massive galaxy population as they do at $z \lesssim 1$.  Indeed, it
appears that the median massive galaxy at $z \sim 2$ has the infrared
luminosity of a LIRG or ULIRG \citep{Reddy2006}.  Moreover, whereas
the number density of massive galaxies at $z \sim 1$ is $\gtrsim 50$
\% of the local value \citep{Juneau2005, BorchEtAl, Scarlata2007}, at
$z \gtrsim 2$ it is inferred to be $\lesssim 15$ \%
\citep{Fontana2006, Arnouts2007, PozzettiEtAl, PerezGonzalez}.  This
marks the redshift interval $1 \lesssim z \lesssim 2$ as potentially
being an era of transition in the universe, in which massive galaxies
first begin both to appear in large numbers, and to take on the
appearance of their local antecedents.  This coincides with end of the
period of peak star formation in the universe; while the cosmic star
formation rate rises sharply for $z \lesssim 1$, it appears to plateau
or even peak for $z \gtrsim 2$ \citep[see, \eg,][]{Hopkins2004,
NagamineEtAl, PanterEtAl, TresseEtAl, PerezGonzalez}.  Whatever the
mechanism that quenches star formation in massive galaxies may be, it
is in operation at $1 < z < 2$.

\vspace{0.2cm}

The technical key to gaining access to the $1 \lesssim z \lesssim 2$
universe is deep near infrared (NIR) data \citep{ConnollyEtAl1998},
since at these redshifts the restframe optical features on which both
spectroscopic and photometric redshift and stellar mass determinations
rely are shifted beyond the observer's optical window.  Moreover, the
inclusion of NIR data makes it possible to construct stellar mass
limited samples with high completeness \citep{VanDokkum2006}.  Among
the next generation of NIR--selected cosmological field galaxy
surveys, the MUltiwavelength Survey by Yale--Chile (MUSYC; Gawiser et
al.\ 2006) is among the first to become public.  MUSYC has targeted
four widely dispersed Southern fields, covering a total of one square
degree in the $UBVRIz'$ bands.  Coupled with this optical imaging
programme, MUSYC also has two NIR imaging campaigns: a wide
($K^\mathrm{(AB)} \lesssim 22$) component over three of the four
fields \citep[][hereafter Paper I]{Blanc2008,PaperI}, and a deeper
($K^\mathrm{(AB)} \lesssim 23.5$) component for four 10' $\times$ 10'
fields \citep{QuadriEtAl}.

This present paper focuses on the Extended Chandra Deep Field South
(ECDFS), one of the four $\frac{1}{2} \times
\frac{1}{2}~\square^{\circ}$ MUSYC fields.  Centered on the historical
Chandra Deep Field-South \citep[$\alpha =
03^\mathrm{h}32^\mathrm{m}28\mathrm{s}$, $\delta = -27^\circ48'30''$;
J2000 ---][]{Giacconi2001}, this is one the best studied fields on the
sky, with observations spanning the full electromagnetic spectrum from
the X-ray to the radio.  Notably, this field is also a part of the
COMBO-17 survey \citep{WolfEtAl}, and has received {\em Hubble Space
Telescope} ACS coverage as part of the GEMS project \citep{RixEtAl},
as well as extremely deep {\em Spitzer Space Telescope} imaging from
the SIMPLE \citep{DamenEtAl} project.  Further, the GOODS project
\citep{DickinsonEtAl} covers the 160 $\square$' at the centre of this
field, including supporting NIR data from the ISAAC instrument on the
VLT \citep{GrazianEtAl, WuytsEtAl}.  Complementing these and other
imaging surveys, a wealth of spectroscopic redshifts are available
from large campaigns including the K20 survey \citep{K20}, the VVDS
project \citep{VVDS}, the two GOODS spectroscopic campaigns
\citep{FORS2, VIMOS}, and the IMAGES survey \citep{IMAGES}, among
others.

\vspace{0.2cm}

The plan of this paper is as follows.  We begin in
\textsection\ref{ch:data} by giving a brief overview of the data used
in this paper --- the MUSYC ECDFS dataset itself, as well as the $z
\approx 0$ comparison sample from \citet{BlantonEtAl-lowz}.  Next, in
\textsection\ref{ch:methods}, we describe our basic methods for
deriving redshifts and restframe properties for $z \gg 0$ galaxies;
our analysis of the $z \approx 0$ comparison sample is described
separately in Appendix \ref{ch:z=0}.  In
\textsection\ref{ch:masslimit}, we construct a stellar-mass limited
sample of massive galaxies from our $K$--selected catalog.

Our basic results --- the color--magnitude and color--stellar mass
diagrams for $\zphot < 2$ --- are presented in
\textsection\ref{ch:results}.  We then analyse three separate aspects
of the data: evolution in the color distribution of the massive galaxy
population in (\textsection\ref{ch:bimodality}); the color evolution
of the red galaxy population as a whole (\textsection\ref{ch:redseq});
and the $\zphot \lesssim 2$ evolution in the absolute and relative
numbers of massive red/blue galaxies (\textsection\ref{ch:growth}).
Our final results are in conflict with those from COMBO-17 in the same
field; in Appendix \ref{ch:combo}, we show that this is a product of
calibration errors in the COMBO-17 data, rather than differences in
our analyses.  In \textsection\ref{ch:sens}, we present a series of
sensitivity analyses in which we repeat our analysis a number of
times, while varying individual aspects of our experimental design,
and seeing how these changes affect our results; this tests thus
enable us to identify and quantify the most important sources of 
systematic uncertainty in our main results.

Finally, in \textsection\ref{ch:obcol}, we present a completely
independent consistency check on our results: we measure the $z
\lesssim 2$ evolution of the relative number of bright, red galaxies
based only on directly observed quantities --- that is, without
deriving redshifts or stellar masses for individual galaxies.  A
summary of our results and conclusions is given in
\textsection\ref{ch:conc}.


\begin{table}[t] \begin{center} 
\caption{Summary of the data comprising \\ the MUSYC ECDFS catalog
\label{tab:observations}}
\begin{tabular}{l c c c c r c}
\hline 
\hline 
Band & $\lambda_0$ & Instrument & Exp. Time & Area & FWHM & $5 \sigma$ depth \\
     & [\AA] & & [min.] & [$\square$'] & & (AB) \\
(1)  & (2) & (3) & (4) & (5) & (6) & (7) \\
\hline
$U$      & 3560  & WFI       & 1315 & 973 & $1\farcs1$ & 26.5 \\
$U_{38}$ & 3660  & WFI       & 825  & 942 & $1\farcs0$ & 26.0 \\
$B$      & 4600  & WFI       & 1157 & 1011 & $1\farcs1$ & 26.9 \\
$V$      & 5380  & WFI       & 1743 & 1020 & $1\farcs0$ & 26.6 \\
$R$      & 6510  & WFI       & 1461 & 1016 & $0\farcs9$ & 26.3 \\
$I$      & 8670  & WFI       & 576  & 976 & $1\farcs0$ & 24.8 \\
$z'$      & 9070  & MOSAIC-II & 78   & 997 & $1\farcs1$ & 24.0 \\
$J$      & 12500 & ISPI      & $\approx$ 80   & 882 & $< 1\farcs5$ & 23.3 \\
$H$      & 16500 & SofI      & $\approx$ 60   & 650 & $< 0\farcs8$ & 23.0 \\
$K$      & 21300 & ISPI      & $\approx$ 60   & 887 & $< 1\farcs0$ & 22.5 \\
\hline
\hline
\end{tabular} \end{center}
\tablecomments{For each band, we give the filter identifier (Col.\ 1)
  and effective wavelength (Col.\ 2), as well as the detector name
  (Col.\ 3).  For the NIR data, exposure times (Col.\ 4) are given per
  pointing; the effective seeing (Col.\ 6) is given for the pointing
  with the broadest PSF.  The $5 \sigma$ limiting depths (Col.\ 7) are
  as measured in $2\farcs5$ diameter apertures; the smallest apertures
  we use.  The references for each set of imaging data are given in
  the main text (\textsection \ref{ch:musyc}).}
\end{table}


Throughout this work, magnitudes are expressed in the AB system;
exceptions are explicitly marked.  All masses have been derived
assuming a `diet Salpeter' IMF \citep{BellDeJong}, which is defined to
be 0.15 dex less massive than a standard \citet{Salpeter} IMF.  In
terms of cosmology, we have assumed $\Omega_\Lambda = 0.70$, $\Omega_m
= 0.30$, and $H_0 = 70~h_{70}$ km s$^{-1}$ Mpc$^{-1}$, where $h_{70} = 1$.


\section{Data} \label{ch:data}

\subsection{An Overview of the MUSYC ECDFS Dataset} \label{ch:musyc}

This work is based on a $K$--selected catalog of the ECDFS from the
MUSYC wide NIR imaging programme; these data are described and
presented in Paper I.  We will refer hereafter to this dataset as
`the' MUSYC ECDFS catalog, although it should be distinguished from
the optical ($B$+$V$+$R$)--selected catalog, and the narrow band
(5000 \AA)--selected photometric catalogs described by
\citet{GronwallEtAl}, and the spectroscopic catalog described by
\citet{TreisterEtAl}.

The vital statistics of the imaging data that have gone into the MUSYC
ECDFS catalog are summarized in Table \ref{tab:observations}.  Unlike
the three other MUSYC fields, the ECDFS dataset was founded on
existing, publicly available optical imaging: specifically, archival
$UU_{38}BVRI$ WFI data,
including those taken as part of the COMBO-17 survey \citep{WolfEtAl}
and ESO's Deep Public Survey \citep[DPS;][]{ArnoutsEtAl2001}, which
have been re-reduced as part of the GaBoDS project \citep{ErbenEtAl,
HildebrandtEtAl}.  We also include $H$ band imaging (P Barmby, priv.\
comm.)\ from SofI on the 3.6m NTT, covering $\sim 80$ \% of the field,
taken to complement the ESO DPS data, reduced and described by
\citet{MoyEtAl}.

\subsubsection{Original Data Reduction and Calibration}

These existing data have been supplemented with original $z'$ band
imaging from MOSAIC-II, reduced as per \citet{GawiserEtAl}, as well as
$J$ and $K$ band imaging from ISPI; both instruments are mounted on
the Blanco 4m telescope at CTIO.  To cover the full $\frac{1}{2}
\times \frac{1}{2}~ \square^{\circ}$ ECDFS in the $JK$ bands, we have
constructed a mosaic of nine $\sim 10 \times 10 ~ \square$' subfields
(the size of the ISPI field of view).  The data reduction for the $JK$
imaging closely follows \citet{QuadriEtAl} and \citet{Blanc2008}, and
is described in detail in Paper I, where we present the MUSYC ECDFS
catalog.

In brief, to facilitate multiband photometry, each reduced image has
been shifted to a common astrometric reference frame ($0\farcs267$
pix$^{-1}$).  The relative astrometry has been verified to $0\farcs15$
(0.56 pix).  To combat aperture effects (\ie\ similar apertures
capturing different fractions of light, due to variable seeing across
different images), we have PSF-matched our images to the one with the
worst effective seeing.  Among the $K$ band pointings, the worst
effective seeing is $1\farcs 0$ FWHM; this sets our limits for
detection and for total $K$ band flux measurements.  Among the other
bands, the worst seeing is $1\farcs 5$ FWHM in the Eastern $J$
subfield; this sets the limit for our multi-color photometry.  After
PSF matching, systematic errors due to aperture effects are estimated
to be $\lesssim 0.006$ mag for the smallest apertures we use.

We have tested the photometric calibration through comparison with the
COMBO-17 catalog of the ECDFS \citep{WolfEtAl}, and with the FIREWORKS
catalog of the GOODS-CDFS region \citep{WuytsEtAl}.  While there are
significant differences between the COMBO-17 and MUSYC photometry, the
comparison to FIREWORKS validates our photometry and photometric
calibration to $\lesssim 0.02$ mag in most cases, particularly for the
redder bands.  Further, we have tested the relative calibration of all
bands using the observed colors of stars; this test validates the
photometric cross-calibration to $\lesssim 0.05$ mag.  (See
\textsection\ref{ch:zps} for a discussion of how sensitive our main
results are to photometric calibration errors.)  


\subsubsection{Photometry}

The photometry itself was done using SExtractor \citep{BertinArnouts}
in dual image mode, using the $1\farcs0$ FWHM $K$ mosaic as the
detection image.  Note that, as we were unable to find a combination
of SExtractor background estimation parameters (for the detection
phase) that were fine enough to map real spatial variations in the
background level, but still coarse enough to avoid being influenced by
the biggest, brightest sources, we were forced to perform our own
background subtraction for the NIR images.  Total fluxes were measured
from this $1\farcs0$ $K$ image, using SExtractor's FLUX\_AUTO.  In
Paper I, we that (in the photometry phase) SExtractor systematically
overestimates the background flux level by $\sim 0.03$ mag; we have
taken steps to correct for this effect.  Following \citet{LabbeEtAl},
we also apply a minimal correction to account for missed flux beyond
the finite AUTO aperture, treating each object as though it were a
point source.  We quantify the impact these two corrections have on
our final results in \textsection\ref{ch:background} and
\textsection\ref{ch:totalmags}.

Multicolor spectral energy distributions (SEDs) were constructed for
each object using the larger of SExtractor's ISO aperture and a
$2\farcs 5$ diameter circular aperture, measured from the $1\farcs5$
FWHM $U$---$K$ images; we then normalize each object's SED using the
total $K$ flux.  This flexibility in aperture size is important to
compromise between using apertures that are small enough to optimize
S:N (the $2\farcs5$ diameter aperture is close to optimal in terms of
S:N for a point source in the $1\farcs5$ FWHM $J$ band image), but
also large enough to account for color gradients, which are important
for the nearest, brightest objects.  (See \textsection\ref{ch:seds}
for a discussion of how our results vary using only fixed aperture
photometry to construct SEDs.)

Photometric errors (accounting for sky noise, imperfect background
subtraction, etc., as well as the pixel--pixel correlations introduced
at various stages in the reduction process) were derived empirically
by placing large numbers of `empty' apertures on each image \citep[see
also][]{LabbeEtAl, GawiserEtAl, QuadriEtAl}. For the $J$ and $K$
bands, this was done for each subfield individually.

\subsubsection{Completeness and Reliability}

We have assessed the completeness of the MUSYC catalog of the ECDFS
in two ways (Paper I).  First, we have tested our ability to recover
synthetic, $R^{1/4}$ profile sources of varying total flux and size
introduced into empty regions of the data, using procedures identical
to `live' detection.  This analysis suggests that, at $K = 22$, the
MUSYC catalog should be 100 \% complete for point sources, dropping
to 64 \% for $R_\mathrm{eff} = 0\farcs6$ (2.25 pix, or 4.8 kpc at $z =
1$) ellipticals.  Secondly, we have compared our catalog to the much
deeper FIREWORKS \citep{WuytsEtAl} catalog of the CDFS-GOODS region.
In the region of overlap, for $21.8 < K \le 22.0$ bin, $\gtrsim 85$ \%
of FIREWORKS detections are found in the MUSYC catalog; all MUSYC
detections in this bin are confirmed in the FIREWORKS catalog.
Taken together, these two analyses suggest that, at our limiting
magnitude of $K = 22$, the MUSYC catalog is primarily magnitude
(\cf\ surface brightness) limited, $\gtrsim 85$ \% complete, and $\sim
100$ \% reliable.

\subsubsection{Sample Selection}

In constructing our main galaxy sample, we have identified stars on
the basis of the ($B-z'$)--($z'-K$) color--color diagram.  This
selection performs extremely well in comparison with both COMBO-17's
SED classification, and with GEMS point sources (Paper I).  We also
make three further selections.  First, to protect against false
detections in regions with lower weights (\eg\ the mosaic edges, and
exposure `holes' in the Eastern $K$ subfield), we require the
effective weight in $K$ to be greater than 75 \%, corresponding to an
effective exposure time of 45 min or more.  Secondly, we have masked
out, by hand, regions around bright stars where the SEDs of faint
objects may be heavily contaminated; this problem is most severe in
the $z'$ band, where the PSF has broad wings.  With these two
selections, the effective area of the catalog becomes 816
$\square$'.  Thirdly, to protect against extremely poorly constrained
redshift solutions, we will limit our analysis to those objects with
S:N $> 5$ for the $K$ SED point.  (See
\textsection\ref{ch:sncompleteness} for a discussion of how this
selection impacts our results.)

Of the 16910 objects in the MUSYC ECDFS catalog, these selections
produce 10430 reliable $K \le 22$ detections, of which 9520 have
reliable photometry in the ISPI, MOSAIC-II, and WFI band passes.  Of
these, 8790 cataloged objects have $K$ S:N $> 5$; 950 of these objects
are excluded as stars, leaving 7840 galaxies in our main sample.

\subsection{The $z \approx 0$ Comparison Sample}

We will investigate the $\zphot \lesssim 2$ evolution in the massive
galaxy population by comparing the situation at $z \gg 0$ to that for
$z \approx 0$ galaxies in the Sloan Digital Sky Survey
\citep[SDSS;][]{YorkEtAl}; specifically, we use the `low-z' sample
from the New York University (NYU) Value Added Galaxy Catalog (VAGC)
of the SDSS presented by \citet{BlantonEtAl-lowz}.  The (Data Release
4) low-z catalog contains $ugriz$ photometry for 49968 galaxies with
$10 < D < 150$ Mpc ($\zspec \lesssim 0.05$), covering an effective
area of 6670 $\square^{\circ}$; 2513 of these galaxies have $M_* >
\masslimit$ M\sun .  Our analysis of these data closely follows that
of the $z \gg 0$ sample, and is described separately in Appendix
\ref{ch:z=0}.


\section{Photometric Redshifts \\ and Restframe Properties}
\label{ch:methods}


\subsection{Photometric Redshifts} \label{ch:photzs}

The technical crux on which any photometric lookback study rests is
the determination of redshifts from broadband SEDs.  We have computed
our photometric redshifts using EAZY \citep{eazy}, a new,
fully--featured, user--friendly, and publicly--available photometric
redshift code.  By default, the $\zphot$ calculation is based on all
ten bands, although we do require that the effective weight in any
given band is greater than 0.6; in practice, this requirement only
affects the $H$ band, where we do not have full coverage of the field.
The characteristic filter response curves we use account for both
atmospheric extinction and CCD response efficiancy as a function of
wavelength.

For our fiducial or default analysis, we simply adopt the recommended
default settings for EAZY: \viz , we adopt the EAZY default redshift
and wavelength grids, template library, template combination method,
template error function, $K$ luminosity prior, etc.\ \citep[See][for a
complete description.]{eazy} Note that, by default, EAZY assigns each
object a redshift by marginalizing over the full probability
distribution rather than, say, through $\chi^2$ minimization.  For
this work, a key feature of EAZY is the control it offers over how the
SED fitting is done: the user is able to specify whether and how the
basic template spectra are combined, whether or not to include
luminosity prior and/or a template error function, and how the output
redshift is chosen.  We will make use of EAZY's versatility in
\textsection\ref{ch:photzsens} to explore how these particular choices
affect our results.

\begin{figure}[t]
\centering \includegraphics[width=8.2cm]{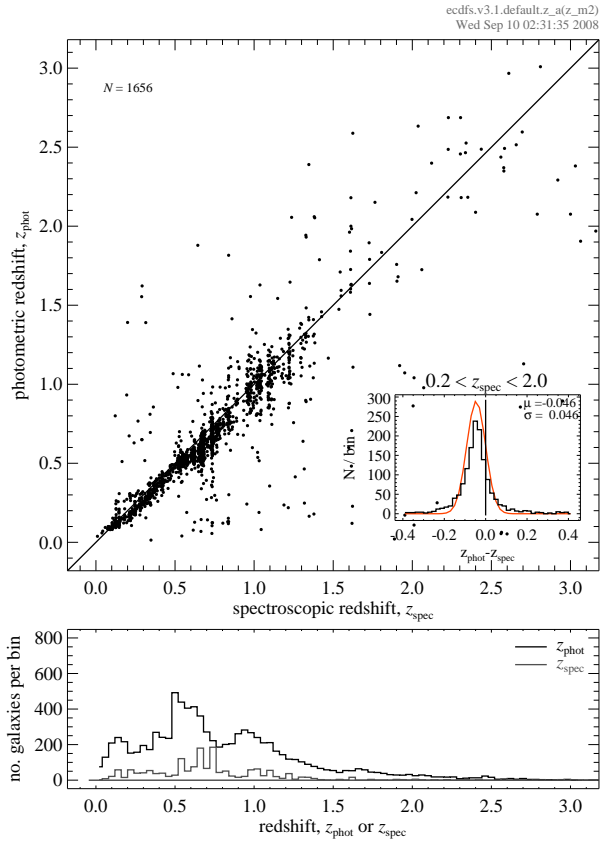}
\caption{Validating the MUSYC catalog photometric redshift
  determinations---{\em Main panel}: the $\zspec$--$\zphot$ diagram
  for the 1656 galaxies from our main $K < 22$ sample, using a
  compendium of `robust' spectroscopic redshifts from the literature
  (see Paper I).  {\em Inset}: the distribution of $\Delta z = \zphot
  - \zspec$ for the same set of galaxies; the curve shows a Gaussian
  fit to this distribution, with parameters as given.  {\em Lower
  panel}: the redshift distributions of the main and spec-$z$ samples;
  our photometric redshifts appear to mildly underestimate the
  redshift of the three overdensities at $0.5 < \zspec < 0.8$.
  Quantitatively, we find the median and NMAD of $\Delta z/(1+\zspec)$
  for the full spec-$z$ sample to be -0.029 and 0.036, respectively
  (See also Figure \ref{fig:derived}); for $\zspec > 1$, we find these
  numbers to be -0.023 and 0.060; for the K20 sample, which is 92 \%
  complete for $K^{\mathrm{(AB)}} < 21.8$, these numbers are -0.028
  and 0.033.
  \label{fig:specz}}
\end{figure}

One of the unique aspects of the ECDFS is the high number of publicly
available spectroscopic redshift determinations, which can be used to
validate and/or calibrate our photometric redshifts.  In Paper I, we
describe a compilation of spectroscopic redshifts for 2914 unique
objects in our catalog, including `robust' redshifts for 1656 galaxies
in our main $K < 22$ sample.  These redshifts come from some of the
many literature sources available in the ECDFS, including those large
surveys referred to in \textsection\ref{ch:intro}, as well as the
X-ray selected spectroscopic redshift catalogs of \citet{Xray} and
\citet{TreisterEtAl}, a new survey by \citet{KoposovEtAl}, and a
number of smaller projects.  K20 is particularly useful in this
regard, given its exceptionally high spectroscopic completeness,
albeit over a very small area: 92 \% of $K^{(\mathrm{Vega})} < 20$
sources over 52 $\square $'.

The main panel of Figure \ref{fig:specz} shows our $\zspec$--$\zphot$
plot.  We prefer to quantify the photometric redshift quality in terms
of the normalized median absolute deviation (NMAD\footnote{The NMAD is
defined as $1.48 \times \mathrm{median}[|x - \mathrm{median}(x)|]$; the
normalization factor of 1.48 ensures that the NMAD of a Gaussian
distribution is equal to its standard deviation.}) in $\Delta
z/(1+\zspec)$, which we will abbreviate as $\sigma_z$; for this
comparison sample, $\sigma_z=0.035$.  Further, the outlier fraction is
acceptably small: 5.9 \%.  Comparing only to the 241 redshifts from
K20, we find $\sigma_z = 0.033$; for the \citet{vdwel05} sample of 28
$z \sim 1$ early type galaxies the figure is 0.022.  For $1 < \zspec <
2$, we find $\sigma_z = 0.059$.  For the 20 \% (269/1297) of $0.2 <
\zphot < 1.8$ galaxies from our mass limited sample defined in
\textsection\ref{ch:masslimit} that have spectroscopic redshifts, we
find $\sigma_z = 0.043$.

Based on their catalog of the GOODS ACS and ISAAC data,
\citet{GrazianEtAl} have achieved a photometric redshift accuracy of
$\left<\Delta z/(1+\zspec)\right>$ of 0.045.  For comparison to
\citet{GrazianEtAl}, the inset panel shows the distribution of $\Delta
z = (\zphot-\zspec)$ for $0 < \zspec < 2$; although offset by -0.046,
the best fit Gaussian to the distribution has a width of 0.046, as
opposed to 0.06 for \citet{GrazianEtAl}.  For an identical sample of
938 galaxies with $\zspec$s, we find $\sigma_z = 0.043$ for the
\citet{GrazianEtAl} $\zphot$s and $\sigma_z = 0.035$ for ours.  In
other words, our photometric redshift determinations are at least as
good as the best published for $K$--selected samples at high
redshifts.  We also note in passing that our $\zphot \lesssim 1$
photometric redshifts agree very well with those from COMBO-17
\citep{WolfEtAl}; a detailed comparison to both these catalogs is
presented in Paper I.

The lower panel of Figure \ref{fig:specz} shows the redshift
distributions for both our main galaxy sample (based on $\zphot$), and
the spectroscopic comparison sample (based on $\zspec$).  Note the
presence of three prominent redshift spikes at $0.5 < \zspec < 0.8$
\citep[see also][]{FORS2}; it appears that our redshift determinations
may slightly underestimate the redshifts of these structures.  The
structures at $z \sim 1.0$, 1.1, 1.2, 1.3, and 1.4 \citep{FORS2} are
also visible in the $\zspec$ distribution, but are `washed out' in the
$\zphot$ distribution.

\subsection{Restframe Photometry and Stellar Masses} \label{ch:rfprops}

The many degeneracies between SED shape and the intrinsic properties
of the underlying stellar population, which are actually a help when
deriving photometric redshifts, make the estimation of such properties
from SED fitting highly problematic.  Systematic uncertainties
associated with parameterisations of the assumed star formation
history are at the level of 0.1 dex \citep{PozzettiEtAl}, while
uncertainties in the stellar population models themselves are
generally accepted to be $\lesssim 0.3$ dex; this is comparable to the
uncertainty associated with the choice of stellar IMF.  For these
reasons, we have opted for considerably simpler means of deriving
restframe parameters.

Once the redshift is determined, we have interpolated restframe fluxes
from the observed SED using a new utility dubbed InterRest, which is a
slightly more sophisticated version of the algorithm described in
Appendix C of \citet{RudnickEtAl}, and is described in detail in Paper
I.  InterRest is designed to dovetail with EAZY, and is also freely
available.\footnote{Code and documentation can be found at:
http://www.strw.leidenuniv.nl/\~{}ent/InterRest/.} We estimate the
systematic errors in our interpolated fluxes (\cf\ colors) to be less
than 2 \% (Paper I).

We then use this interpolated restframe photometry to estimate
galaxies' stellar masses using a prescription from \citet{BellDeJong},
which is a simple linear relation between restframe $(B-V)$ color and
stellar mass-to-light ratio: $M_*/L_V$:
\begin{equation}
\log_{10}~M_*/L_V = -.734 ~+~ 1.404~\times ~(B-V + 0.084) ~ ,
\label{eq:bellmass} \end{equation}
assuming $M_{V,\odot} = 4.82$.  (Here, the factor of 0.084 is to
transform from the Vega magnitude system used by \citet{BellDeJong} to
the AB system used in this work.)  This prescription assumes a `diet
Salpeter' IMF, which is defined to be 0.15 dex less massive than the
standard \citet{Salpeter} IMF, and is approximately 0.04 dex more
massive than that of \citet{Kroupa2001}.  Further, to prevent the most
egregious overestimates of stellar masses, we limit $M_*/L_V \le 10$
(see also Figure 4 of Borch et al.\ 2006).  Although this limit
affects just 1.2 \% of our main sample, we found it to be important
for getting the high-mass end of the $z \approx 0$ mass function right
(Appendix \ref{ch:z=0}).

It is not immediately obvious whether using these color-derived $M/L$s
is significantly better or worse than, say, from using stellar
population synthesis to fit the whole observed SED.  The prescription
we use has been derived from SED-fit $M/L$s; the scatter around this
relation is on the order of 0.1 dex.  By comparison, the precision of
SED-fit $M/L$ determinations is limited to 0.2 dex by degeneracies
between, \eg, age, metallicity, and dust obscuration \citep[see,
\eg,][]{PozzettiEtAl, ConseliceEtAl}.  Thus, the increase in the
random error in $M_*$ due to the use of color-derived rather than
SED-fit stellar masses is $\sim 10$ \%.  

In addition to being both simpler and more transparent, however, the
use of color--derived $M/L$s has the major advantage of using the same
restframe information for all galaxies, irrespective of redshift.
This is especially important when it comes to the comparison between
the high-z and low-z samples, where the available photometry samples
quite different regions of the restframe spectrum.  Since Equation
\ref{eq:bellmass} has ultimately been derived from SED-fit mass
estimates, however, our color--derived mass estimates are still
subject to the same systematic uncertainties.  To the extent that such
systematic effects are independent of color, redshift, etc., they can
be accommodated within our results by simply scaling our limiting
mass.  On the other hand, if there is significant evolution in the
color--$M/L$ relation with redshift, then there is the risk that the
use of color-derived $M/L$s may introduce serious systematic errors
with redshift.  We investigate this issue further in
\textsection\ref{ch:masses}, in which we also demonstrate that our
results are essentially unchanged if we use conventional SED-fitting
techniques to derive $M/L$s.

\subsection{The Propagation of Redshift Errors} \label{ch:derived}

\begin{figure*}[t]
\centering \includegraphics[width=16cm]{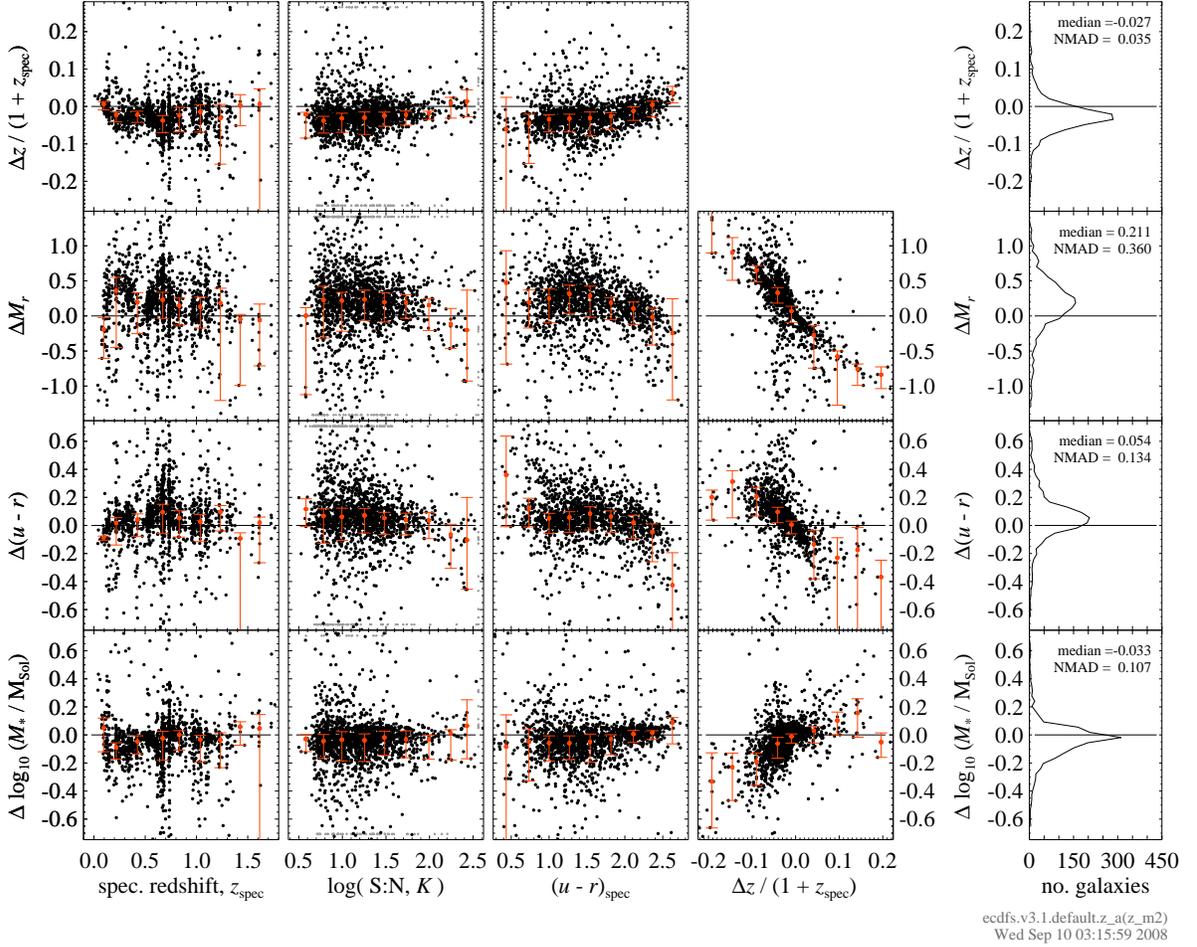}
\caption{Photometric redshift errors, and their effect on other
  derived quantities---In each panel, the abscissa shows the
  difference in a derived quantity, derived assuming the spectroscopic
  or photometric redshift; in all cases, `$\Delta$' should be
  understood as the difference between the $\zphot$-- minus
  $\zspec$--derived values.  We show: ({\em top to bottom, in rows})
  redshift, absolute magnitude, restframe color, and stellar mass, as
  a function of ({\em left to right, in columns}) redshift, observed
  signal--to--noise, restframe color, and photometric redshift error.
  The black points are for the spectroscopic sample shown in Figure
  \ref{fig:specz}; the red points show the median offset in bins; the
  error bars reflect the 15/85 percentiles.  In each panel, points
  that fall outside of the plotted range are shown as small grey
  plusses.  The separate panels at right show the distribution of
  $\Delta$s for all galaxies in the $\zspec$ sample; the median and
  scatter (NMAD) in the difference between $\zphot$-- and
  $\zspec$--derived quantities are as shown.  In all cases, the
  clearest systematic effects are as a function of redshift, with some
  systematic errors for the reddest and bluest galaxies.  The way that
  redshift errors propagate mean that the uncertainties {\em due to
  redshift errors} are much smaller for stellar masses than they are
  for either absolute magnitudes or restframe colors.
  \label{fig:derived} }
\end{figure*}

A primary concern in this paper is the importance of systematic
errors.  To address this concern in the context of our photometric
redshift determinations, we show in the top panels of Figure
\ref{fig:derived} the $\zspec$--$\zphot$ agreement as a function of
redshift, S:N in the $K$ `color' aperture, and restframe color.  In
each panel, the points with error bars show the median and 15/85
percentiles in discrete bins.


Looking first at our photometric redshifts: the first panel of Figure
\ref{fig:derived}, shows the photometric redshift error, $\Delta z /(1
+ \zspec)$, as a function of $\zspec$.  We see that there is a
systematic offset between $\zphot$ and $\zspec$ for $\zspec \lesssim
1$, such that our $\zphot$s tend to be slightly too low (see also
Figure \ref{fig:specz}); for $z \gtrsim 1$, this effect appears to be
less.  At least for $K$ S:N $\gtrsim 10$, random errors in the
photometric redshifts do not appear to be a strong function of
S:N.

There is a clear systematic effect as a function of restframe color.
For galaxies redder than ($u-r$) $\approx 2$ (approximately the lower
limit for $z \approx 0$ red sequence galaxies), the agreement between
$\zphot$ and $\zspec$ is very good.  For galaxies with $(u-r) \lesssim
2$, however, it seems that we systematically underestimate the true
redshift by approximately $\Delta z \lesssim 0.02 (1 + \zspec)$.  It
is plausible that this is in fact the driver of the weak apparent
systematic effect with redshift, coupled with there being a greater
proportion of blue galaxies in the spectroscopic redshift sample at
lower redshifts.

\vspace{0.2cm}

How do these errors in redshift estimation play out in the derivation
of restframe properties?  We address this question with reference to
the lower panels of Figure \ref{fig:derived}, which illustrate how
redshift uncertainties affect our derivation of three basic restframe
quantities ({\em top to bottom}): absolute luminosity, restframe
color, and stellar mass.  In each panel, we plot the difference
between the values derived adopting the spectroscopic or photometric
redshift, as a function of ({\em left to right}) redshift, $K$ band
S:N, and restframe color, as well as photometric redshift error.

Quantitatively, for our $\zspec$ comparison sample, the random
photometric redshift error of $\Delta z/(1 + \zspec) = 0.035$
translates into a 0.360 mag error in absolute magnitude, 0.134 mag
error in restframe color, and a 0.107 dex error in stellar mass.  (By
contrast, the typical uncertainty in for a $K$ S:N = 10 galaxy is
$\Delta K$ = 0.12--0.16 mag $\approx$ 0.05--0.07 dex.)  Just as for
the redshifts themselves, the clearest systematic effects in the
derived quantities is with restframe color: there appear to be mild
systematics with redshift for $z \lesssim 1$, and no clear trend with
S:N, at least for S:N $>$ 10.

It is straightforward to understand why redshift errors play a larger
role in the derivation of magnitudes rather than colors.  When
calculating magnitudes, the primary importance of the redshift is a
distance indicator.  For the $\zspec$ sample shown in Figure
\ref{fig:derived}, the random scatter in $\Delta M_r$ due to distance
errors alone (calculated by taking the difference in the distance
modulus implied by $\zphot$ versus that by $\zspec$) is 0.28 mag; \ie\
$\sim 75$ \% of the scatter seen in Figure \ref{fig:derived}.  On the
other hand, colors are distance independent, and this element of
uncertainty is cancelled out.

What is surprising is the relative insensitivity of our stellar mass
estimates to redshift errors.  Focusing on the right panels of Figure
\ref{fig:derived}, it can be seen that where the photometric redshift
underestimates the true redshift/distance, we will infer both too
faint an absolute luminosity and too red a restframe color.  When it
comes to computing a stellar mass, however, these two effects operate
in opposite directions: although the luminosity is underestimated, the
too--red color leads to an overestimate of the stellar mass--to--light
ratio.  The two effects thus partially cancel one another, leaving
stellar mass estimates relatively robust to redshift errors.

In a photometric redshift survey, the measurement uncertainty {\em due
to random photometric redshift errors} is considerably less for
stellar masses than it is for absolute magnitudes.  This conclusion
remains unchanged using SED--fit stellar masses, rather than our
favored color--derived ones.  Conversely, we can say that (random)
photometric redshift errors are not a dominant source of uncertainty
in our stellar mass estimates.  Indeed, as we have already noted, the
size of these errors is comparable to the uncertainties in our total
flux measurements.


\section{Constructing a \\ Stellar Mass Selected Sample}
\label{ch:masslimit}

\begin{figure*}
\centering \includegraphics[width=8cm]{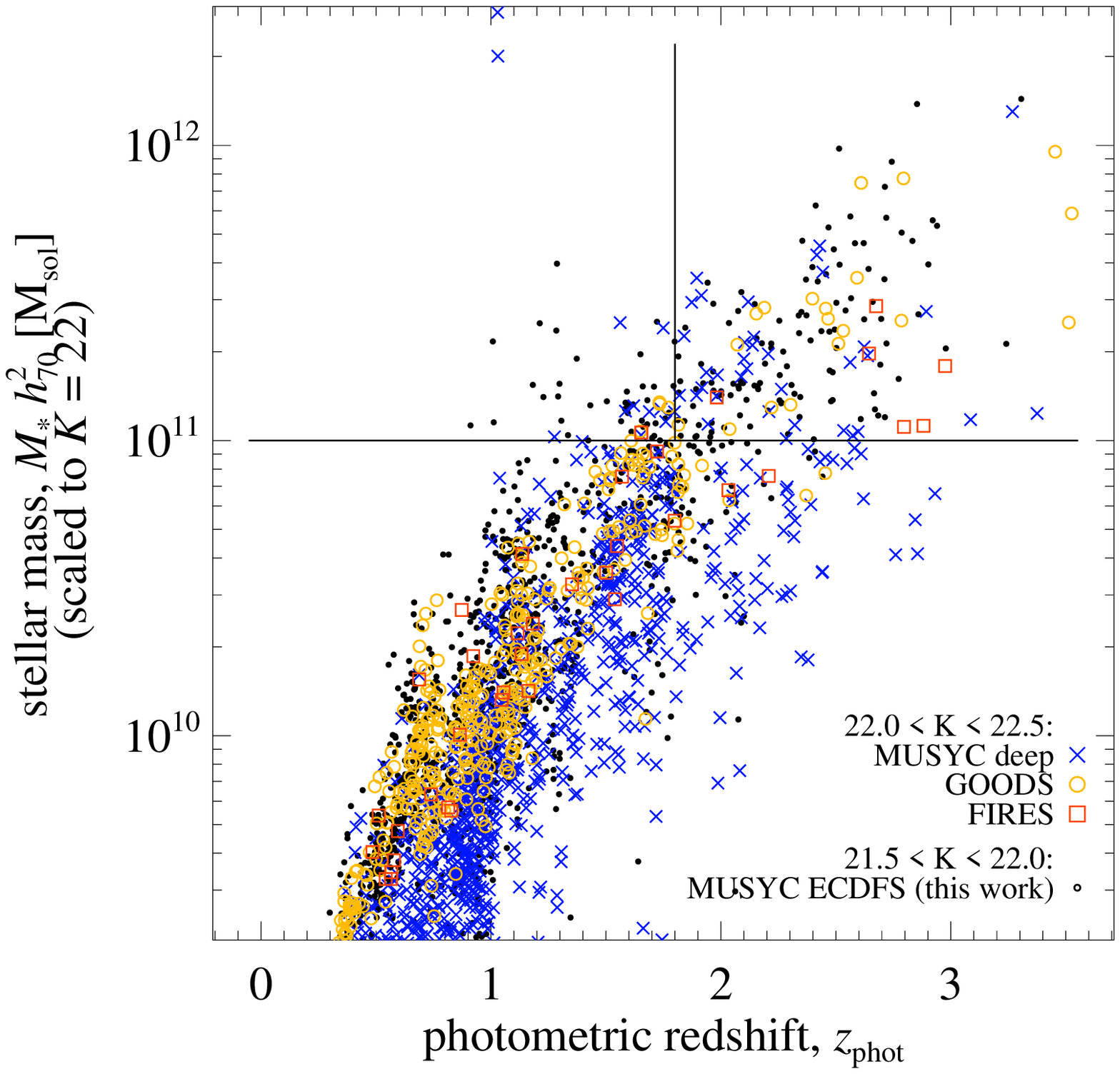}
\includegraphics[width=8cm]{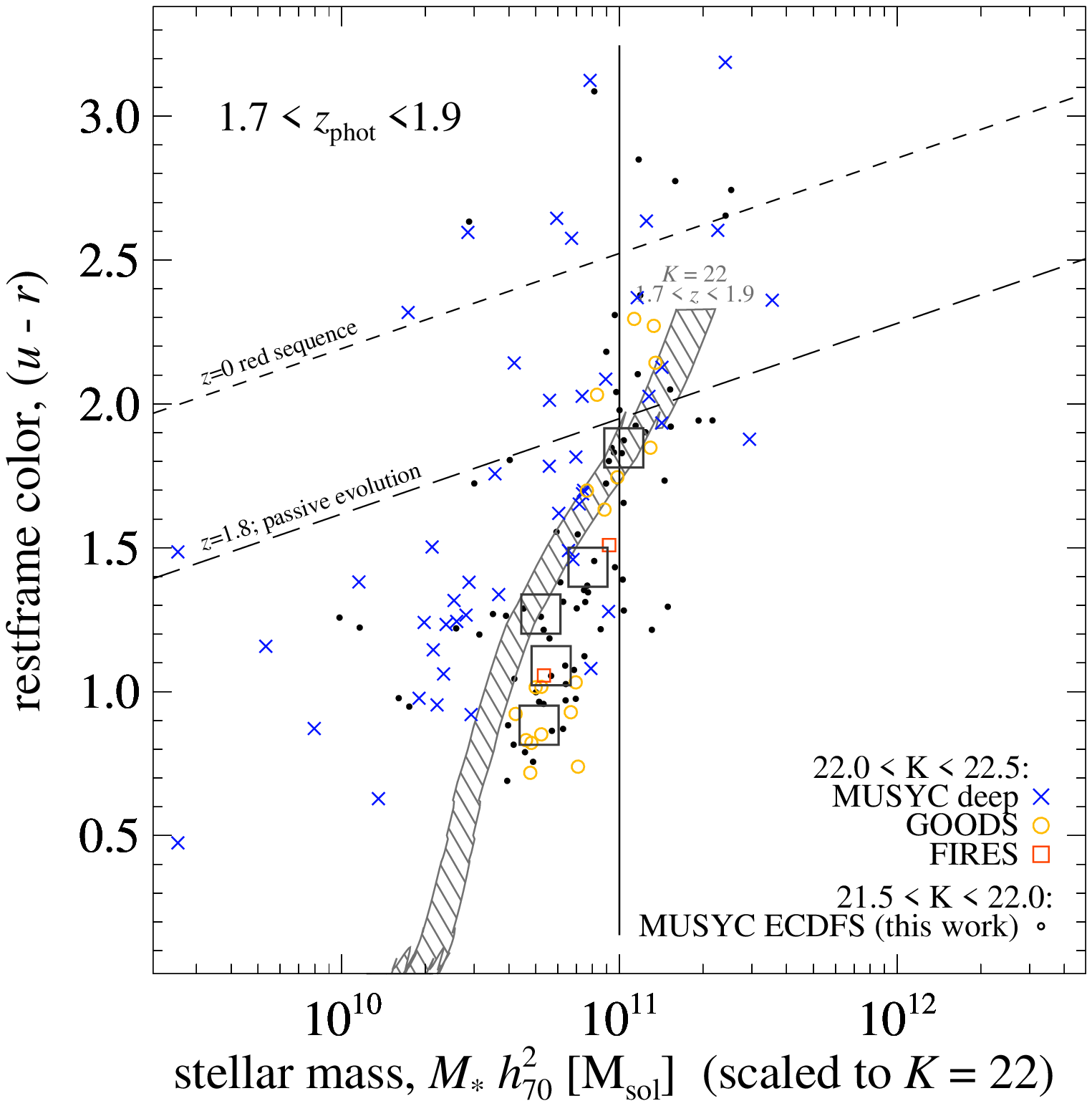}
\caption{Empirically determining our mass completeness limit as a
  function of redshift---{\em Left panel}--- The black points show
  stellar masses for MUSYC ECDFS galaxies with $21.5 < K < 22.0$,
  scaled down in flux to match our $K = 22$ detection limit, and
  plotted as a function of photometric redshift.  The other symbols
  show stellar masses for $22.5 > K > 22.0$ galaxies, scaled up in
  flux to $K = 22$; these galaxies are drawn from the MUSYC deep
  fields ({\em blue crosses}), the FIREWORKS catalog ({\em yellow
  circles}), and the FIRES catalogs ({\em red squares}).  Each sample
  has been analyzed in exactly the same manner.  The upper envelope of
  these points effectively defines, as a function of redshift, the
  limiting stellar mass corresponding to our observed $K$ flux limit.
  For $M_* > \masslimit $ M\sun , we are nearly complete ($\gtrsim 90
  \%$) to $\zphot = 1.8$.  {\em Right panel}---the color--stellar mass
  diagram for $K \approx 22$ galaxies at $1.7 < \zphot < 1.9$---the
  large squares show the median values for the MUSYC ECDFS points,
  binned by color; all other symbols are as in the left panel.  Here,
  the right envelope of the colored points defines our mass
  completeness limit at $\zphot \approx 1.8$ as a function of color.
  For comparison with Figures \ref{fig:cmr} and \ref{fig:mstar}, the
  hatched region shows estimated completeness limits based on
  synthetic SSP spectra.  While we may well miss galaxies considerably
  redder than the predicted red sequence (see
  \textsection\ref{ch:redseq}), at $\zphot \approx 1.8$, this
  empirical argument suggests that we are approximately complete
  ($\gtrsim 85$ \%) for galaxies with $M_* > 10^{11}$ M\sun\ and
  $(u-r) < 2$.
\label{fig:masslimit} }
\end{figure*}

For moderate to high redshifts, NIR selection has the key advantage of
probing the restframe optical light, which is a reasonable tracer of
stellar mass.  In this section, we empirically relate our observed
flux detection/selection limit to an approximate completeness limit in
terms of stellar mass and redshift.

To this end, we have taken galaxies with $K$ fluxes immediately below
our detection limit from three significantly deeper $K$--selected
catalogs; \viz\ the MUSYC deep NIR catalogs \citep{QuadriEtAl},
the FIREWORKS catalog \citep{WuytsEtAl}, and the FIRES catalogs
\citep{LabbeEtAl, NFS}.  By taking objects from these catalogs that lie
immediately below our detection threshold, and scaling their fluxes
(and so stellar masses) to match our $K = 22$ limit, it is then
possible to empirically determine the stellar mass--redshift relation
for $K \approx 22$ galaxies.  The upper envelope of points in
($M_*,~\zphot$) space thus represents the most massive galaxies at our
observed flux limit, and so directly provides a redshift-dependent
mass completeness limit.

This is illustrated in the left panel of Figure \ref{fig:masslimit}.
In this panel, the large, open, colored symbols represent $22.0 < K <
22.5$ objects from the deeper catalogs, scaled up in flux to $K =
22$; \viz\ the MUSYC deep NIR catalogs ({\em blue crosses}), the
FIREWORKS catalog ({\em yellow circles}), and the FIRES catalogs
({\em red squares}).  Again, these points represent objects
immediately at our detection limit; the upper envelope of these points
therefore represents the most massive galaxies that might escape
detection/selection in our analysis.  This suggests that for $M_* >
\masslimit$ M\sun , we are approximately complete for $\zphot < 1.8$.

It is possible to do the same thing using the faintest detections
from our own catalog, scaled down in flux to our selection limit.
Specifically, we have taken galaxies with $21.5 < K < 22.0$ and scaled
their fluxes (and masses) down to $K=22$.  For this test, we also
restrict our attention to galaxies with well constrained redshifts, by
requiring that the EAZY `odds' parameter be greater than 0.95.

These points are shown as the closed circles in the left panel of
Figure \ref{fig:masslimit}.  While the results of this `internal' test
are broadly consistent with the previous `external' one, they do
suggest slightly higher incompleteness.  Of the $21.5 < K < 22.0$
sources with $1.6 < \zphot < 1.8$, 23 \% (20/87) would have $M_* >
\masslimit$ M\sun\ when scaled down to $K = 22$, indicating that our
completeness for $K = 22$, $M_* = 10^{11}$ M\sun\ galaxies is $\sim
75$ \% for $1.6 < \zphot < 1.8$.  However, the $21.5 < K < 22.0$
subsample shown here represents only 30 \% of our full $K < 22$
sample in this mass and redshift range, suggesting that the overall
completeness is more like $> 90$ \%.  

\vspace{0.2cm}

As a second and complimentary check on this conclusion, the right
panel of Figure \ref{fig:masslimit} shows the color--stellar mass
diagram for a narrow redshift slice at $1.7 < \zphot < 1.9$.  Here,
the large squares show the median values from the MUSYC ECDFS points,
binned by color; the other symbols are the same as in the other panel
of this Figure.  As before, the points in this panel represent objects
at our detection limit; the right envelope of these points thus
describes our mass completeness limit for $\zphot \approx 1.8$, this
time as a function of restframe color.

Again, the `internal' and `external' analyses broadly agree.  For blue
galaxies, both tests suggest that MUSYC should be very nearly complete
for $M_* > \masslimit$ M\sun\ and $\zphot < 1.8$.  For $(u-r) > 1.5$
galaxies, however, the down--scaled MUSYC points again suggest
slightly lower completeness than those scaled up from deeper
catalogs: 45 \% (13/29) of these galaxies would fall foul of one of
our selection criteria if their masses were scaled down to $M_* =
\masslimit$ M\sun .  Using an argument analogous to that above, this
suggests that our completeness fraction for galaxies with $M_* >
\masslimit$ M\sun , $\zphot = 1.8$, and $(u-r) > 1.5$ is at least
$\sim 85$ \%.

\vspace{0.2cm}


\begin{figure*}
\centering \includegraphics[width=16cm]{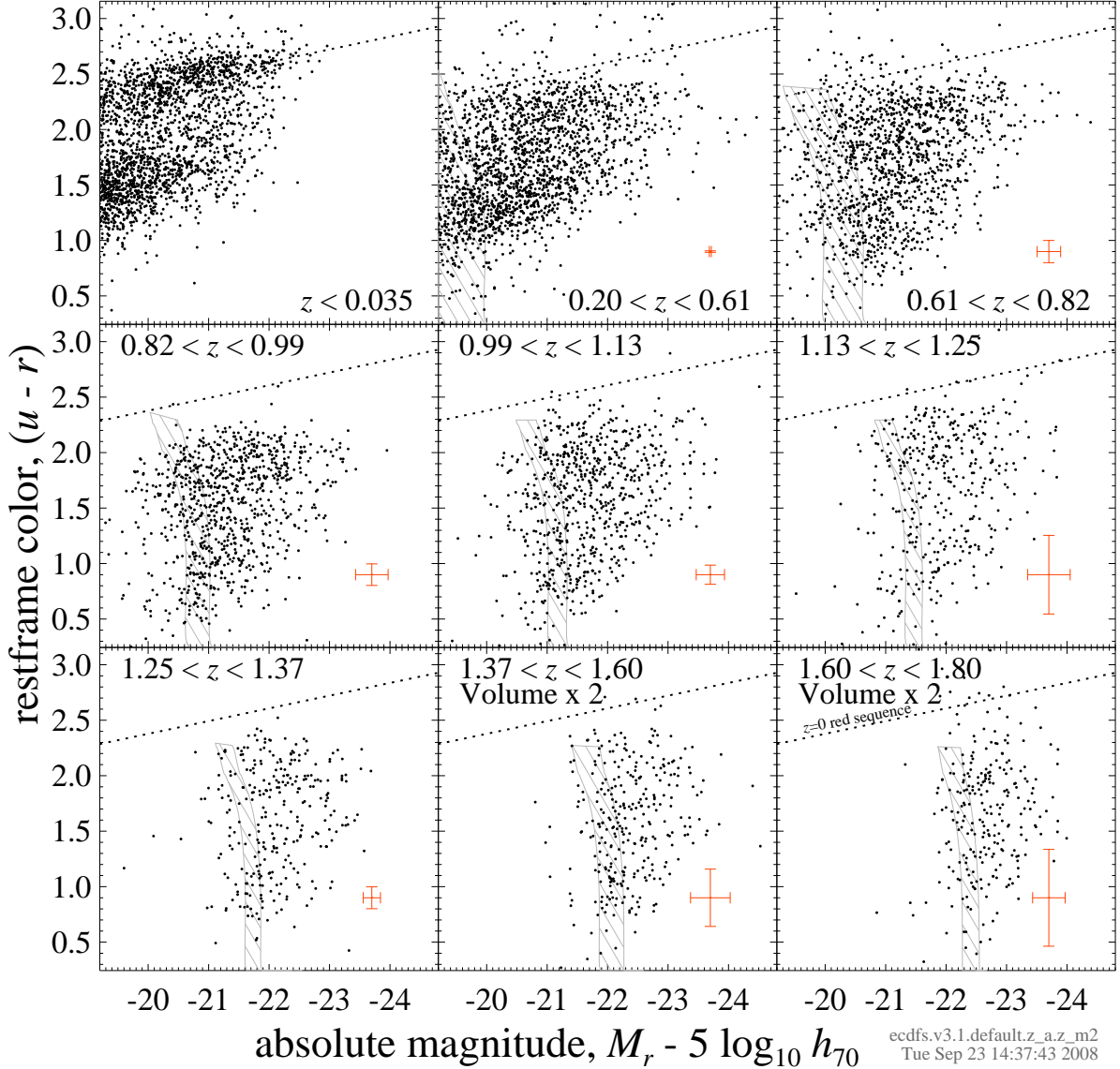}
\caption{The color--magnitude diagram (CMD) for galaxies with $\zphot
  \lesssim 2$ --- The first panel shows a random selection from the
  NYU VAGC's `low-z' sample, based on DR4 of the SDSS
  \citep{BlantonEtAl-lowz}, discussed in Appendix \ref{ch:z=0}; the
  other panels show the MUSYC ECDFS data, discussed in the main text.
  Except where marked, bins are of equal comoving volume; for the
  low-z sample, we have plotted a random sub--sample to yield the same
  effective volume: the density of points is thus directly related to
  bivariate comoving number density.  The shaded area shows the
  approximate $K = 22$ detection/selection limits, based on synthetic
  spectra for an SSP; the error bars show representative errors for a
  $M_* \approx 10^{11}$ M\sun , $(u-r) \approx 2.0$ galaxy at the mean
  redshift of each bin.  The dotted line in each panel shows our fit
  to the CMR for bright, red sequence galaxies at $z \approx 0$,
  derived in Appendix \ref{ch:z=0}.  In this work, we prefer to use
  stellar mass, rather than absolute magnitude, as a basic
  parameter---accordingly, we focus our attention on the CM$_*$D,
  presented in Figure \ref{fig:mstar}.  \label{fig:cmr}}
\end{figure*}

We therefore adopt $M_* > \masslimit$ M\sun\ and $\zphot < 1.8$ as our
approximate completeness limits, corresponding to our $K < 22$
detection/selection limit.  In \textsection\ref{ch:undetected}, we
apply simple completeness corrections to determine the extent to which
our results may be affected by incompleteness.
As a final caveat, however, there remains the concern of additional
incompleteness due to our $K$ S:N $> 5$ criterion, which we will
address in \textsection \ref{ch:sncompleteness}.



\section{The Color--Magnitude and \\Color-Stellar Mass Diagrams for $\zphot \lesssim 2$} \label{ch:results}


\begin{figure*}
\centering \includegraphics[width=16cm]{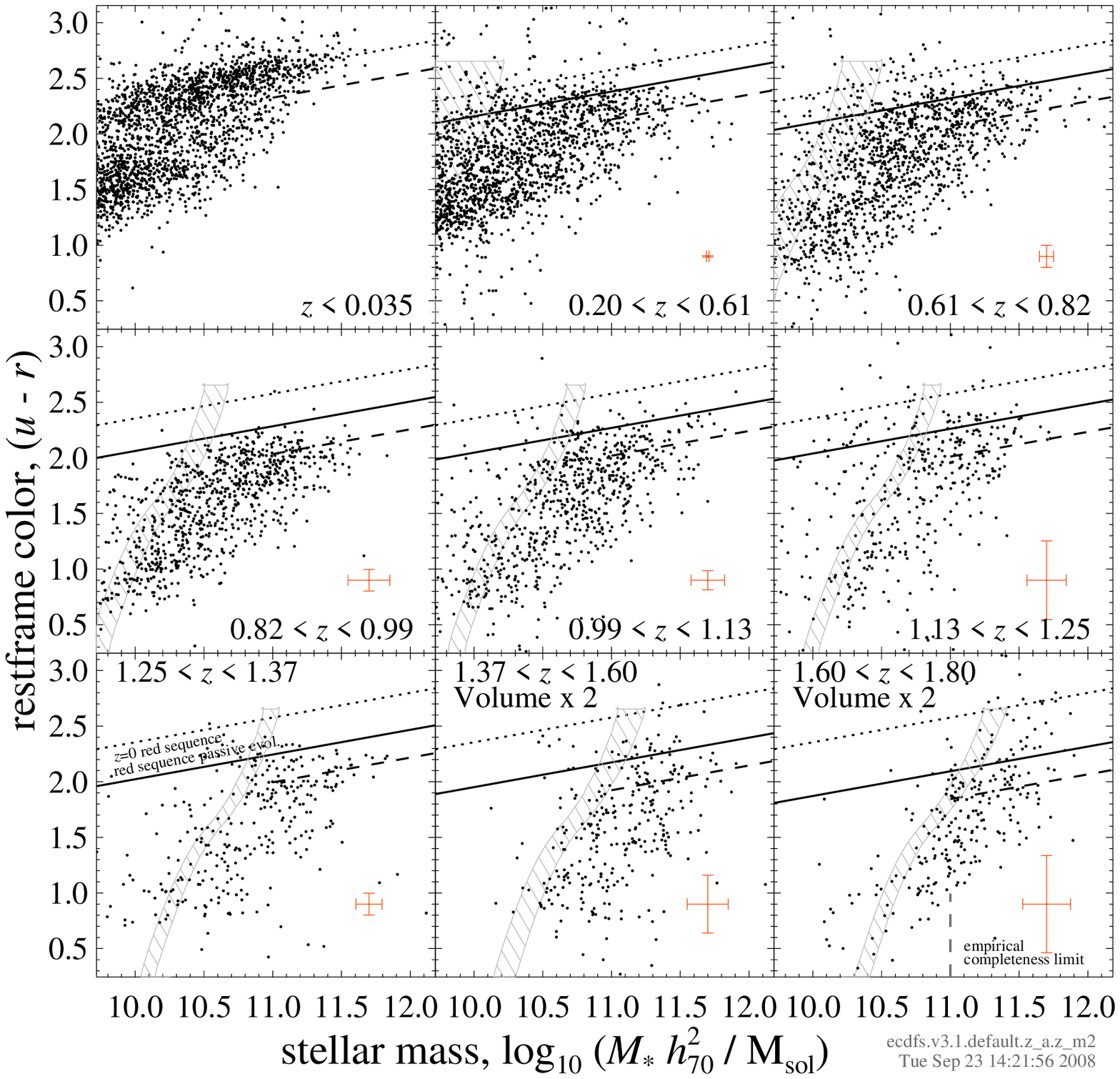}
\caption{The color--stellar mass diagram (CM$_*$D) for galaxies with
  $\zphot \lesssim 2$ --- As in Figure \ref{fig:cmr}, the $z \approx
  0$ bin is based on the low-z sample of SDSS galaxies, discussed in
  Appendix \ref{ch:z=0}; the $\zphot \gg 0$ points based on the MUSYC
  ECDFS data, described in the main text.  The hatched area shows
  approximate selection limits, based on synthetic spectra for an SSP;
  our empirical completeness limit is marked in the last bin.  The
  error bars show representative errors for a $M_* \approx 10^{11}$
  M\sun , $(u-r) \approx 2.0$ galaxy near the bin's mean redshift,
  based on 100 Monte Carlo realizations of the catalog data,
  including photometric redshift errors.  Within each panel, the
  dotted line shows our fit to the CM$_*$R for bright red sequence
  galaxies at $z \approx 0$, derived as per Appendix \ref{ch:z=0}; for
  the $\zphot \gg 0$ bins, the solid lines show our fit to the color
  evolution of the massive red galaxy population, derived in
  \textsection\ref{ch:redseq}; the dashed line shows our red galaxy
  selection criterion, introduced in \textsection\ref{ch:growth}.  We
  analyse the key features of this diagram further in Figures
  \ref{fig:hists}, \ref{fig:redseq}, \ref{fig:redseq2}, and
  \ref{fig:evo}. \label{fig:mstar}}
\end{figure*}


In this section, we present our basic observational results: the
color--magnitude and color--stellar mass diagrams for $\zphot \lesssim
2$.


\subsection{The Color--Magnitude Diagram for $\zphot \lesssim 2$} \label{ch:cmr}

Figure \ref{fig:cmr} shows the color--magnitude diagram (CMD), plotted
in terms of ($u-r$) color and absolute $r$ magnitude, $M_r$, for
$\zphot \lesssim 2$.

The first panel of Figure \ref{fig:cmr} is for $z \approx 0$ galaxies
from the `low-$z$' comparison sample; these data and our analysis of
them are described in Appendix \ref{ch:z=0}.  The basic features of
the CMD --- the red sequence, blue cloud, and green valley --- are all
easily discernible.  The dotted line shows our characterization of the
$z \approx 0$ CMR for red galaxies (Equation \ref{eq:cmr}), also
discussed in Appendix \ref{ch:z=0}.

The other eight panels show the $0.2 < \zphot < 1.8$ MUSYC ECDFS data.
Except for the two highest redshift bins, which are twice as large,
the $\zphot \gg 0$ bins have been chosen to have equal comoving
volume.\footnote{The exact redshift limits we have used are $\zphot =
0.200$, 0.609, 0.825, 0.987, 1.127, 1.254, 1.373, 1.486, 1.595, 1.700,
1.804, 1.906, and 2.000; elsewhere we will round these values as
convenient.}  For the $z \approx 0$ bin, we plot only a random
sub-sample of the low-z catalog, chosen to effectively match the
volume of the higher redshift bins.  Thus, the density of points in
the color--magnitude plane is directly related to changes in the
bivariate comoving number density.

In the bottom-right corner of each panel, we show representative error
bars for a $M_* \approx 10^{11}$ M\sun\ galaxy with $(u-r) \approx
2.0$, near the mean redshift of each bin.  In order to derive these
errors, we have created 100 Monte Carlo realizations of our catalog,
in which we have perturbed the catalog photometry according to the
photometric errors, and repeated our analysis for each: the error bars
show the scatter in the values so derived.  The shaded grey regions
show approximately how our $K < 22$ completeness limit projects onto
color--magnitude space through each redshift bin, derived using
synthetic single stellar population (SSP) spectra.

Examining this diagram it is clear that, in the most general terms
possible, bright/massive galaxies were considerably bluer in the past.
At a fixed magnitude, the entire $z \sim 1$ galaxy population is a few
tenths of a magnitude bluer than at $z \approx 0$.  At the same time,
particularly for $z \gtrsim 1$, there is a growing population of
galaxies with $M_r < -22$ and $(u-r) < 2$ that has no local analogue.
While there are some indications of a red sequence within the $z \gg
0$ data, particularly for $\zphot \lesssim 1$, it is certainly not so
easily distinguishable as locally.


\subsection{The Color--Stellar Mass Diagram for $\zphot \lesssim 2$} 
\label{ch:mstar}

There are a number of advantages to using stellar mass as a basic
parameter, rather than absolute magnitude.  Principal among these is
the fact that stellar mass is more directly linked to a galaxy's
growth and/or assembly: while a galaxy's brightness will wax and wane
with successive star formation episodes, a galaxy's evolution in
stellar mass is more nearly monotonic.  On the other hand, it must be
rememebered that the necessary assumptions in the derivation of
stellar mass estimates produce greater systematic uncertainties than
for absolute luminosities.  As bursts of star formation and other
aspects of differing star formation histories among galaxies are
likely greater at higher redshifts, however, we will focus on the
color--stellar mass diagram in this and following sections.

\begin{figure*}
\includegraphics[width=16cm]{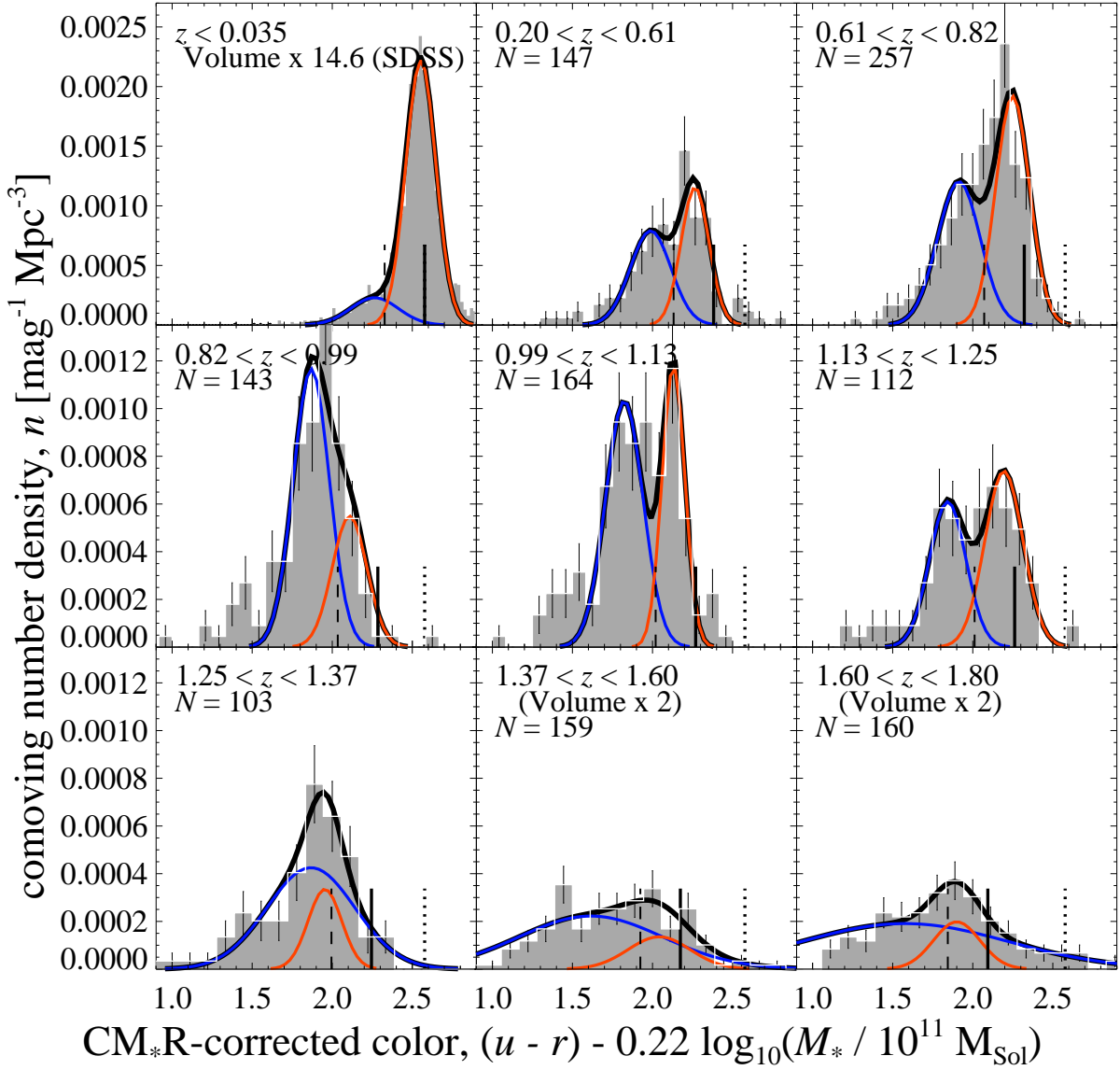}
\caption{Color distributions for $M_* > \masslimit$ M\sun\ galaxies at
  $z \lesssim 2$ --- In each panel, the histograms show the color
  distribution for $M_* > \masslimit$ M\sun\ galaxies, after
  subtracting out the slope of the local CM$_*$R, normalized at $M_* =
  10^{11}$ M\sun .  The shaded distributions show the data themselves.
  The smooth curves show double Gaussian fits to the observed
  distributions.  For $\zphot \gtrsim 1.2$, the Gaussian fits to the
  observed color distributions are not robust; our inability to
  reliably distinguish separate red and blue populations for $\zphot
  \gtrsim 1.2$ is likely due to insufficient S:N in our NIR data (see
  Figure \ref{fig:delur}).  In each panel, the vertical dotted line
  shows the location of the $z \approx 0$ red sequence; the vertical
  solid lines show our fit to the observed color evolution of the red
  sequence, derived in \textsection\ref{ch:redseq}; the dashed lines
  show our red galaxy selection criterion, introduced in
  \textsection\ref{ch:growth}.
 \label{fig:hists} }
\end{figure*}

Figure \ref{fig:mstar} shows the color--stellar mass diagram (CM$_*$D)
for $z \lesssim 2$.  As in Figure \ref{fig:cmr}, the first panel shows
a random sub-sample of the low-z sample; the other panels show the
MUSYC ECDFS data.  The dotted line in each panel shows the $z \approx
0$ color--stellar mass relation (CM$_*$R), which we have derived in
Appendix \ref{ch:z=0}, given in Equation \ref{eq:cmr}.

Each of the basic features of the CMD are also seen in the CM$_*$D.
We see an increasing number of galaxies with $\zphot \gtrsim 1$ and
with $M_* > 10^{11}$ M\sun\ and $(u-r) \lesssim 2$ which have no
analogues in the local universe.  For a SSP, the colors of these
galaxies would suggest ages of $\lesssim 1$ Gyr: these massive
galaxies appear to be in the throes of their final star formation
episodes.  For $\zphot \gtrsim 1.2$, these galaxies may even dominate
the massive galaxy population.  Some evidence for a distinct red
sequence is visible in the CM$_*$D for $\zphot \lesssim 1$, but not
much beyond.

The next three sections are devoted to more quantitative discussion of
each of the following three specific observations:
\begin{enumerate}
\item We see evidence for a red galaxy sequence for $\zphot \lesssim
  1.2$; beyond this redshift, whether due to physical evolution or to
  observational errors, it becomes impossible to unambiguously
  identify a distinct red sequence on the basis of the present data
  (\textsection\ref{ch:bimodality}).

\item At a fixed mass, a red sequence galaxy at $z \sim 1$ is a few
  tenths of a magnitude bluer than its $z \sim 0$ counterpart
  (\textsection\ref{ch:redseq}).

\item At higher redshifts, there appear to be fewer massive galaxies on
  the red sequence.  Further, it appears that the proportion of blue
  cloud galaxies among the most massive galaxies increases;
  conversely, the red fraction is lower at higher redshifts.
  (\textsection\ref{ch:growth}).
\end{enumerate}


\section{The Color Distribution \\of Massive Galaxies for $\zphot < 2$} \label{ch:bimodality}

In an attempt to quantitatively separate the massive galaxy population
into distinct red and blue sub-populations, Figure \ref{fig:hists}
plots the color distribution of the $M_* > \masslimit$ M\sun\ galaxy
population, after subtracting out the slope of the $z \approx 0$
CM$_*$R, and using the same redshift bins as in Figures \ref{fig:cmr}
and \ref{fig:mstar}.  The grey histograms in each panel show the data
themselves.  Note that for the $z \approx 0$ panel of this plot, we
have used the full low-z sample.  Also, recall that for this mass
regime, we are approximately complete (volume limited) to $\zphot =
1.8$.

\subsection{The Massive Galaxy Red Sequence at $\zphot \lesssim 1.2$}

Locally, red sequence galaxies totally dominate the massive galaxy
population: what bimodality exists between the red and blue
populations is weak.  (As the name suggests, `bimodality' implies two
distinct local maxima in the distribution.)  This is a reflection of
the apparent `transition mass' between red and blue galaxies observed
by \citet{Kauffmann2003}; the bimodality is stronger in a luminosity
limited sample, including a greater proportion of bright but less
massive blue cloud galaxies (see also Figure \ref{fig:sdss}; Appendix
\ref{ch:z=0}).  At slightly higher redshifts, where some progenitors
of $z \approx 0$ red sequence galaxies are still forming stars in the
blue cloud, we may then expect the bimodality to actually become
stronger, before weakening again as the fraction of those galaxies
already on the red sequence becomes small at moderate--to--high
redshifts.  In general, however, the color distributions shown in
Figure \ref{fig:mstar} are not clearly bimodal.

With this in mind, as a simple means of separating red from blue
galaxies, we have fit the observed distributions in each redshift bin
with double Gaussian functions.  These fits are shown by the smooth
curves in each panel of Figure \ref{fig:hists}. For the most part,
these fits provide a reasonable description of the $0.2 < \zphot
\lesssim 1.2$ data \citep[see also][]{BorchEtAl}.

\subsection{A Massive Galaxy Red Sequence at $\zphot \gtrsim 1.3$?}

In contrast to lower redshifts, for $\zphot \gtrsim 1.3$, we are no
longer able to reliably fit the color distributions in this manner: on
the basis of the sensitivity tests presented in
\textsection\ref{ch:sens}, neither the red/blue separation nor the
fits to the distributions of these populations is robust.  Looking at
the typical errors shown in each panel of Figure \ref{fig:mstar}, the
measurement errors on the ($u-r$) colors of red galaxies rises sharply
from $\Delta (u-r) \lesssim 0.1$ mag for $\zphot \lesssim 1.2$ to 0.2
mag for $\zphot \sim 1.3$, 0.3 mag for $\zphot \sim 1.5$, and 0.4 mag
for $\zphot \sim 1.7$.  This would suggest that our inability to
reliably distinguish both a red and a blue population at these higher
redshifts may very well be due to the large errors in restframe colors
for higher redshift galaxies.

In order to help interpret our $\zphot \gtrsim 1$ results, we have
tested our ability to recover a single--color galaxy population, given
our observational and analytical errors.  To this end, we have
generated a mock photometric catalog containing only red sequence
galaxies: beginning with our main galaxy sample, we have replaced each
galaxy's photometry using synthetic spectra for a SSP formed over a
short period beginning at $\zform = 5$, assuming the catalog values
of $\zphot$ and $M_*$, and then adding typical MUSYC ECDFS photometric
errors.  We have then re--analyzed this catalog using the same
methods and procedures as for the main analysis, including
re-computing photometric redshifts, restframe colors, and stellar
masses.


\begin{figure}[b] \centering
\includegraphics[width=8cm]{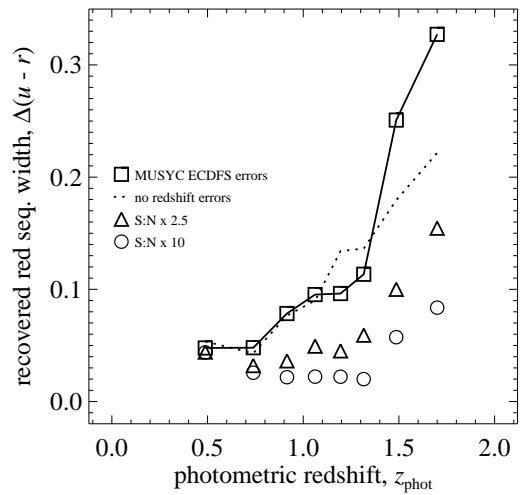}
\caption{The effects of photometric errors on our ability to recover a
  (red) single--color galaxy population---These results are based on
  mock galaxy catalogs containing only passively evolving galaxies,
  taking redshifts and stellar masses of galaxies from the main
  sample, which we have analyzed in the same manner as the actual
  data: we are thus testing our ability to recover a single--color
  galaxy population as a function of redshift.  The squares,
  triangles, and circles show the width of the recovered color
  distributions assuming typical errors for the MUSYC ECDFS data
  divided by 1, 2.5, and 10, respectively; the dotted line shows the
  results assuming MUSYC ECDFS photometric errors, but no redshift
  errors.  Even with spectroscopic redshifts, the depth of the MUSYC
  ECDFS NIR data precludes the detection of a $\zphot \gtrsim 1.4$ red
  sequence.
  \label{fig:delur}}
\end{figure}

The results of these tests are shown in Figure \ref{fig:delur}, which
plots the observed width of the color distribution for this
intrinsically single color population, as a function of photometric
redshifts, assuming photometric errors typical for the MUSYC ECDFS
(squares).  The measurement errors on the $(u-r)$ colors of red
galaxies rise sharply from 0.05---0.07 mag for $\zphot \lesssim 1$ to
0.10 mag for $\zphot \approx 1$, and then continue to increase for
higher redshifts.  In order to demonstrate that this is a product of
photometric errors {\em per se} and not redshift errors, we have also
repeated this analysis holding the redshifts of each object fixed; the
results of this test are shown as the dotted line.  Even with
spectroscopic redshifts, the depth of our NIR data would seem to
preclude the detection of a distinct red sequence at $z \gtrsim 1.3$.

What then would be required in order to confirm the non/existence of a
red sequence at $\zphot \gtrsim 1.5$?  We have also constructed mock
galaxy catalogs with S:N that is 2.5 and 10 times greater than
typical values for the MUSYC ECDFS catalog; the results of these
tests are shown as the triangles and circles, respectively.  Even
pushing a full magnitude deeper, it would be difficult to identify a
red sequence at $z \sim 1.5$, assuming that observational errors of
$\Delta (u-r) \lesssim 0.1$ mag would be required to robustly identify
a red sequence.  In order to probe $z \gtrsim 1.5$, an order of
magnitude improvement is required.  This would suggest that the
detection of a red sequence at $z \gtrsim 1.5$ would require a $J$
band ($5 \sigma$ point source) limit of $\sim 25.8$, roughly the final
target depth for the Ultra Deep component of the UKIRT Infrared Deep
Sky Survey \citep{LawrenceEtAl}.

It is clear from this analysis that we cannot confirm or exclude the
existence of a red galaxy sequence at $z \gtrsim 1.3$ on the basis on
the present data.  This is in good accord with the recent detections
of a $z \lesssim 1.5$ red galaxy sequence by \citet{CirasuoloEtAl} and
Franzetti et al.\ (2007 --- see also Kriek et al.\ 2008).  Moreover,
we note in passing that if we were to subtract away the broadening
effect of photometric errors, as derived from the test described
above, then the implied intrinsic width of the red sequence is
$\approx 0.1$ mag for all $\zphot \lesssim 1.2$, consistent with the
$\zspec < 1.0$ findings of \citet{Ruhland2008}.


\section{The Color Evolution \\ of Massive Red Galaxies} \label{ch:redseq}


\begin{figure}[t] \centering 
\includegraphics[width=8cm]{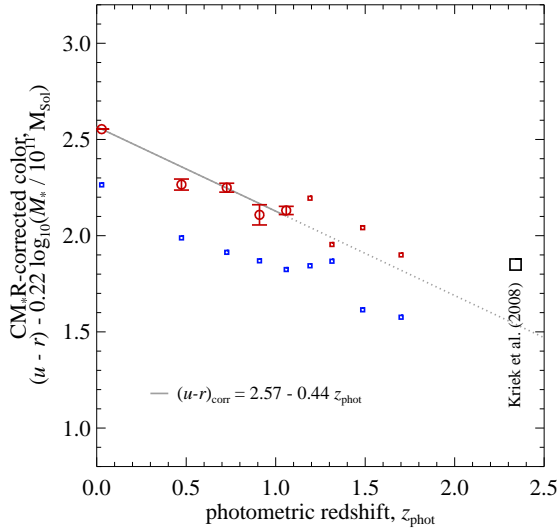}
\caption{The color evolution of massive galaxies for $z \lesssim 2$
  --- Points show the fit centers of the color distributions for the
  red and blue galaxy subpopulations (see Figure \ref{fig:hists}).
  Only the $\zphot \lesssim 1.2$ points ({\em circles with error
    bars}) were used when fitting for the color evolution of the red
  galaxy sequence ({\em solid line}); the errors were derived from
  bootstrap resampling.  The large square at $z = 2$ shows the
  approximate equivalent ($u-r$) color of the $3.3 \sigma$ detection
  of a $(U-B)$ red sequence among $z \sim 2$ galaxies from
  \citet{Kriek2008}, based on NIR spectra.  We see rather a rather
  smooth reddening of the red sequence from $z \sim 1.2$ to the
  present day, which is well described by the linear fit given.
  \label{fig:redseq}}
\end{figure}

\begin{figure}[t] \centering 
\includegraphics[width=8cm]{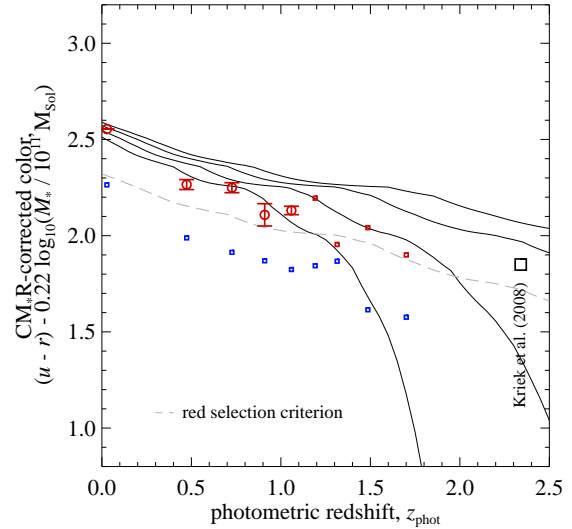}
\caption{Comparing the color evolution of massive galaxies for $z
  \lesssim 2$ to passively evolving stellar population models --- The
  data in this panel are identical to Figure \ref{fig:redseq}.  The
  overlaid curves show simple expectations for the passive evolution
  of a SSP, formed in a short burst ($e$-folding
  time of 100 Myr) beginning at $\zform = 2$, 3, 5, and 10 ({\em
  bottom to top}).  While the $\zform \gtrsim 3$ curves provide a good
  qualitative description of the observed evolution, they have
  substantially lower metallicity than would be expected from the
  local mass-metallicity relation.  Moreover, these very simple models
  make no attempt to account for progenitor bias, and other important
  effects.
  \label{fig:redseq2}}
\end{figure}

Our next task is to quantify the color evolution of the massive red
galaxy population.  We have addressed this question based on the
double Gaussian fits to the (CM$_*$R--corrected) color distributions
of $M_* > \masslimit$ M\sun\ galaxies shown in Figure \ref{fig:hists}.
In Figure \ref{fig:redseq}, we plot the fit centers of the blue ({\em
lower points}) and red ({\em upper points}) galaxy color distributions
as a function of redshift; the errors on the locus of the massive red
galaxy population shown in this Figure have been obtained by bootstrap
resampling.

From this plot, it is clear that the red galaxy population as a whole
has become progressively redder by $\sim 0.4$ mag in $(u-r)$ over the
past 9 Gyr; the evolution in the blue cloud is similar.  Making a
linear fit to the (robust) $\zphot < 1.1$ measurements, we find
$\Delta (u-r)_{\mathrm{corr}} = 2.57 - (0.44 \pm 0.02) ~ \zphot$.
(Note that this fit is constrained to match the $z \approx 0$ point.)
These results do not change significantly if we also include the point
$\zphot \sim 1.2$, but it is clear that if we were to fit to the
$\zphot \gtrsim 1.3$ points, we would find slightly less strong
evolution.



\citet{Kriek2008} report a $3.3 \sigma$ detection of a red sequence in
the spectrally derived $(U-B)$ color distribution of a mass-selected
sample of 36 $\zphot \gtrsim 2$ galaxies, 12 of which lie in the
ECDFS.  The square at $z = 2$ in Figure \ref{fig:redseq} shows the
approximate ($u-r$) color equivalent of their red sequence detection.
While the \citet{Kriek2008} point is slightly redder than an
extrapolation of our linear fit, the two results agree rather well.

In Figure \ref{fig:redseq2}, we compare the observed color evolution
of the red galaxy population with na\" ive expectations from passive
evolution of synthetic spectra.  For this purpose, we have used P\'
egase V2.0 \citep{pegase} models with an initial metallicity of $Z =
0.004$, and assuming a short burst of star formation ($e$--folding
time of 100 Myr), beginning at $\zform = 2$, 3, 5, or 10 ({\em bottom
to top}).

The $\zform \gtrsim 3$ tracks do an adequate job of describing the
amount of observed evolution for $\zphot < 1.3$.  However, to get this
level of agreement, the model (luminosity weighted) metallicities at
$z = 0$ must be roughly solar ($Z \approx 0.2$), whereas the local
mass--metallicity relation would suggest that these abundances should
be super--solar by a factor of 3 or more \citep{TremontiEtAl}.  If we
were to adopt $Z=0.2$ initially, leading to final metallicities that
are super--solar by approximately 50 \%, the model colors would
become nearly 0.5 mag too red.  The problem is even worse for SSP
models: while solar metallicities lead to colors which are too red by
0.25 mag, an acceptable description of the data is only possible
assuming roughly half solar abundances.  (See also Bell et al. 2004.)

Nevertheless, these simple models do provide an acceptable description
of the rate or amount of color evolution of the red galaxy population,
$\Delta (u-r)$, if not the $(u-r)$ colors {\em per se}.  Using these
extremely simple models to interpret the color evolution of the red
galaxy population, the observations suggest that the bulk of the stars
in massive red galaxies were formed at $z \gtrsim 3$; including the
point from \citet{Kriek2008} would suggest a formation redshift
$\gtrsim 5$.


\begin{figure*}[t]
\centering 
\includegraphics[width=18cm]{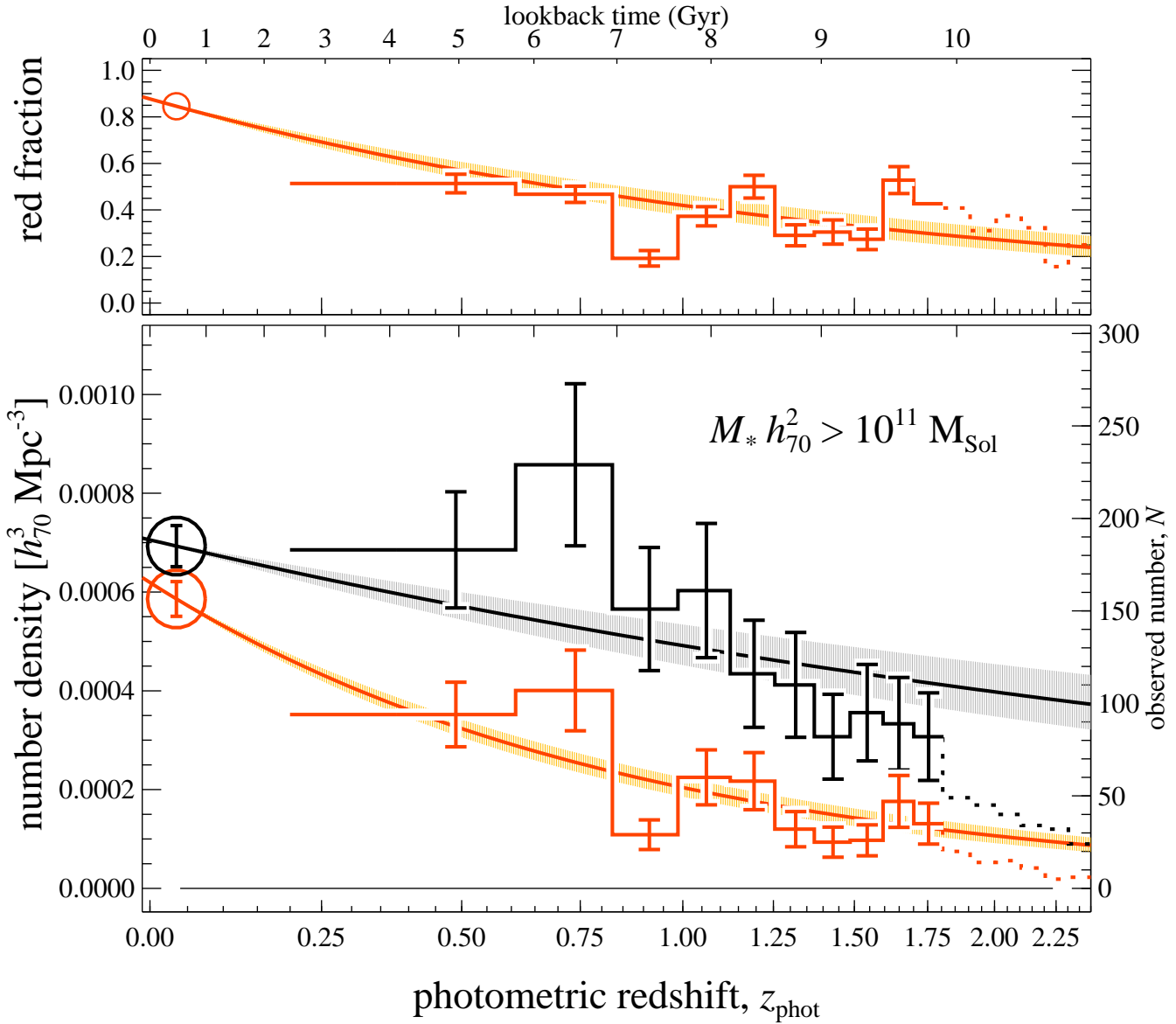}
\caption{The rise of massive, red galaxies over cosmic time---{\em
    Lower panel}: evolution in the number density of all galaxies
    ({\em black histograms}) and of red galaxies ({\em red
    histograms}) with $M_* > 10^{11}$ M\sun\ in the MUSYC ECDFS
    catalog; The error bars shown reflect the expected
    field--to--field variation, calculated as per
    \citet{SomervilleEtAl}.  {\em Upper panel}: evolution in the red
    galaxy fraction among $M_* > \masslimit$ M\sun\ galaxies, as a
    function of photometric redshift.  In this panel, the error bars
    have been derived by bootstrap resampling.  In both panels, the
    dotted histograms show where we are significantly affected by
    incompleteness.  The $z \approx 0$ points ({\em circles}) are
    derived from the low-z sample, discussed in Appendix \ref{ch:z=0}.
    The smooth curves show our fits to the observations; the shaded
    regions show the $1 \sigma$ uncertainties in the fits: we find
    $\ptot = -0.52 \pm 0.12$, $\pred = -1.70 \pm 0.14$, and $\pfrac =
    -1.17 \pm 0.18$.
\label{fig:evo}}
\end{figure*}

This is not to say, however, that the observations are consistent with
all massive red galaxies being formed at $z \gtrsim 5$, or even 3:
even among $M_* > \masslimit$ M\sun\ galaxies, there are simply not
enough stars at $z \gtrsim 2$ to build the $z \approx 0$ red sequence
population \citep[][see also Figure \ref{fig:evo}]{Fontana2006,
ConseliceEtAl}.  Instead, what we see is that the colors of massive
red galaxies are consistent with being dominated by ancient stars at
all redshifts.  This implies both extended star formation histories
among red galaxies (and/or their progenitors), as well as a large
spread in the times at which galaxies of a given (stellar) mass join
the red population --- a kind of long migration of galaxies, occurring
over many Gyr \citep[\cf][]{BrownEtAl2008}.  A proper description of
red sequence evolution would therefore have to account for, among
other things, progenitor bias \citep{VanDokkumFranx}: the continual
skewing of the population by new additions \citep[see
also][]{Faber2007, Ruhland2008}.  This is beyond the scope of this
work.






\section{The Rise of Red Galaxies Over $\zphot \lesssim 2$} 
\label{ch:growth}

In this section, we turn our attention to the third of our basic
results.  Whereas our focus until now has been on the properties of
the red galaxy population as a whole, we now look at how the number of
red galaxies within the total massive galaxy population---\ie\ the red
galaxy fraction---evolves with time.

\subsection{Defining a Red Galaxy Selection Criterion}  \label{ch:redsel}

Our inability to robustly distinguish separate red and blue galaxy
populations on the basis of the observed color distributions for
$\zphot \gtrsim 1.2$ forces us to devise some alternate means of
separating `red' galaxies from the general field population. 

We have already seen that the color evolution of the red galaxy
population is roughly consistent with ancient stars at all redshifts
(Figure \ref{fig:redseq2}).  Our simple solution is therefore to use
the predicted color evolution for a passively evolving stellar
population formed at high redshift to define a redshift-dependent
`red' selection criterion, \viz :
\begin{eqnarray}
  (u-r) > 2.57 &+& 0.24 \times \log _{10} (M_* / 10^{11} \text{M\sun}) 
\nonumber \\
  &+& \delta(\zphot) - 0.25 ~ , \label{eq:redseq}
\end{eqnarray}
where $\delta(\zphot)$ is the $(u-r)$ color evolution of a SSP with
$z_\mathrm{form}=5$, as shown in Figure \ref{fig:redseq2}.  This
selection limit is shown as the dashed lines in each of Figures
\ref{fig:mstar}, \ref{fig:hists}, and \ref{fig:redseq}.

How does this definition of `red' relate to things like membership of
the red sequence, star formation rate and/or history, etc.?  As we
remarked in the first paragraph, a galaxy's optical color is a
reflection of its mean stellar age, modulo the complicating factors of
metallicity and dust extinction.  In addition to `red and dead'
galaxies, therefore, simply selecting `red' galaxies can potentially
catch a significant number of star forming galaxies with high dust
obscuration.  In other words, while all passive galaxies are red, not
all red galaxies are passive.

In this sense, it is not unreasonable to interpret the redshift
evolution of the number and fraction of red, massive galaxies as
placing an upper limit on the numbers of `fully formed' (in the sense
that they have essentially completed their star formation and/or
assembly) massive galaxies.  These results can thus be used to
constrain the epoch at which the star formation quenching mechanism
operates.


\subsection{The Number Density Evolution of Massive Galaxies for $\zphot < 1.8$}

Figure \ref{fig:evo} shows the evolving number density of $M_* >
\masslimit$ M\sun\ galaxies for $\zphot < 1.8$.  As before, the $z
\approx 0$ point comes from our analysis of the low-z sample discussed
in Appendix \ref{ch:z=0}; the histograms are for the main MUSYC ECDFS
sample.  For the $z \gg 0$ galaxies, since we have used bins of equal
comoving volume, the observed numbers ({\em right axes}) can be
directly related to a comoving number density ({\em left axes}),
modulo uncertainties in the cosmological model.  The black
point/histograms refer to the total $M_* > \masslimit$ M\sun\
population; the red point/histograms refer to red galaxies only; the
dotted lines show where our results are significantly affected by
incompleteness.

The error bars on the $z \gg 0$ histograms include the estimated
measurement uncertainty due to field--to--field variation, derived as
in \citet{SomervilleEtAl}, but modified for cuboid rather than
spherical volumes (R Somerville, priv.\ comm.).  For any single
measurement, this is the dominant source of uncertainty: typically
$\sim 30$ \%, as compared to random photometric errors, which are at
the $\sim 10$ \% level.  Note, however, that each of the $\zphot
\gtrsim 1$ bins contains its own `spike' in the $\zspec$ distribution
\citep{FORS2}; the $0.6 < \zphot < 0.8$ bin contains two, and the $0.8
< \zphot < 1.0$ bin none.  Further, note that for $\zphot \gtrsim 1$,
our redshift errors are comparable to the size of the bins themselves;
in this sense, the densities in neighboring bins are correlated, and
the variations due to large scale structure are to a certain extent
masked.

The observed evolution in the number density of massive galaxies is
moderate: the number density of $M_* > \masslimit$ M\sun\ galaxies at
$1.5 < \zphot < 1.8$ is $52 \pm 8$ \% of the $z \approx 0$ value.  We
see a very similar trend in the mass density, although a handful of
galaxies with inferred $M_* \gg 10^{12}$ M\sun\ make this measurement
considerably noisier.

In order to quantify the observed evolution, we have made a parametric
fit to our measured number densities of the form:
\begin{equation}
  n_{\mathrm{tot}}(z) = n_{0} ~ (1 + z)^{\ptot} ~ . \label{eq:ptot}
\end{equation}
In practice, since the dominant source of uncertainty in any single
point is from field--to--field variance, we perform a linear fit in
$\log n$--$\log (1+z)$ space, weighting each point according to its
uncertainty as shown in Figure \ref{fig:evo}.  Further, we constrain
the fits to pass through the $z \approx 0$ point, effectively
eliminating $n_0$ as a free parameter.  In this way, we find $\ptot =
-0.52 \pm 0.12$.  The fit itself is shown in Figure \ref{fig:evo} as
the smooth black curve; the shaded region shows the $1 \sigma$
uncertainty in the fit.

\subsection{Evolution in the Red Galaxy Fraction from $\zphot = 1.8$ to the Present Day}

\label{ch:redfrac}
We now turn our attention to the red galaxy population.  From Figure
\ref{fig:evo}, it is clear that the observed evolution is much
stronger for red galaxies than it is for the total population: the
number density of red galaxies at $1.5 < \zphot < 1.8$ is $18 \pm 3$
\% of the $z \approx 0$ value.  Making a fit to the red galaxy number
densities to quantify this evolution, in analogy to the previous
section, we find $\pred = -1.60 \pm 0.14$ ({\em smooth red curve}).

A complementary way of characterizing the rise of massive red galaxies
is to look at the evolution of the red galaxy fraction.  There are
several advantages to focusing on the red galaxy fraction, rather than
the number density of red galaxies {\em per se}.  First and foremost,
the uncertainty in the red fraction due to large scale structure and
field--to--field variation should be considerably smaller than those
in the number density, especially at $z \gtrsim 1$ \citep{CooperEtAl}.

We show the red fraction as a function of photometric redshift in the
upper panel of Figure \ref{fig:evo}.  Fitting these results ({\em
smooth red curve}) with the same functional form as in Equation
\ref{eq:ptot}, we find $\pfrac = -1.06 \pm 0.16$.  Taken together, the
results encapsulated in Figure \ref{fig:evo} suggest that $\lesssim
20$ \% of $M_* > \masslimit$ M\sun\ galaxies in the local universe
were already on the red sequence by $\zphot \approx 1.6$ (9.5 Gyr
ago).  By the same token, approximately 50 \% of these galaxies only
(re-)joined the red sequence after $z = 0.5$ (5.0 Gyr ago).

\subsection{The Importance of the $z \approx 0$ Comparison Point} \label{ch:z=0impact}

It is clear from Figure \ref{fig:evo} that almost all the signal for
$z \gg 0$ evolution in the red fraction comes from the $z \approx 0$
comparison point \citep[see also][]{BorchEtAl}.  Fitting to the $z \gg
0$ data alone, we find $f_\mathrm{red} = (0.53 \pm 0.05) ~ (1 +
\zphot)^{-0.29 \pm 0.17}$; that is, less than a 2$\sigma$ signal of
evolution.  The same is also true for the number density measurements
\citep[see also][]{BorchEtAl}; the reasons for this are not just the
relatively mild evolution, but also to the relative size of the error
bars on the low-- and high--redshift measurements.  Fitting only to
the $z \gg 0$ points, we find $\ptot = -1.55 \pm 0.84$ and $\pred =
-1.77 \pm 1.84$; these fits `overpredict' the $z \approx 0$ number
densities by factors of 2.2 and 1.3, respectively.

In this sense, then, rather than quantifying the absolute evolution in
the $z \gg 0$ population, we are performing a difference measurement
between the situations at $z \approx 0$ and $z \gg 0$.  For this
reason, it is imperative that care is taken when deriving the $z
\approx 0$ comparison values to ensure that the low-- and
high--redshift samples have been analyzed in a uniform way (Appendix
\ref{ch:z=0}).  

In comparison to more sophisticated analyses by \citet{Bell2003} and
\citet{ColeEtAl}, both of which combine the 2MASS and SDSS datasets,
our $z \approx 0$ number densities are approximately 10 \% higher, and
15 \% lower, respectively.  Adopting these values in place of our own
determinations, we find $\ptot = -0.72 \pm 0.12$ using the
\citet{Bell2003} mass function, and $\ptot = -0.42 \pm 0.12$ using the
\citet{ColeEtAl} mass function, changes of $-0.2$ and +0.1 with
respect to our default analysis.  These rather large discrepancies
underline the importance of uniformity in the analysis of high-- and
low--redshift galaxies.


\subsection{Comparison with Other Works} \label{ch:comparisons}

\begin{figure}[t]  \centering 
\includegraphics[width=8.5cm]{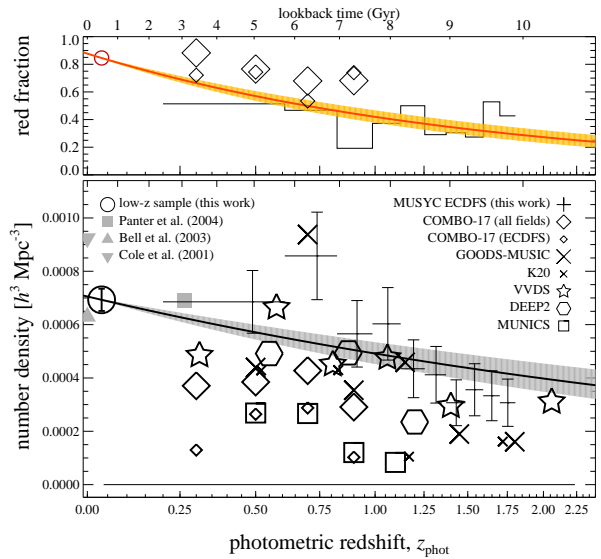}
\caption{Comparison with other works---The different symbols
  correspond to different authors and surveys as marked: the filled
  symbols refer to other authors' analyses of SDSS data; references
  for other $z \gg 0$ surveys are given in the main text.  As in
  Figure \ref{fig:evo}, 
  the smooth curves show
  our fits to the MUSYC ECDFS data; the shaded regions show the
  $1 \sigma$ errors on the fits.  Excepting the MUSYC ECDFS data points,
  all points have been derived by integrating up Schechter function
  fits to the observed mass functions, in redshift bins.  Apart from
  the MUNICS and the COMBO-17 ECDFS data points, there is good
  agreement between the many different groups' results; the cause for
  the discrepancy between the COMBO-17 and MUSYC results in the ECDFS
  is investigated in Appendix \ref{ch:combo}.
  \label{fig:comp}}
\end{figure}

In Figure \ref{fig:comp}, we compare our results to a selection of the
steadily growing number of similar measurements; we show results from:
the COMBO-17 survey \citep{BorchEtAl}, the GOODS-MUSIC catalog
\citep{Fontana2006}, the K20 survey \citep{Fontana2004}, the VVDS
\citep{PozzettiEtAl}, the DEEP2 survey \citep{Bundy2006}, and MUNICS
\citep{Drory2004}.  (Note that all of these results have come within
the last five years.)  In all cases, the points shown in Figure
\ref{fig:comp} have been obtained by integrating up a fit mass
function, taking into account different choices of cosmology and IMF.
We note that the strong redshift spike at $\zphot \sim 0.7$ is also
present in the GOODS-MUSIC results, which are based on the CDFS.

The comparison with the COMBO-17 results deserves some further
comment.  While our results agree reasonably well with the combined
results from three of the COMBO-17 fields ({\em large diamonds}),
there is significant disagreement between our results and the COMBO-17
results in the ECDFS ({\em small diamonds}).  In Appendix
\ref{ch:combo}, we present a detailed comparison between the MUSYC and
COMBO-17 datasets, and demonstrate that the discrepancy between the
two results is due to systematic calibration errors in the COMBO-17
photometry, rather than differences in our analyses.  We note in
passing that these calibration errors have only a small effect on the
COMBO-17 photometric redshifts, and affect the ECDFS only; not the
other COMBO-17 fields \citep{combocalib}.

Particularly given the significant uncertainties in these
measurements, and with the possible exception of the MUNICS results,
the degree of agreement between these surveys is impressive.  However,
in all of these surveys, field--to--field variance is a---if not
the---major source of uncertainty.  More and larger fields are
necessary to lock down the growth rate of massive galaxies.

\section{Quantifying Potential Systematic Errors} \label{ch:sens}

Having now described our experiment in full, in this section we
describe a wide array of sensitivity analyses, in order to determine
how sensitively our results depend on specific choices in our
experimental design.  The basic idea is to vary individual aspects of
our analysis, and see what effects these changes have on the results
presented in \textsection\ref{ch:growth}; in particular, we will focus
on how the best-fit values of $\ptot$ and $\pfrac$ depend on our
experimental design.  The results of many of these tests are
summarized in Tables \ref{tab:zps} and \ref{tab:sens}.  Note that we
have used the same red galaxy selection criterion (Equation
\ref{eq:redseq}) throughout all of these tests.

We will consider, in turn, how sensitive or robust our results are to
basic photometric calibration errors (\textsection\ref{ch:zps}), the
methods used for the photometry itself
(\textsection\ref{ch:photometry}), incompleteness
(\textsection\ref{ch:completeness}), and variations on our photometric
redshift computation (\textsection\ref{ch:photzsens}).  We then
discuss possible systematic effects arising from our method for
estimating stellar masses in \textsection\ref{ch:masses}, and other
concerns in \textsection\ref{ch:others}.  Our findings in the section
are summarized in \textsection\ref{ch:systematics}.


\subsection{Photometric Calibration} \label{ch:zps}

How sensitive or robust are our results to errors in the basic
photometric calibration?  And how accurate are the calibrations of
each of the different bands, relative to one another, and in an
absolute sense?  We address these two questions in this section.


\subsubsection{The Effect of Perturbing Individual Bands}

In order to gauge the effect of calibration errors on our final
results, we have perturbed the photometry in each band in turn by $\pm
0.05$ mag, and repeated our full analysis, from the derivation of
photometric redshifts and stellar masses to fitting the $\gamma$s.
Roughly speaking, these perturbations can affect our results in two
separate ways: either through direct changes in the SEDs themselves,
or indirectly, through changes in the derived photometric redshifts,
and so the transformation from observed to restframe quantities.  In
order to disentangle these two effects, we have also repeated our
analysis with these $\pm 0.05$ mag shifts, but while holding the
photometric redshifts fixed to their default values.  The results of
these tests are summarized in Table \ref{tab:zps}.

For the $UBV$ bands, the effect of changing the photometric
calibration on our main results is driven almost exclusively by
changes in the photometric redshifts.  This is simply because our
stellar mass estimates typically do not depend directly on the
observed $UBV$ photometry: the restframe $B$ has already shifted past
the observed $V$ by $z \approx 0.25$.  (Recall that we use the $B-V$
color to infer $M_*/L_V$, and thus $M_*$.)  Similarly, the effect that
changing the reddest $H$ and $K$ bands has on $\ptot$ is small: $\pm <
0.02$ for both.  The situation changes for the $RIz'J$ bands, where
the direct effect comes to dominate over the indirect effect from
changes in the redshifts.

In terms of the SED photometry, our results are most sensitive to
errors in the $R$ and $J$ band photometric calibrations: a $\pm 0.05$
mag shift in the $R$ or $J$ zeropoint changes the value of $\ptot$ by
$\mp 0.08$ or $\pm 0.14$, respectively.  For the $R$ band, the effect
on restframe properties is focused on the $\zphot \approx 0.7$ bin,
where we happen to measure the highest density; this is also the $z
\gg 0$ bin that has the single greatest effect on our measurement of
$\ptot$.  The critical importance of the $J$ band stems from the fact
that it plays a role in the derivation of restframe $V$ photometry,
and thus $M_*$, for all $\zphot \gtrsim 1$; \ie , roughly two thirds
of our surveyed volume.  It seems plausible that our sensitivity in
this regard might be substantially reduced if we were to use a
different method for estimating stellar masses (see also
\textsection\ref{ch:masses}).

While the SED modeling results are not particularly sensitive to the
$K$ band calibration, it still plays a critical role in normalizing
each SED through the total $K$ flux measurement.  Although changing
the $K$ calibration by $\pm 0.05$ mag effectively rescales the stellar
masses by just $\mp 0.02$ dex, this can change the inferred number
densities by as much as $\mp 10$ \%.  In terms of $\ptot$, the effect
is $\mp 0.07$.


\begin{table*}[t] \begin{center} 
\caption{Photometric Calibration Sensitivity Tests \label{tab:zps}}
\begin{tabular}{ c  r l  r l r l r l c  c }
\hline
\hline
& \multicolumn{2}{c}{$\zphot$ fixed} &
\multicolumn{6}{c}{Recalculating $\zphot$} & \multicolumn{2}{c}{Photometric offset with respect to:} \\
& \multicolumn{2}{c}{change in $\ptot$} &
\multicolumn{2}{c}{change in $\ptot$} &
\multicolumn{2}{c}{change in $\pred$} &
\multicolumn{2}{c}{change in $\pfrac$} &  stellar SEDs & FIREWORKS  \\
Band & +0.05 mag & $-$0.05 mag & +0.05 mag & $-$0.05 mag 
& +0.05 mag & $-$0.05 mag & +0.05 mag & $-$0.05 mag & & \\
(1) & (2) & (3) & (4) & (5) & (6) & (7) & (8) & (9) & (10) & (11) \\
\hline
$U$ & +0.00  & +0.00  & +0.01  &$-$0.01 & +0.02  & +0.02  &+0.001   & +0.041       &$-$0.004 & +0.013 \\
$U_{38}$ & +0.00  & +0.00  & +0.00  &+0.01   & +0.01  & +0.02  &$-$0.002 & +0.009  &$-$0.051 & $-$0.020 \\
$B$ & +0.00  & +0.00  & +0.00  & +0.00  & +0.03  & +0.01  & +0.061  &$-$0.006      &$-$0.017 &  +0.050 \\
$V$ & +0.01  & +0.00  &$-$0.02 & +0.00  & +0.04  &$-$0.10 & +0.037  &$-$0.068      &$-$0.006 &  +0.038 \\
$R$ & +0.05  &$-$0.03 & +0.08  &$-$0.08 & +0.06  &$-$0.05 & +0.071  &$-$0.053      &  +0.017 &  +0.016 \\
$I$ & +0.04  &$-$0.03 & +0.01  &$-$0.02 & +0.03  &$-$0.03 &$-$0.051 & +0.066       &  +0.023 &  +0.055 \\
$z'$ &$-$0.02 & +0.01  &$-$0.03 & +0.02  &$-$0.01 &$-$0.01 &$-$0.011 &$-$0.024     &$-$0.011 &$-$0.004 \\
$J$ &$-$0.10 & +0.09  &$-$0.13 & +0.10  &$-$0.14 & +0.11  & +0.022  &$-$0.033      &  +0.032 &  +0.015 \\
$H$ &$-$0.03 & +0.03  &$-$0.02 & +0.01  &$-$0.02 & +0.05  &$-$0.009 & +0.039       &$-$0.032 &$-$0.012 \\
$K_{\mathrm{SED}}$ & +0.00 & +0.02 &$-$0.02 & +0.01 &$-$0.07 & +0.08 &$-$0.078 & +0.096  & --- & --- \\
\hline
$K_{\mathrm{tot}}$ &$-$0.07 & +0.07 &$-$0.07 & +0.07 &$-$0.03 & +0.03 & +0.057 &$-$0.025 & --- & $-$0.017 \\
\hline
\hline \end{tabular} \end{center}

\tablecomments{For each band (Col.\ 1), Col.s (2)---(9) give how our
  parametric fits to the $\zphot \lesssim 2$ evolution of the $M_* >
  \masslimit$ M\sun\ popuation changes with $\pm 0.05$ mag
  perturbations to individual bands' zeropoints: Col.s (2) and (3)
  show the effect on the total number density evolution of massive
  galaxies due to the change in the photometry only.  Col.s (4) and
  (5) show the same information, but including the effect of changes
  in the photometric redshifts that come with changing the
  calibration; similarly, Col.s (6) and (7) show the effect on the
  number density measurements for red, massive galaxies, and Col.s (8)
  and (9) show the effect n the red galaxy fraction.  Col.s (10) and
  (11) give zeropoint offsets for each band based on stellar colors
  and comparison with the FIREWORKS catalogue \citep{WuytsEtAl},
  respectively.}

\end{table*}

\subsubsection{Testing the MUSYC ECDFS Photometric Calibration} \label{ch:calib}

We now turn our attention to identifying and quantifying potential
calibration errors in the MUSYC ECDFS dataset.  We will then be able
to use this information to estimate the extent to which our results
may actually be affected by such errors.

To address this question, we have used EAZY to fit main sequence
stellar spectra from the BPGS stellar atlas \citep{bpgs}, fixing
$\zphot = 0$, to the observed photometry for stars.  Assuming that
whatever calibration errors do exist do not affect the choice for the
best fit template spectrum, we can then interpret the mean difference
between the best fit and the observed photometry as being the product
of calibration errors.  We note that since this test is concerned
exclusively with SED shapes, it cannot comment on the absolute
calibration of any given band; instead it assesses the relative-- or
cross--calibration of the ten band photometry that makes up the MUSYC
ECDFS dataset.

The offsets we derive in this way are given in Col.\ (10) of Table
\ref{tab:zps}.  The most notable offsets derived in this way are
$-$0.048 in $(U-U_{38})$, $0.033$ in $(I-z')$, 0.064 in ($J-H$), and
$0.032$ in $(J-K)$.  These offsets give an indication of the
plausible size of any calibration errors or inconsistencies in the
MUSYC SED photometry.  If we recalibrate our photometry to eliminate
these apparent offsets (holding the $K$ band fixed), and repeat our
full analysis, we find that the best fit values of $\ptot$, $\pred$,
and $\pfrac$ change by $+0.00$, $+0.02$, and $+0.04$, respectively.

\vspace{0.2cm}

We have also checked the absolute calibration of each band in
comparison to the FIREWORKS catalog \citep{WuytsEtAl} in the region
of overlap (Paper I).  Using $(B-z')$--$(z'-K)$ selected stars, we have
constructed empirical `color transforms' between FIREWORKS and MUSYC
filters, as a function of HST color.  Comparing these diagrams to
predictions derived from the BPGS stellar spectral atlas \citep{bpgs},
we treat any offset between the predicted and observed stellar
sequences as a calibration error in the MUSYC photometry.  (This is
the same method we used to identify the disagreement between the
COMBO-17 and MUSYC calibrations discussed in Appendix \ref{ch:combo}.)

The offsets we have derived in this way are given in Col.\ (11) of
Table \ref{tab:zps}.  The biggest offsets derived in this way are
$\approx 0.05$ mag in ($B$-F435W) and ($I-F850LP$).  The offset in
($K$-$K_\mathrm{ISAAC}$) is $-0.017$ mag; for the crucial $J$ band,
the offset is just $+0.015$ mag.  In agreement with the stellar SED
fitting test, this analysis also finds an inconsistency between the
$I$ and $z$ bands at the level of 0.05 mag.  If we recalibrate our
photometry to match the FIREWORKS catalog, we find that our values
of $\ptot$, $\pred$, and $\pfrac$ change by $-0.03$, $-0.13$, and
$-0.08$ with respect to our default results.  The sizes of these
changes are in excellent agreement with the results of the previous
section; more than half of these changes can be ascribed just to the
0.02 mag rescaling of total $K$ magnitudes.

We estimate that the systematic uncertainty in our main results due to
photometric calibration errors is at the level of $\Delta \ptot
\lesssim 0.05$ and $\Delta \pfrac \sim 0.10$.

\subsection{Photometric Methods} \label{ch:photometry}



While we rely on SExtractor for our basic photometry, we have
introduced three sophistications in our analysis.  In this section, we
investigate the effects that these three changes have on our results.

\subsubsection{Background Subtraction} \label{ch:background}

We have applied a correction to account for the tendency of SExtractor
to overestimate the background level; for individual objects, this
correction is typically on the order of $-$0.03 mag. To test the
sensitivity of our results to background subtraction errors, we have
repeated our analysis without applying this correction.  Repeating our
analysis relying on our own background estimation we find changes in
$\ptot$ and $\pred$ of $-0.06$ and $-0.05$, respectively; the change
in $\pfrac$ is just $-0.02$.

\subsubsection{Total Magnitudes} \label{ch:totalmags}

Our total flux measurements are based on SExtractor's AUTO flux
measurement.  It is well known that the AUTO aperture misses a
significant amount of flux, especially for the faintest sources
\citep{BertinArnouts, LabbeEtAl, BrownEtAl}.  Following
\citet{LabbeEtAl}, we partially redress this by applying a minimal
correction to correct for light laying beyond the AUTO aperture,
treating each object as if it were a point source.  For individual
sources, this correction amounts to as much as 20 \%.  Experiments
with synthetic $r^{1/4}$ sources placed in our own image suggest that
for the specific example of a $K = 22$ elliptical galaxies with $R_e =
0\farcs4$ ($\approx 3$ kpc at $z = 1$), this correction reduces the
missed flux from 0.33 mag to 0.15 mag.

In terms of the measured number densities, the effect of adopting this
correction is 1---5 \% for $\zphot \lesssim 1.5$, rising to 5---10 \%
for $1.5 < \zphot < 1.8$.  The use of uncorrected AUTO fluxes as total
flux measurements thus produces a rather mild spurious evolutionary
signal.  Repeating our analysis without this correction, we find
slightly steeper evolution in the number densities: the values of
$\ptot$, $\pred$ change by $-0.07$ and $-0.06$, respectively.  By
comparison, the red fraction measurements remain almost unchanged:
$\pfrac$ changes by $-$0.01.

\subsubsection{SED Apertures and Colour Gradients} \label{ch:seds}

When constructing multi-color SEDs, we have used the larger of
SExtractor's ISO aperture (based on the $1\farcs0$ FWHM $K$ mosaic)
and a fixed $2\farcs5$ diameter aperture.  This aperture flexibility
is intended to guard against potential biases due to color gradients
in individual galaxies.  This problem is presumably the most severe
for the largest, relatively low redshift galaxies with significant
bulge components, leading to overestimates in both ($B-V$) and $M_*$.

Consistent with this expectation, when repeating our analysis relying
exclusively on the fixed $2\farcs5$ aperture fluxes to construct SEDs,
we find the measured number densities of massive galaxies increases by
$\sim 5$ \% for $\zphot < 1.1$.  The increases in these low redshift
bins brings them closer to the $z \sim 0$ point, leading to a slight
decrease in the measured evolution: $\ptot$ and $\pred$ change by
+0.05 and +0.13, respectively; $\pfrac$ changes by just +0.02.

We estimate that the systematic uncertainty in our results associated
with our photometric methods are on the order of $\ptot \lesssim 0.07$
and $\pfrac \approx 0.00$.




\subsection{Correcting for Incompleteness} \label{ch:completeness}

In \textsection\ref{ch:masslimit}, we have argued that we are
approximately complete for $M_* > \masslimit$ M\sun\ galaxies for
$\zphot < 1.8$.  In this section, we examine the effects of
incompleteness due to both the $K < 22$ detection threshold and the
$K$ S:N $> 5$ `analysis' selection, by deriving simple completeness
corrections.

\subsubsection{Incompleteness Due to Low S:N} \label{ch:sncompleteness}

In addition to incompleteness due to our $K < 22$ detection limit,
discussed below, there is the concern of incompleteness due to the $K$
S:N $> 5$ selection limit, which we have imposed to ensure against
extremely poorly constrained photometric redshifts.  This cut affects
4.5 \% of all $K \le 22$ sources in the ECDFS catalog, with 43 \% of
those galaxies laying in the slightly shallower Eastern pointing.  At
a fixed magnitude, there is not a strong dependence of the fraction of
S:N $< 5$ detections as a function of $J-K$ color or---with the caveat
that this cut is designed to remove poorly constrained redshifts---as
a function of $\zphot$.

In order to assess the impact of this cut on our results, in this
section we attempt to correct for this additional source of
incompleteness.  Our procedure is as follows: we determine the
completeness fraction in the face of this selection, as function of
total $K$ magnitude, $f(K)$; this can then be used to weight each
retained galaxy according to its $K$ magnitude, $w(K)=1/f(K)$.  Of
galaxies in our main sample, 8 \% of all galaxies, and 11 \% of $M_* >
\masslimit$ M\sun\ galaxies at $1 \le \zphot \le 2$ are given $w(K) >
1.25$.

Recalculating our number density measurements using these weights, we
find that the number density of massive galaxies at $1.2 < \zphot <
1.8$ changes by less than 5 \%; the same is true for red galaxies
alone.  Repeating our analysis with these corrections, we find $\ptot
= -0.49$, $\pred = -1.56$, and $\pfrac = -1.05$, amounting to
differences of +0.03, +0.04, and +0.01 with respect to our fiducial
analysis.  We get similar results adopting a more stringent S:N $> 10$
criterion.  To put this change into perspective, it is comparable to
that due to uncertainty in the background subtraction for our basic
photometry.


\subsubsection{Correcting for Undetected Sources}  \label{ch:undetected}

We have corrected for incompleteness due to our $K < 22$ detection
limit as follows: taking the $M_* > \masslimit$ M\sun\ galaxy
population in a given redshift bin, we then predict what the observed
$K$ magnitude for each galaxy would be if the galaxy were shifted
through the next most distant redshift bin.  In other words, we
effectively `K correct' each galaxy's observed $K$ flux from its
fiducial $\zphot$ to higher redshifts, using the same machinery as for
the interpolation of restframe fluxes.  This makes it possible to
determine the fraction (in terms of volume) of the next redshift bin
over which a given galaxy would remain detectable; the overall
completeness is then simply the average of this quantity for all
galaxies in the bin.  We have performed this correction for red and
blue galaxies separately.

Based on this analysis, we are indeed 100 \% complete for $\zphot <
1.25$; for $1.7 < \zphot < 1.8$, we are more than 80 \% complete
overall, and at least 75 \% complete for red galaxies.  This agrees
reasonably well with our completeness estimates in
\textsection\ref{ch:masslimit}.  For higher redshifts, our estimated
completeness drops to 75 \% for $1.9 < \zphot < 2.0$, and to 65 \% for
$2.1 < \zphot < 2.2$.  The estimated incompleteness correction factor
to the measured number densities is thus less than 1.5 for all $\zphot
< 2.2$. Using these completeness corrections to extend our analysis to
$\zphot < 2.2$, we find $\ptot = -0.47 $ and $\pfrac = -0.90$; changes
of $-0.05$ and $-0.16$ with respect to our default values for $\zphot
< 1.8$.

Fitting to these incompleteness--corrected measurements for the number
density of massive galaxies with $\zphot < 1.8$ (\ie , repeating our
default analysis, but with a completeness correction), we find $\ptot
= -0.37$, $-1.42$, and $\pfrac = -1.06$.  In comparison to our
uncorrected results for $\zphot < 1.8$, these represent changes of
$-0.15$, $-0.18$, and $-0.00(3)$, respectively.  However, we also note
that if we were to assume that we are 100 \% complete for $\zphot <
1.4$, these changes become $-0.10$, $-0.12$, and $+0.01$; that is, the
uncertainties on these incompleteness corrections are comparable to in
size to the corrections themselves.

\subsection{Photometric Redshifts} \label{ch:photzsens}

The next major aspect of our analysis that we will explore in detail
is systematic effects associated with the derivation of photometric
redshifts; we split this discussion into two parts.  In the first part
part (\textsection\ref{ch:templates}) we explore how our results
depend on the choice of templates used in the $\zphot$ calculation.
Then, in the second (\textsection\ref{ch:zdissect}), we investigate
how our results depend on the exact method used for deriving
photometric redshifts, by varying individual aspects of the EAZY
algorithm.  Some illustrative results from a selection of these
sensitivity tests are given in Figure \ref{fig:photzplots}.

\subsubsection{Trialing Different Template Sets} \label{ch:templates}


\begin{figure*} \centering
\includegraphics[width=14cm]{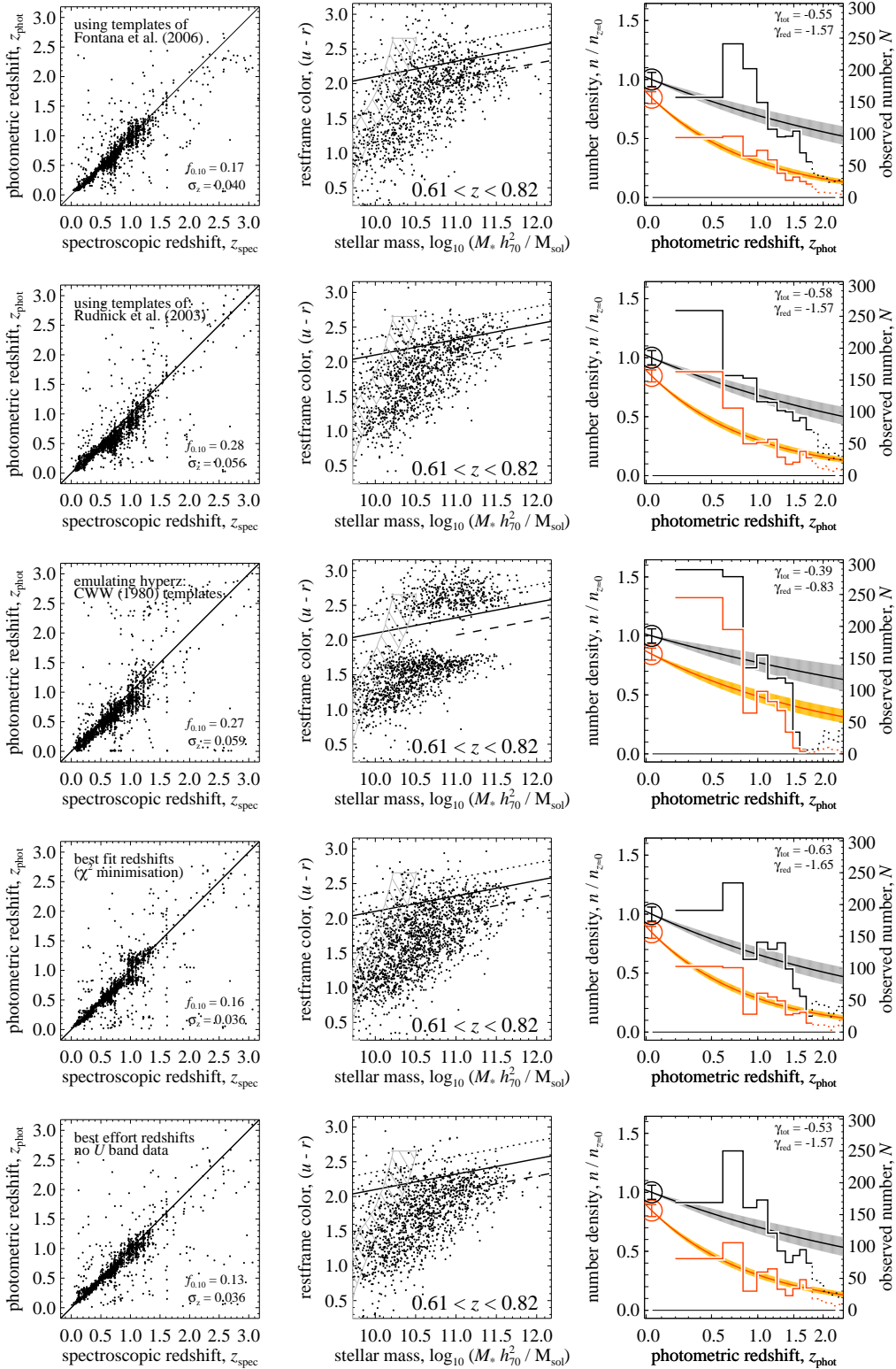}
\caption{The effect of different photometric redshift analyses on our
  results---In rows, we show how our results would change ({\em top to
  bottom}): adopting the template set used by \citet{RudnickEtAl};
  adopting the template set described by \citet{FontanaEtAl2006};
  emulating hyperz in its default configuration; adopting the maximum
  likelihood (\ie $\chi^2$ minimisation) photometric redshift value;
  neglecting the $U$ band photometry.  In each row, we show ({\em left
  to right}) the analogues of Figure \ref{fig:specz}, the third panel
  of Figure \ref{fig:mstar}, and Figure \ref{fig:evo}.
  \label{fig:photzplots}}
\end{figure*}

{\em Using the \citet{Fontana2006} template set}---Using a library of
$\sim 3000$ synthetic P\'egase V2.0 \citep{pegase} spectra described
by \citet{Fontana2006}, \citet{GrazianEtAl} obtain $\sigma_z \approx
0.045$ for $K$ selected galaxies from their catalog of the GOODS-CDFS
data.  In fact, this library provides the parent catalog for EAZY's
default template set \citep[see][]{eazy}.  If we use this template
library in place of the EAZY default, and do not allow template
combination, we find $\sigma_z = 0.039$; the fraction of objects
having $\Delta z/(1+z) > 0.1$, $f_{0.1}$, is 16 \% (\cf\ 12.4 \% for
our default redshifts).  Adopting these redshifts in place of our
default determinations, we find $\ptot = -0.55$ and $\pfrac =
-1.00$; differences of $-0.03$ and $-0.06$, respectively
(see Figure \ref{fig:photzplots}).

{\em Using the \citet{RudnickEtAl} template set}---In the past, our
group has tended to determine photometric redshifts as per
\citet{RudnickEtAl}, which considers linear combinations among: four
empirical template spectra from \citet{cww}; two starburst spectra
from \citet{Kinney}; and two \citet{BruzualCharlot} synthetic spectra
for SSPs with ages of 10 Myr and 1 Gyr.  If we use these templates
with EAZY, we find $\sigma_z$ = 0.055, with $f_{0.1} = 27 $ \%.
Looking at Figure \ref{fig:photzplots}, these photometric redshift
determinations suffer from a number of systematics whereby there are
clear preferred redshift solutions at, \eg , $\zphot \sim 0.4$, as
well as a larger systematic underestimate of the true redshift: in
addition to the larger random error, we also find a systematic offset
in $\Delta z/(1+z)$ of $-0.054$.  The effect this has on the $\zphot
\lesssim 0.8$ number densities is strong, and is dominated by the
shifting of many of the galaxies from the $0.6 \lesssim \zphot
\lesssim 0.8$ bin down into the $0.2 \lesssim zphot \lesssim 0.6$ bin.
The net effect of this change on our final results is considerably
less, however: using these redshifts, we find $\ptot = -0.58$
and $\pfrac = -1.11$; changes of $-0.06$ and $-0.05$ with
respect to our fiducial results.

{\em Emulating hyperz with empirical templates}---As a final test for
this section, we have also repeated our photometric redshift analysis
emulating the popular hyperz \citep{hyperz} code in its default
configuration; that is, $\chi^2$ minimization between the observed
photometry and synthetic photometry using four \citet{cww} empirical
template spectra, and making no allowance for dust extinction beyond
what is included in the empirical spectra.  Using the spectroscopic
redshift sample shown in Figure \ref{fig:specz}, we find $\sigma_z =
0.060$, and again a substantial systematic offset: $\Delta z/(1+z)$ =
$-0.051$.

One very striking consequence of using so few template spectra is a
very pronounced biasing of galaxy restframe colors, corresponding to
the colors of the template spectra themselves (see Figure
\ref{fig:photzplots}).  This of course has a very strong effect on our
stellar mass estimates.  For this reason, it is critical when relying
on photometric redshifts to derive restframe properties to use a
template basis set that spans the full range of galaxy colors.

\subsubsection{Variations in our Default Photometric Redshift Calculation} \label{ch:zdissect}

{\em Adopting the best fit redshift}---By default, EAZY assigns each
object a redshift by taking a probability weighted integral over the
full redshift grid.  However, it also outputs the single most likely
redshift, determined by $\chi^2$ minimization.  For these redshifts,
we find $\sigma_z = 0.035$, and $f_{0.1} = 14.7$ \%.  For $1 < \zspec
< 2$, however, we find $\sigma_z = 0.086$.  We also see evidence for
rather strong systematic effects in the photometric redshifts, such
that there are preferred photometric redshift solutions, for example
at $\zphot \sim 0.8$ and $1.2$ (see Figure \ref{fig:photzplots}),
corresponding to the points where the optical breaks fall between the
observed filters.  With these redshifts, the values of $\ptot$,
$\pred$, and $\pfrac$ change by $-0.11$, $-0.05$, and $-0.14$,
respectively.

{\em No Luminosity Prior}---Our default analysis makes use of a
Bayesian $K$ band luminosity prior.  If we do not include this
feature, we find $\sigma_z = 0.038$ and $f_{0.1} = 13.5$ \%.  The best
fit values of $\ptot$, $\pred$, and $\pfrac$ change by $-0.11$,
$-0.38$, and $-0.54$, respectively.  These changes are large; however,
we note that these results are not internally consistent, in that it
is not true that $\pfrac \approx \pred / \ptot$.  Further, these
results are not consistent with the test we describe in
\textsection\ref{ch:obcol}.

{\em No Template Error Function}---A novel feature of the EAZY code is
the inclusion of a template error function; \ie\ a systematic error,
as a function of restrame wavelength, designed to down-weight those
parts of the spectra like the restframe near-UV where galaxies show an
greater intrinsic variation.  If we do not include this feature, we
find $\sigma_z = 0.034$, and $f_{0.1} = 15.3$ \%; the best fit values
of $\ptot$, $\pfrac$, and $\pfrac$ change by $-0.04$, $-0.14$, and
$-0.13$, respectively.

{\em Two Template Combination}---By default, we allow non-negative
combinations of all six EAZY templates when fitting to the observed
SEDs to determine the redshift.  If we instead allow combinations of
only two (but any two) of the template spectra, we find $\sigma_z =
0.038$ and $f_{0.1} = 14.4$ \%.  The fit values of $\ptot$ and $\pfrac$
change by +0.02 and $-0.02$, respectively. 

{\em `Best Effort' Photometric Redshifts}---It is well known that the
WFI $U$ filter suffers from a red leak beyond 8000 \AA .  We find that
we get our best $\zphot$--$\zspec$ agreement if we ignore the $U$ band
photometry in the computation of photometric redshifts.  For the
spectroscopic sample shown in Figures \ref{fig:specz} and
\ref{fig:derived}, we find $\sigma_z = 0.035$ and $f_{0.1} = 12.0$ \%;
this translates into an uncertainty in the stellar mass estimates due
to redshift uncertainties of just 0.1 dex.  This only modest
improvement in our photometric redshift estimates changes the fit
values of $\ptot$ and $\pfrac$ change by $-0.01$ and $+0.05$,
respectively.

\vspace{0.2cm}

These tests show that our results do depend rather sensitively on the
details of the photometric redshift calculation, typically at the
level of $\Delta\gamma \lesssim$ 0.15.  In particular, the inclusion
of a luminosity prior is crucial in shaping our results.  However, we
note that in most cases there are objective reasons to favor our
default redshifts and/or results.  Similarly, the use of different
template sets changes our results at the level of $\Delta\gamma
\approx 0.06$.  The effect of minor refinements to our photometric
redshift calculation (two template combination versus unrestricted;
the inclusion/omission of the $U$ band data) is small: $\Delta\gamma
\lesssim 0.05$.  Note that, in any case, changing the photometric
redshift calculation cannot change our qualitative results: a gradual
increase in the red galaxy fraction with time.

\begin{table*}[t] \begin{center} 
\caption{Summary of Sensitivity Test Results \label{tab:sens}}
\begin{tabular}{l c c c c c c c }
\hline
\hline
& 
\multicolumn{2}{c}{$1.5 < \zphot < 1.8$} &
\multicolumn{3}{c}{$\zphot < 1.8$} &
\multicolumn{2}{c}{$z \approx 2$} \\
Test description 
& 
$n_{\mathrm{tot}}/n_{\mathrm{0}}$ & $n_{\mathrm{red}}/n_{\mathrm{0}}$ &
$\ptot$ & $\pred$ & $\pfrac$ & 
$n_{\mathrm{tot}}/n_{\mathrm{0}}$ & $n_{\mathrm{red}}/n_{\mathrm{0}}$ \\
(1) & (2) & (3) & (4) & (5) & (6) & (7) & (8) \\ 
\hline
\\ 
{\em Fiducial Results} &&&&&&& \\
Default analysis ($\zphot < 1.8$) 
&  $ 0.52  \pm  0.08 $ &  $ 0.18  \pm  0.03 $ &  $-0.52 \pm 0.12$ & $-1.60 \pm 0.14$ & $-1.06 \pm 0.16$   &  0.57  &  0.15 \\
Correcting for incompleteness ($\zphot < 2.2$)   &  $ 0.57  $ &  $ 0.20  $ &   $-0.47\pm0.09$ & $-1.39\pm0.10$ & $-0.90\pm0.11$   &  0.61  &  0.19
\\
\\ 
{\em Photometric Calibration} (\textsection\ref{ch:calib}) &&&&&&& \\
Recalibration to match stellar SEDs 
&  $ 0.54  $ &  $ 0.18  $ &   $-0.52\pm0.12$ & $-1.58\pm0.14$ & $-1.02\pm0.17$   &  0.58  &  0.16 \\
Recalibration to match FIREWORKS$^{(*)}$ 
&  $ 0.49  $ &  $ 0.15  $ &   $-0.55\pm0.12$ & $-1.73\pm0.14$ & $-1.14\pm0.15$   &  0.56  &  0.13 \\ 
\\ 
{\em Photometric Methods} (\textsection\ref{ch:photometry}) &&&&&&& \\
No correction for background oversubtraction
 &  $ 0.49  $ &  $ 0.16  $ &   $-0.58\pm0.12$ & $-1.65\pm0.14$ & $-1.08\pm0.17$   &  0.54  &  0.15 \\
No correction for missed flux 
&  $ 0.49  $ &  $ 0.16  $ &   $-0.59\pm0.12$ & $-1.66\pm0.14$ & $-1.07\pm0.17$   &  0.53  &  0.14 \\
Fixed $2\farcs5$ apertures for SEDs 
  &  $ 0.52  $ &  $ 0.17  $ &   $-0.47\pm0.12$ & $-1.47\pm0.14$ & $-1.04\pm0.14$   &  0.61  &  0.18 \\
\\ 
{\em Incompleteness Corrections} (\textsection\ref{ch:completeness}) &&&&&&& \\
Correcting for incompleteness ($\zphot < 1.8$) 
&  $ 0.57  $ &  $ 0.20  $ &   $-0.37\pm0.12$ & $-1.42\pm0.13$ & $-1.06\pm0.15$   &  0.67  &  0.19 \\
Correcting for $K$ S:N $<5$ incompleteness 
&  $ 0.54  $ &  $ 0.18  $ &   $-0.49\pm0.12$ & $-1.56\pm0.14$ & $-1.05\pm0.16$   &  0.59  &  0.16 \\ 
\\ 
{\em Using Other $\zphot$ Template Sets} (\textsection\ref{ch:templates}) &&&&&&& \\
Using \citet{RudnickEtAl} templates 
&  $ 0.52  $ &  $ 0.14  $ &   $-0.58\pm0.12$ & $-1.57\pm0.14$ & $-1.11\pm0.15$   &  0.54  &  0.16  \\
Using \citet{Fontana2006} templates$^{(*)}$ 
&  $ 0.53  $ &  $ 0.21  $ &   $-0.55\pm0.12$ & $-1.57\pm0.14$ & $-1.00\pm0.09$   &  0.56  &  0.16 \\
Emulating hyperz with CWW (1980) templates 
&  $ 0.48  $ &  $ 0.24  $ &   $-0.39\pm0.14$ & $-0.83\pm0.15$ & $-0.69\pm0.15$   &  0.66  &  0.35 \\
\\ 
{\em Variations on our Default $\zphot$ Analysis} (\textsection\ref{ch:zdissect}) &&&&&&& \\
Adopting the best fit redshift 
&  $ 0.55  $ &  $ 0.22  $ &   $-0.63\pm0.13$ & $-1.65\pm0.14$ & $-0.92\pm0.14$   &  0.51  &  0.15 \\
No $K$ luminosity prior 
&  $ 0.41  $ &  $ 0.09  $ &   $-0.63\pm0.12$ & $-1.98\pm0.15$ & $-1.59\pm0.22$   &  0.51  &  0.10 \\
No template error function 
  &  $ 0.53  $ &  $ 0.15  $ &   $-0.56\pm0.12$ & $-1.74\pm0.14$ & $-1.19\pm0.20$   &  0.55  &  0.13 \\ 
Two template combinations 
&  $ 0.54  $ &  $ 0.16  $ &   $-0.50\pm0.12$ & $-1.56\pm0.14$ & $-1.08\pm0.15$   &  0.59  &  0.16 \\
Neglecting $U$ data (`best effort' $\zphot$s) 
&  $ 0.46  $ &  $ 0.19  $ &   $-0.53\pm0.12$ & $-1.59\pm0.14$ & $-1.01\pm0.16$   &  0.57  &  0.16 \\
\\ 
{\em Other Variations}  (\textsection\ref{ch:masses} and \textsection\ref{ch:others}) &&&&&&& \\
SED--fit stellar masses 
&  $ 0.48  $ &  $ 0.19  $ &   $-0.49\pm0.12$ & $-1.62\pm0.14$ & $-1.11\pm0.14$   &  0.59  &  0.15  \\ 
Including spectroscopic redshifts 
&  $ 0.51  $ &  $ 0.18  $ &   $-0.54\pm0.12$ & $-1.60\pm0.14$ & $-1.06\pm0.14$   &  0.56  &  0.15 \\ 
Excluding X-ray detections 
&  $ 0.50  $ &  $ 0.18  $ &   $-0.59\pm0.12$ & $-1.63\pm0.14$ & $-1.03\pm0.15$   &  0.53  &  0.15 \\
GOODS area only 
  &  $ 0.47  $ &  $ 0.22  $ &   $-0.63\pm0.17$ & $-1.37\pm0.24$ & $-1.00\pm0.25$   &  0.51  &  0.20 \\
\\
\hline \hline
\end{tabular}
\tablecomments{Each row describes a different sensitivity test,
  designed to determine how and how much our main results depend on
  our choice of experimental design.  These tests are described in
  \textsection\ref{ch:sens}.  Each row gives an identifier (Col.\ 1),
  a different test; references are given for a description of each.
  For the {\em `Dissecting our Default $\zphot$ Analysis} tests,
  additional $\zphot$ information is given in
  \textsection\ref{ch:zdissect}.  For each test, we give the measured
  number density of $M_* > \masslimit$ M\sun\ galaxies with $1.5 <
  \zphot < 1.8$: these values are given for the total massive galaxy
  samples (Col.\ 2), and the red subpopulation (Col.\ 3), and are
  normalised relative to the $z \approx 0$ value for all galaxies.  We
  also give the best fit values for the total $0.2 < \zphot < 1.8$
  evolution in the number density of all massive galaxies (Col.\ 4),
  in the number density of red, massive galaxies (Col.\ 5), and in the
  red fraction among massive galaxies (Col.\ 6).  Finally, we give
  values for the number density of massive galaxies (Col.\ 7) and of
  red, massive galaxies (Col.\ 8), obtained by extrapolating the fits
  to $z = 2$.  Dominant effects are marked with an asterisk; a summary
  of ourinterpretations of these results can be found in
  \textsection\ref{ch:systematics}.}
\end{center} 

\end{table*}



\subsection{Stellar Mass Estimates} \label{ch:masses}

We have relied on a rather simple method for estimating stellar
mass--to--light ratios, based on restframe ($B-V$) colors.  Certainly
this method does not make use of the full amount of information
available in the full, ten-band SED; on the other hand, it does ensure
that the same information is used for all galaxies, irrespective of
redshift.

Nevertheless, we have trialed reanalyzing our data using standard
population synthesis SED fitting techniques to estimate stellar masses
for our main sample, in order to assess the potential scale of biases
arising from this aspect of our experimental design.  Specifically, we
have fit \citet{BruzualCharlot} synthetic spectra to the observed
photometry, with redshifts fixed to the fiducial $\zphot$, using the
hyperzspec utility, which is a part of the hyperz v1.2 release package
(M Bolzonella, priv.\ comm.).  

While the random scatter in comparison to the color-derived stellar
mass estimates is 0.18 dex, after correcting for IMF, cosmology, etc.,
the SED--fit masses are systematically lower by 0.3 dex.  This is also
true for both the FIREWORKS \citep{WuytsEtAl} or GOODS-MUSIC
\citep{GrazianEtAl} catalogs of the CDFS-GOODS data.  Besides this
offset, we do not see evidence for strong evolution in the
normalization of the color relation in comparison to the SED fit
masses \citep[but see][]{LinEtAl}.  Since we have not been able to
identify the source of this offset, we have simply corrected for it.
Note, however, that we have not refit the $z \approx 0$ SEDs for this
test.

Using these SED--fit stellar masses, we find $\ptot = -0.49$ $\pred =
-1.62$, and $\pfrac = -1.11$.  These represent differences of just
+0.03, $-0.02$, and $-0.05$ with respect to our fiducial values.
Similarly, using the prescription for $M/L$ as a function of $(B-V)$
from \citet{Bell2003} rather than that from \citet{BellDeJong}, we
find $\ptot = -0.41$, $\pred = -1.55$, and $\pfrac = -1.12$; changes
of +0.11, +0.05, and $-0.06$, respectively.

This suggests that our results are not grossly effected by our choice
of method for estimating stellar masses, and especially not by the use
of color-- as opposed to SED--derived stellar masses.

\subsection{Other Potential Systematic Effects} \label{ch:others}

{\em Large Scale Structure and Field--to--Field Variance}---One
interesting test of the effect of field--to--field variance is to
restrict our attention only to the MUSYC ECDFS coverage of the
GOODS-CDFS region.  For this 143 $\square$' sub-field, we find $\ptot
= -0.62$, $\pred = -1.37$, and $\pfrac = -1.00$, differences of
--0.10, +0.23, and +0.06 with respect to the full 816 $\square$'
ECDFS.  In comparison to the full ECDFS, the dearth of galaxies at
$\zphot \sim 0.8$ is even more pronounced, and seems to extend over
the range $0.8 < \zphot < 1.2$; this appears to be the main driver of
the strong change in $\pred$.  Conversely, if we exclude the
significantly underdense GOODS area, we find $\ptot = -0.49$,
$\pred = -1.60$, and $\pfrac = -1.10$; differences of
$+0.03$, $-0.00$, and $-0.04$, respectively.  Again, this analysis
underscores the fact that measurements of the red fraction are more
robust against the effects of large scale structure than those of the
number densities themselves.

{\em The $z \approx 0$ value}---In \textsection\ref{ch:z=0impact}, we
have shown that the $z \approx 0$ comparison point is critical in
providing most of the signal for $z \gg 0$ evolution in the massive
galaxy population.  We have also considered  our results vary if we
change the way we treat the $z \approx 0$ point.  In our default
analysis, we constrain the fits so that they pass through the $z
\approx 0$ points, effectively using the $z \approx 0$ point to
normalize the high--redshift measurements.  Since the (Poisson
statistical) errors on the $z \approx 0$ points are just a few
percent, our results do not change significantly if we include the $z
\approx 0$ points in the fits.

For our default analysis, we have approximately accounted for the
effect of photometric redshift errors, as they apply to the $z \gg 0$
sample, on the $z \approx 0$ measurements; we have done this by
randomly perturbing the masses and colors of $z \approx 0$ galaxies
using the typical uncertainties for $z \gg 0$ galaxies, due to the use
of photometric redshifts, and given in Figure \ref{fig:derived}.
Systematic and random errors in the number densities both of all and
of red galaxies are on the order of 5 \%.  If we do not account for
the Eddington bias in this way, we find that $\ptot = -0.46$ and
$\pfrac = -1.16$, changes of 0.06 and -0.10, respectively.

{\em Correcting for Galactic Dust Extinction}---Note that we have not
included specific corrections for Galactic foreground dust extinction.
The CDFS was specifically chosen for its very low Galactic gas and
dust column density; the suggested corrections for the optical bands
are $\lesssim 0.05$ mag.  These corrections are typically as large or
larger than the uncertainties on the photometric zeropoints
themselves; it is for this reason that we have chosen not to apply
these corrections.  If we were to include these corrections, however,
our derived values of $\ptot$ and $\pfrac$ change by $-0.03$ and
$+0.04$, respectively.

{\em Including Spectroscopic Redshift Determinations}---From among the
thousands of spectroscopic redshift determinations that are publicly
available in the ECDFS, we have robust $\zspec$s for just over 20 \%
(1669/7840) of galaxies in our main sample, and for 20 \% (269/1297)
of those with $0.2 < \zphot < 2.0$ and $M_* > \masslimit$ M\sun .
Repeating our analysis with this additional information included, our
results do not change significantly: we find $\ptot = -0.50$
and $\pred = -1.70$, differences of just +0.02 and +0.00,
respectively.  Similarly, we show in Appendix \ref{ch:combo} that our
$z \lesssim 0.8$ results do not change by more than a few percent if
we adopt photometric redshifts from COMBO-17 (Wolf et al. 2004;
$\sigma_z = 0.035$).

{\em Excluding X-ray--Selected Galaxies}---Note that we have made no
attempt to exclude Type I AGN or QSOs from our analysis.  In order to
discover how significant an omission this is, we have repeated our
analysis excluding all those objects appearing in the X-ray selected
catalogs of \citet{Xray} and \citet{TreisterEtAl}.  This excludes
3.2 \% (250/7840) of galaxies from our main sample, and 6.5 \%
(96/1482) from our $M_* > \masslimit$ M\sun\ sample.  Given these
numbers, it is perhaps unsurprising that the exclusion of these
objects does not greatly affect our results: $\ptot$ and $\pfrac$
change by $-0.07$ and $+0.03$, respectively.

\subsection{Quantifying Potential Systematic Errors --- Summary} \label{ch:systematics}

How, and how much, do our results depend on our analytical methods and
experimental design?  We have now described a rather large number of
sensitivity tests, designed both to identify which aspects of our
experimental design are crucial in determining our results, as well as
to estimate the size of lingering systematic errors in our results.
The results of many of these tests are summarized in Table
\ref{tab:sens}.  Clearly, these sorts of tests can provide a
staggering array of `metadata', offering insights to guide not only
the interpretation of the present data and results, but also the
design of future experiments.

As we have shown in \textsection\ref{ch:z=0impact}, the systematic
uncertainties in the $\gamma$s due to discrepancies between the
treatment of $z\approx0$ and $z \gg 0$ galaxies is on the order of
$\pm 0.2$; this is the single greatest potential source of systematic
uncertainty.  Since we have treated these two samples in a uniform
manner, we do not consider this as a lingering source of systematic
uncertainty.

The dominant sources of systematic uncertainty are, in order of
importance: the method of deriving photometric redshifts ($\Delta
\ptot \approx 0.15$, $\Delta \pfrac \approx 0.15$); photometric
calibration errors ($\Delta \ptot \approx 0.05$; $\Delta \pfrac
\approx 0.10$); incompleteness ($\Delta \ptot \approx 0.1$, $\Delta
\pfrac \approx 0$); the choice of photometric redshift template set
($\Delta \ptot \approx 0.06$; $\Delta \pfrac \approx 0.06$);
photometric methods ($\Delta \ptot \approx 0.07$; $\Delta \pfrac
\approx 0$); and stellar mass estimates ($\Delta \ptot \approx 0.05$;
$\Delta \pfrac \approx 0.05$).  We can also, for example, dismiss
incompleteness due to our $K$ S:N criterion, contamination of the
sample by QSOs, and minor details of our photometric redshift
calculation as significant sources of systematic uncertainty.

In summary, we estimate the systematic errors on the measured values
to be $\Delta \ptot = 0.21$ and $\Delta \pfrac = 0.20$; these values
have been obtained by adding the dominant sources of systematic error
in quadrature.  We note that previous studies have not generally taken
(all) sources of systematic error into account in their analysis.


\section{A Final Independent Consistency Check: \\
  Quantifying the Evolution of \\ 
  Bright, Red Galaxies Without Redshifts} \label{ch:obcol}

Our goal in this paper has been to quantify the growth of massive
galaxies in general, and of massive red galaxies in particular, over
the past 10 Gyr, or since $z \sim 2$.  We have now seen---not
completely unexpectedly---that the use of photometric redshifts is
potentially a major source of systematic error in such measurements.
For this reason, in this section we present a complementary and
completely independent measurement which does not rely on redshift
information {\em at all}.


\subsection{Selecting Red Galaxies in Redshift Intervals}
\label{ch:bctracks}

\begin{figure}[t]
\centering \includegraphics[width=8.2cm]{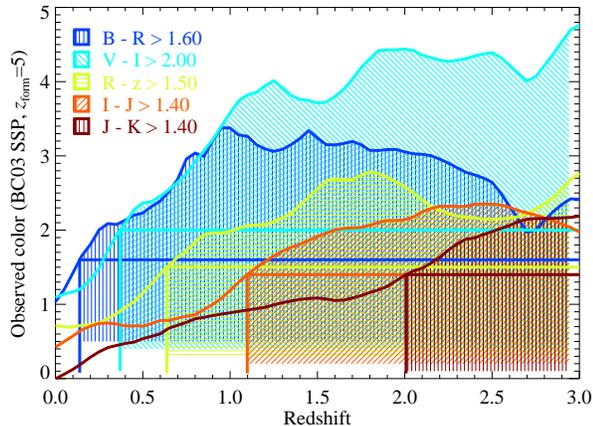}
\caption{Evolutionary tracks, in terms of observed colors, for a
  passively evolving stellar population --- Each track is for a
  synthetic P\'egase V2.0 \citep{pegase} stellar population that
  evolves passively from $\zform = 5$; each of these tracks describe
  an approximate `red envelope' for the true $z \lesssim 2$
  color--redshift distribution.  The sharp rise in each track is
  due to the Balmer and/or 4000 \AA\ breaks becoming redshifted
  between the two observed filters.  The shaded regions show where
  each of the four observed color criteria given will select red
  galaxies ; taken together, they therefore provide a means for
  selecting passive galaxies in rough redshift bins (see also
  \textsection\ref{ch:bcselect}).
\label{fig:bctracks} }
\end{figure}

Figure \ref{fig:bctracks} shows evolutionary tracks for a passively
evolving stellar population, in terms of observed colors in the MUSYC
ECDFS bands.  Specifically, we have used P\'egase V2.0 \citep{pegase}
models with a short star formation episode ($e$-folding timescale of
100 Myr), beginning with $Z = 0.004$ at $\zform = 5$, and ending with
$Z \approx$ Z\sun .  This (approximately) maximally old model
describes a `red envelope' for the colour--redshift relation for
observed galaxies: at a given redshift only extreme dust extinction
can produce observed colors redder than these tracks.


As is evident in this figure, as the Balmer and/or 4000 \AA\ breaks
are first redshifted between the observed bands, there is a sharp rise
in the observed colour --- by selecting objects that are very red in a
certain colour, it is thus possible to select galaxies that lie beyond
a certain redshift.  This is completely analogous to the ERO
\citep[and references therein]{McCarthy} or DRG \citep{FranxEtAl}
selection criteria for red galaxies at moderate--to--high redshifts.
The particular selections we have adopted in Figure \ref{fig:bctracks}
(and given in Col.\ 1 of Table \ref{tab:colresults}) translate to
minimum redshifts of approximately 0.16, 0.40, 0.66, 1.12, and 2.05.

By applying several of these colour criteria in concert, with the
reddest criteria given primacy, it is then possible to select red
galaxies in rough redshift intervals: the highest redshift galaxies
are selected by the criterion $(J - K) > 1.4$; then, the next highest
redshift galaxies are then selected from the remaining set (\ie\
$(J-K) < 1.4$) by the criterion $(I - J) > 1.4$, and so on.  In the
current context, this can be thought of as a two-colour, binned
photometric redshift.

The prevalence of objects selected in this way are given in Col.\ 4 of
Table \ref{tab:colresults}.  To our limit of $K < 22$, we find 386,
655, 690, 304, and 286 objects selected by this tiered set of colour
criteria.


\begin{table}[t]
\caption{Quantifying the evolution of bright, red galaxies without
redshifts \label{tab:colresults}}
\begin{center}
\begin{tabular}{c c c c c r }
\hline
\hline
Color Criterion & 
\multicolumn{2}{c}{Redshift} &
\multicolumn{3}{c}{Prevalence} \\
& $z_{\mathrm{low}}$ & 
$\left<z\right>$ & Observed &  Expected    &  Relative \\
(1) & (2) & (3) & (4) & (5) & (6) \\
\hline
&&&&\\
$B - R > 1.6$  &  0.1 & 0.3  &  0.47 / $\square$''  & 0.76 / $\square$'' & 62 \% \\
$V - I > 2.0$  &  0.4 & 0.6  &  0.80 / $\square$''  & 1.80 / $\square$'' & 45  \% \\
$R - z > 1.5$  &  0.6 & 0.9  &  0.84 / $\square$''  & 1.94 / $\square$'' & 43  \% \\
$I - J > 1.4$  &  1.1 & 1.4  &  0.37 / $\square$''  & 1.29 / $\square$'' & 29   \% \\
$J - K > 1.4$  &  2.0 & 2.6  &  0.35 / $\square$''  & 1.55 / $\square$'' & 23   \% \\
&&&&\\
\hline
\hline
\end{tabular}
\end{center}
\tablecomments{We have used a tiered set of color selection criteria
  (Col.\ 1) to select red galaxies in approximate redshift intervals.
  Here, it is understood that redder criteria are given primacy; for
  example, objects counted in the second row may or may not satisfy
  the first criterion, but do not satisfy any of the three subsequent
  criteria.  We give the minimum and mean redshifts of color-selected
  galaxies from the synthetic, `passive evolution' catalog in Col.s
  (2) and (3), respectively.  The observed prevalence of
  color-selected objects in the MUSYC ECDFS catalog are given in
  Col.\ (4); Col.\ (5) gives the same quantity for the synthetic
  catalog.  The ratio of the two is given in Col.\ (6), which gives
  an estimate of the relative number of (potentially) passively
  evolving galaxies at different redshifts.} \end{table}

\subsection{Interpreting the Numbers of Colour Selected Galaxies}
\label{ch:bcselect}




It is clear that the exact redshift range over which a galaxy might
satisfy a given color criterion depends on that galaxy's SED: as well
as passively evolving galaxies, these selections will also include
galaxies whose red observed colors are due to, \eg , dust obscuration
or considerably higher redshifts.  Thus, the details of the evolving,
bivariate color--magnitude distribution is folded into the numbers of
color-selected galaxies.  For this reason, we are forced to interpret
the numbers of color-selected galaxies with reference to a particular
model for the evolution of red galaxies.

This is done as follows: we construct a mock catalog in which
galaxies' luminosities are distributed according to the $z \approx 0$
luminosity function for red galaxies, which we have determined from
the low-z sample, as analyzed in Appendix \ref{ch:z=0}.  This
luminosity function is nearly identical to that of
\citet{BlantonEtAl-lf}; our results do not change significantly
adopting the red galaxy luminosity functions from either
\citet{Bell2003} or \citet{ColeEtAl}.  Galaxies are placed randomly
(\ie\ uniform comoving density) in the volume $z < 5$.  Next, we
generate synthetic photometry for each object in the catalog using
the set of P\'egase V2.0 \citep{pegase} models shown in Figure
\ref{fig:bctracks}, which are essentially passively evolving from
$\zform = 5$.  We also include typical errors for the MUSYC ECDFS
catalog, to approximately account for photometric scatter.  Finally,
we apply our colour selection criteria to this catalog, exactly as
for the observed ECDFS catalog.  The predicted prevalence of colour
selected galaxies, to be compared to those observed, are given in the
Col.\ 5 of Table \ref{tab:colresults}.




Whereas all red galaxies are assumed to evolve passively in the
synthetic reference catalog, the real color selected samples will
include dusty or high--redshift star--forming galaxies: while all
passive galaxies are red, not all red galaxies are passive.  As we
argue in \textsection\ref{ch:redsel}, the number of color selected
galaxies can therefore be used to place an upper limit on the number
of passively evolving galaxies, since these color criteria should
select a complete but contaminated sample of genuinely passive
galaxies.

The results presented in Table \ref{tab:colresults} thus suggest that
the number density of bright, passively evolving galaxies has
increased by a factor of at least $\sim 2$ since $z \sim 1$.
Moreover, we find that the number density of passive galaxies in the
range $1 \lesssim z \lesssim 2$ is at most $\sim$ 25 \% of the present
day value.  Taken at face value, this would suggest that at most 1/4
present day red sequence galaxies can have evolved passively from $z
\sim 2$; the remainder were still forming, whether through star
formation or by the hierarchical assembly of undetected, and thus
fainter, subunits, or some combination of the two.

Finally, we emphasize the essential simplicity of this analysis, and
thus its suitability for comparisons between different datasets, which
may use different strategies for the computation of photometric
redshifts, restframe colors, stellar masses, and so on.

\subsection{Tying it all Together}

\begin{figure}[t]
\centering \includegraphics[width=8.2cm]{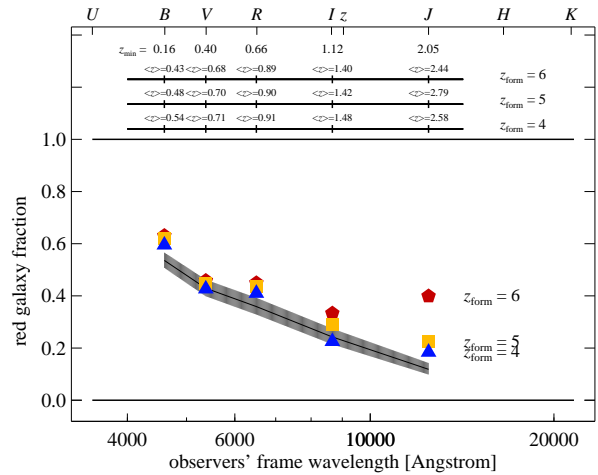}
\caption{Quantifying the number evolution of bright, red galaxies
  without redshifts---The observed numbers of color-selected galaxies,
  relative to the number expected for passive evolution, plotted as a
  function of the effective wavelength of the bluer filter used in the
  selection criterion; the criteria themselves are given in Table
  \ref{tab:colresults}.  The ordinate in this plot is essentially the
  number of galaxies with colors that are consistent with passive
  evolution from $\zform$, relative to the number of $z \approx 0$ red
  sequence galaxies; we have plotted results assuming $\zform$ = 4
  ({\em blue triangles}), 5 ({\em yellow squares}), and 6 ({\em red
    pentagons}).  The mean redshift of color selected galaxies, as
  derived from the synthetic `passive evolution' catalog, does
  depend mildly on the choice of evolutionary scenario.  The mean
  redshift of galaxies selected by each criterion, $\left<z\right>$,
  is given separately for each choice of $\zform$ at the top of the
  Figure.  The minimum redshift is more robust; this is given as
  $z_{\mathrm{min}}$.  Lastly, the solid line and shaded region show
  the fit to the $\zphot \lesssim 2$ evolution in the number density
  of massive red galaxies, derived in \textsection\ref{ch:growth} (see
  Figure \ref{fig:evo}); the shaded region shows the $1 \sigma$
  uncertainty in the fit.  (To put this relation on this plot, we have
  used the $\zform=5$ catalog to map each color criterion to
  $\left<z\right>$.)  The agreement between these two very different
  calculations is impressive.
  \label{fig:obcols} }
\end{figure}

Our final task is to compare the results of this section with those in
\textsection\ref{ch:growth}; we do this in Figure \ref{fig:obcols}.
In this plot, the abscissa is the effective wavelength of the bluer of
the two filters used in each color criterion; the ordinate is the
number of color selected galaxies observed in the MUSYC ECDFS
catalog, relative to the expectation for passive evolution.  The
yellow squares refer to the $\zform = 5$ model shown in Figure
\ref{fig:bctracks} and presented in Table \ref{tab:colresults}.  The
blue triangles and red pentagons show how these results would vary had
we assumed $\zform = 4$ and $\zform = 6$, respectively.

The specific redshift range that is selected by each color criterion
depends on the particular tracks used.  For a given color criterion,
the mean redshift of selected sources (derived from the synthetic
`passive evolution' catalog) thus varies for different choices of
$\zform$: approximate transformations are given for each scenario at
the top of the Figure.

Finally, the solid line shows our fit results for the $\zphot \lesssim
2$ number density evolution of red galaxies, shown in Figure
\ref{fig:evo} .  In order to put this curve on this plot, we have used
the mean redshift of (synthetic, passive) color--selected sources
assuming passive evolution from $\zform=5$.  Given the very different
assumptions and methods lying behind these two results, the agreement
is very impressive.

Both analyses clearly indicate that the number of red galaxies at $z
\approx 0.7$ is approximately 50 \% of the present day value; at $z
\approx 1.5$, it is approximately 25 \%.  In other words, at most,
approximately 1/2 local red sequence galaxies were already `in place'
by $z \approx 0.7$, and have evolved passively since that time; at $z
\approx 1.5$, at most 1/4 were in place.


\section{Summary and Conclusions} \label{ch:conc}


We have presented the color--magnitude (Figure \ref{fig:cmr}) and
color--stellar mass diagrams (Figure \ref{fig:mstar}) for $z \lesssim
2$, based on a sample of $K < 22$ galaxies drawn from the MUSYC
catalog of the ECDFS.  On the basis of the ten band SEDs, we achieve a
photometric redshift accuracy of $\sigma_z = 0.036$ (Figure
\ref{fig:specz}); this figure represents the current state of the art
for broadband photometric lookback surveys.  We have used an empirical
argument to demonstrate that our main galaxy sample is approximately
complete (volume limited) for $M_* > \masslimit$ and $\zphot < 1.8$
(Figure \ref{fig:masslimit}).

Based on the joint color--stellar mass--redshift distribution of this
mass-selected sample, we make the following conclusions:
\begin{enumerate}
\item The color distribution of the massive galaxy population is well
  described by the sum of two separate Gaussian distributions for
  $\zphot \lesssim 1.2$ (Figure \ref{fig:hists}).  Beyond this
  redshift, the depth of our NIR data makes it impossible to identify
  distinct red and blue subgroups from within the general massive
  galaxy population on the basis of the observed color distribution.
  The question as to the existence or otherwise of a $\zphot \gtrsim
  1.5$ red sequence will require data approximately an order of
  magnitude deeper than our own (Figure \ref{fig:delur}).

\item The colors of red sequence galaxies have become progressively
  redder by $\sim 0.5$ mag since $\zphot \approx 1.1$ (Figure
  \ref{fig:redseq}).  Making a linear fit to the observed evolution,
  we find $\Delta (u-r) \propto (-0.44 \pm 0.02) ~ \zphot$.  Simple
  models can only describe the observations assuming sub--solar
  metallicities (Figure \ref{fig:redseq2}): it remains a challenge to
  consistently describe the observed colors and the mass--metallicity
  relation.

\item While the number density of massive galaxies evolves mildly for
  $\zphot \lesssim 2$, we see strong evolution in the red galaxy
  fraction.  That is, the massive galaxy population appears to be {\em
  changing} more than it is {\em growing}.  We have quantified this
  evolution using parametric fits of the form $(1+z)^{\gamma}$.  For
  $\zphot < 1.8$, we find $\ptot = -0.52 \pm 0.12 (\pm 0.20)$, $\pred
  = -1.60 \pm 0.14 (\pm 0.21)$, and $\pfrac = -1.06 \pm 0.16 (\pm 0.21)$.
\end{enumerate}

The systematic errors (given in brackets above) have been derived on
the basis of a whole raft of sensitivity analyses, and are due
primarily to photometric calibration errors and systematic effects
arising from the photometric redshift calculation (\textsection
\ref{ch:sens}).

Finally, in \textsection\ref{ch:obcol}, we showed that these results
are completely consistent with an independent analysis based only on
directly observed quantities; that is, without deriving redshifts,
etc., for individual galaxies.

\vspace{0.2cm}

The two major advances in this work are the quantification of the
$\zphot \lesssim 1.2$ color evolution of the red galaxy sequence as a
whole, and the quantification of the $\zphot \lesssim 2$ evolution of
the red fraction among $M_* > \masslimit$ M\sun\ galaxies.

Knowing that the vast majority of $\masslimit$ M\sun\ galaxies in the
local universe fall on the red sequence (Figure \ref{fig:hists}), we
can identify our massive galaxy sample as the immediate progenitors of
local red sequence galaxies.  Further, by extrapolating the observed
$\zphot < 1.4$ color evolution of the red sequence to higher
redshifts, we can identify our `red' galaxies as those which {\em
potentially} have already found their place on the red sequence: the
high redshift analogues of local red sequence galaxies.  However,
simply selecting `red' galaxies may include a significant number of
galaxies whose red colors are due to heavy dust obscuration, we argue
that our red galaxy measurements provide an approximate upper limit on
the number of passive or quiescent galaxies.

With this assumption, our results suggest that at least 1/2 of all
$M_* \gtrsim \masslimit$ M\sun\ galaxies joined (or rejoined) the red
sequence only after $z \sim 1$, and that at most 1/5 massive galaxies
have resided on the red sequence since $z \sim 2$.  These results
provide new constraints for quenching models, such as quasar and
``radio mode'' AGN feedback \citep{Croton2006, Cattaneo2006}.
Establishing which of our `red' galaxies are genuinely `red and dead'
offers the opportunity to considerably tighten these constraints.

\vspace{0.2cm}

This work was supported through grants by the Nederlandse Organisatie
voor Wetenschappelijk Onderzoek (NWO), the Leids Kerkhoven-Bosscha
Fonds (LKBF), and National Science Foundation (NSF) CAREER grant AST
04-49678.  We also wish to thank the organizers and participants of
the several workshops hosted by the Lorentz Center, where many aspects
of this work were developed and refined.  SW gratefully acknowledges
support from the W M Keck Foundation.



\appendix


\section{The $z=0$ comparison point}
\label{ch:z=0}


In this Appendix, we describe the process by which we have derived the
$z \approx 0$ number density of massive galaxies in general, and of
red galaxies in particular, from the low-z subsample of the New York
University (NYU) Value Added Galaxy Catalog (VAGC) from the Sloan
Digital Sky Survey (SDSS) data release 4 (DR4); this catalog has
been compiled and described by \citet{BlantonEtAl-lowz}.  We will
refer to this sample simply as the `low-z' sample, as opposed to the
`high-z', K-selected sample in the main text.  The overarching concern
is uniformity in the analysis of the low-z and high-z samples, where
possible and appropriate, in order to make as fair as possible a
comparison between the high-z and low-z data.  Our discussion proceeds
in three parts: first, we describe our analysis of low-z galaxies; we
then describe our characterisation of the galaxy red sequence;
finally, we give our derived number density of massive galaxies at $z
\approx 0$, including mimicking the effects of the random photometric
redshift errors typical for the $z \gg 0$ sample.


\begin{figure*}[b]
\centering \includegraphics[width=16cm]{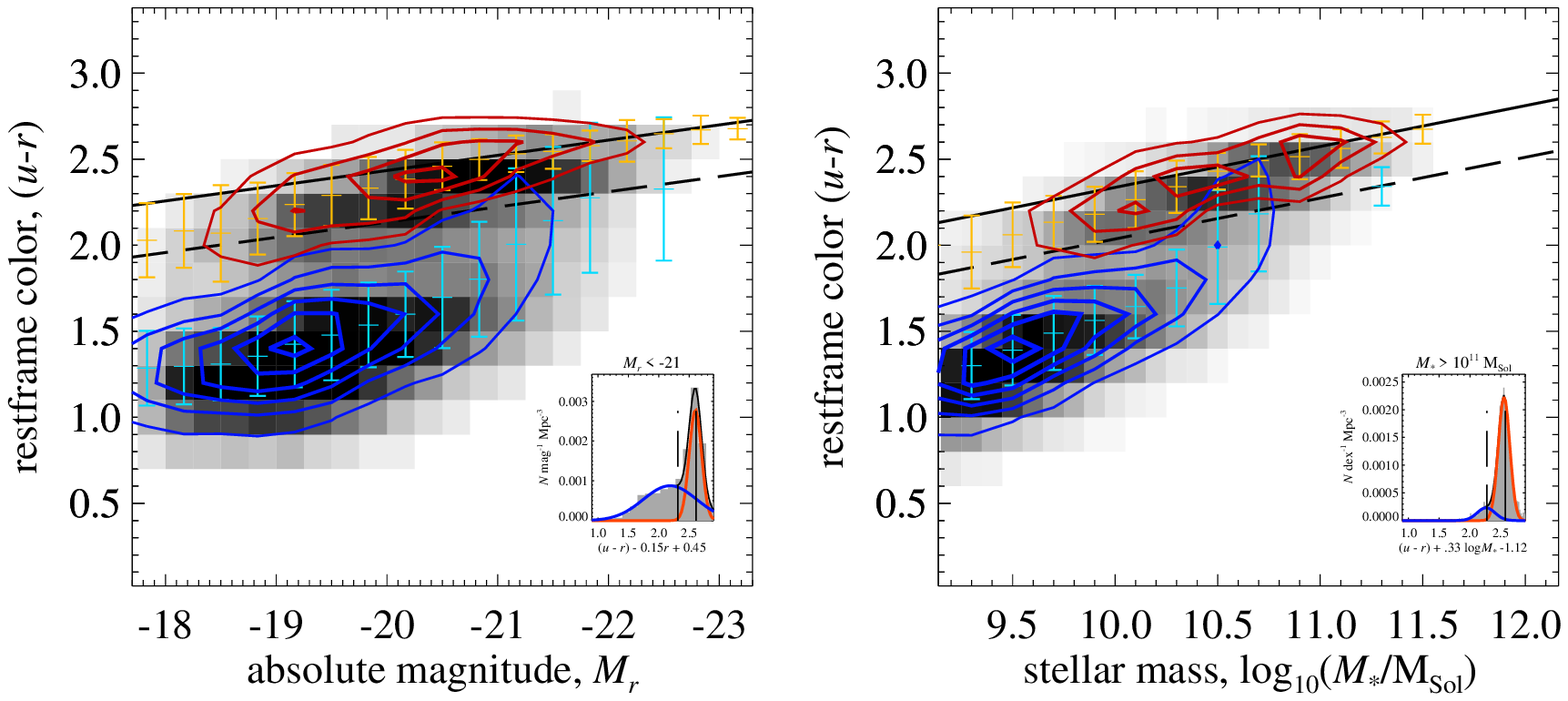}
\caption{The `low z' sample from the NYU VAGC of the SDSS (DR4), used
  to derive the $z \approx 0$ comparison point for the main, `high z'
  sample discussed in the text---Color--magnitude ({\em left}) and
  color--stellar mass ({\em right}) diagrams --- The greyscale shows
  the density of points.  The errors bars show the center and width of
  the color distributions of the red and blue populations, obtained
  from double Gaussian fits in narrow magnitude/mass slices.  These fits
  to the populations are also represented by the logarithmic (0.5 dex)
  contours, to give some idea of the data density of red/blue
  galaxies.
  \label{fig:sdss} }
\end{figure*}


\subsection{Low-z Galaxy Analysis}

For each galaxy in the low-z sample, we have constructed a $ugriz$ SED
using `model' magnitudes (measured by fitting either an exponential or
$r^{1/4}$ profile in each band, with structural parameters derived
from the $i$ band image) as given in the low-z catalog.  In analogue
to high-z galaxies, these SEDs are then scaled up to the `Petrosian'
$r$ magnitude.  The Petrosian apertures are designed to measure a
fixed fraction of a galaxy's light, irrespective of brightness or
distance, but with some dependence on the light distribution:
neglecting the effects of seeing, the Petrosian aperture captures 99
\% of the total light for an exponential profile, compared to $\gtrsim
80$ \% for a de Vaucouleurs profile \citep{BlantonEtAl-lowz}.  We make
no attempt to correct for missed flux, but note that similar effects
are present at a similar level in our own data (see
\textsection\ref{ch:totalmags}).  Following \citep{BlantonEtAl-lowz},
we use the following factors to correct the basic SDSS calibration to
AB magnitudes: (-0.042, +0.036, +0.015, +0.013, -0.002) for
($u,~g,~r,~i,~z$).

We have derived restframe photometry and stellar masses for low-z
galaxies using exactly the same machinery as for the high-z sample,
adopting the heliocentric redshifts given in the low-z catalog.  Our
interpolated restframe $ugriz$ photometry agrees quite well with that
given by \citet{BlantonEtAl-lowz} in the low-z catalog.  (Whereas we
use galaxy color-color relations, based on six empirical galaxy
template spectra, to interpolate restframe photometry,
\citet{BlantonEtAl-lowz} determine restframe fluxes by fitting a
non-negative linear combination of five carefully chosen synthetic
spectra, and then integrating the best fit spectrum.)  On average, in
comparison to \citet{BlantonEtAl-lowz}, our $u-r$ colors are $\sim
0.02$ mag bluer for blue galaxies and $\sim 0.03$ mag redder for red
galaxies.  We calculate the distance modulus using the proper motion
corrected redshifts given in the catalog.  Finally, we have derived
stellar masses based on the interpolated restframe $B$ and $V$ fluxes,
using Equation \ref{eq:bellmass}.


\subsection{The Red Galaxy Sequence at $z=0$}

Figure \ref{fig:sdss} shows our CMD ({\em left}) and CM$_*$D ({\em
right}) diagrams for the general $z \approx 0$ field galaxy
population; in each panel, the logarithmic greyscales show the data
density.  We note that these two plots are very nearly equivalent: the
CM$_*$D can be seen as simply a `sheared' version of the CMD,
with the bluer galaxies dragged further towards the left.  A separate
red galaxy sequence is easily discernible in each panel.

We have characterized this red galaxy sequence, as a function of
magnitude or of mass, by taking a narrow slices in either absolute
magnitude or stellar mass, and fitting double Gaussians to the color
distribution in each bin.  The results of these fits are shown in each
panel of Figure \ref{fig:sdss} as the red and blue contours.  This
gives some indication of the relative numbers of red and blue galaxies
as a function of magnitude/mass.  In order to characterize the
separate red and blue populations, we also show the center and width
of each Gaussian fit as the yellow and cyan points and error bars.

In order to determine the slope of the high luminosity/mass end of the
red sequence CMR and CM$_*$R, we have simply made linear fits to the
centers of the $M_r < -21$ and $M_* > 10^{10.5}$ M\sun\ `red'
distributions.  The results of these fits are:
\begin{equation}
  (u-r) = 2.612 - 0.090 \times (M_r + 22) 
  = 2.573 + 0.237 \times \log _{10} (M_* / 10^{11} \text{ M\sun} ) ~ .
\label{eq:cmr} \end{equation}
These relations are shown as the solid black lines in each panel.
Finally, in order to make some distinction between `red' and `blue'
galaxies which could be easily applied to the high-$z$ sample, we
simply use a cut 0.25 mag bluer than this relation.  In each panel,
this cut can be seen to approximately coincide with the so--called
`green valley' between the red and blue galaxy populations.


Finally, in the inset of each diagram, we show the color distribution
for the brightest ($M_r < -21$) or most massive ($M_* > \masslimit$)
galaxies, after subtracting away the color-magnitude or color-stellar
mass relations given by Equation \ref{eq:cmr}.  (See also
\ref{fig:hists}.)  We also show double Gaussian fits to these
distributions, which are the basis for the $z \approx 0$ points shown
in Figures \ref{fig:redseq} and \ref{fig:redseq2}.  We note in passing
that, at least for $-19 \gtrsim M_r \gtrsim -21.5$, the width of the
color distributions of red and blue galaxies is not a strong function
of magnitude: $\Delta (u-r) \sim 0.12$, and $\Delta (u-r) \approx
0.20$ for red and blue galaxies, respectively.

\subsection{The Number Density of Massive Galaxies at $z \approx 0$}

For our purposes (see \textsection\ref{ch:growth}), the crucial final
quantity is the number density of massive galaxies at $z \approx 0$.
We find that the number density of $M_* > 10^{11}$ M\sun\ galaxies is
$(6.64 \pm 0.38) \times 10^{-4}~ h^3$ Mpc$^{-3}$; for ($u$, $r$) red
galaxies, the number is $(6.07 \pm 0.36) \times 10^{-4}~ h^3$
Mpc$^{-3}$.  This measurement is not significantly affected by either
volume or surface brightness incompleteness.  Compared to the more
sophisticated analyses of \citet{Bell2003} and \citet{ColeEtAl}, our
value for the number density of massive galaxies is 10 \% higher and
25 \% lower, respectively.


As we have repeatedly stressed, our prime concern is uniformity in the
analysis of the low- and high-$z$ samples.  Photometric redshift
errors are a major source of uncertainty for the high-$z$ sample ---
how would comparable errors affect the low$z$ measurements?  Given the
redshift range of our low-$z$ galaxies ($0.003 < z < 0.05$) and the
estimated photometric redshift errors among high-$z$ galaxies ($\Delta
z/(1+z) \sim 0.035$), it would be inappropriate to simply apply typical
redshift errors and repeat our calculations: this would effectively throw
away all distance information.  Instead, what we have done is to apply
the typical uncertainties on $M_r$, ($u-r$) and $M_*$ due to
photometric redshift errors, as shown in Figure \ref{fig:derived}.
For simplicity, we have ignored correlations between these errors.

In line with our earlier findings, including the effects of
photometric redshift errors makes less of a difference to measurements
based on stellar masses than to those based on absolute magnitudes.
Our fits to the red sequence become:
\begin{equation}
  (u-r) = 2.608 - 0.113 \times (M_r + 22) = 2.577 + 0.223 \times \log
  _{10} (M_* / 10^{11} \text{ M\sun} ) ~ .
\label{eq:cmrdelz} \end{equation}
Our measurements of the $z \approx 0$ number density of galaxies more
massive than $\masslimit$ M\sun\ becomes $(6.93 \pm 0.38) \times 10^{-4}~
h^3$ Mpc$^{-3}$ in total, and $(5.86 \pm 0.35) \times 10^{-4}~ h^3$
Mpc$^{-3}$ for red sequence galaxies.  That is, the effects of the
redshift errors expected among $z \gg 0$ galaxies affect the
measurement of the $z \approx 0$ number density by less than 5 \%.

These numbers, approximately accounting for the Eddington bias due to
photometric redshift errors (as they apply to the $z \gg 0$ sample),
are then what have used as a local reference value to compare to the
$z \gg 0$ data.


\section{A Detailed Comparison with COMBO-17} \label{ch:combo}


In the ECDFS, the COMBO-17 and MUSYC datasets represent parallel
analytical paths describing the same physical reality---indeed the
COMBO-17 broadband observations form the foundation of the MUSYC raw
optical data.  Ideally, then, the two surveys' results should agree
identically; instead, in comparison to COMBO-17, the MUSYC results
suggest 40 \% more massive ($M_* > \masslimit$ M\sun ) galaxies at
$0.2 < \zphot < 0.8$---a degree of difference more on a par with that
expected from field--to--field variation.  In this Appendix, we
identify the main cause of this discrepancy.




\subsection{Photometric Redshift Accuracy: MUSYC photometry with COMBO-17 redshifts}

\begin{figure}[b] \centering 
\includegraphics[width=12cm]{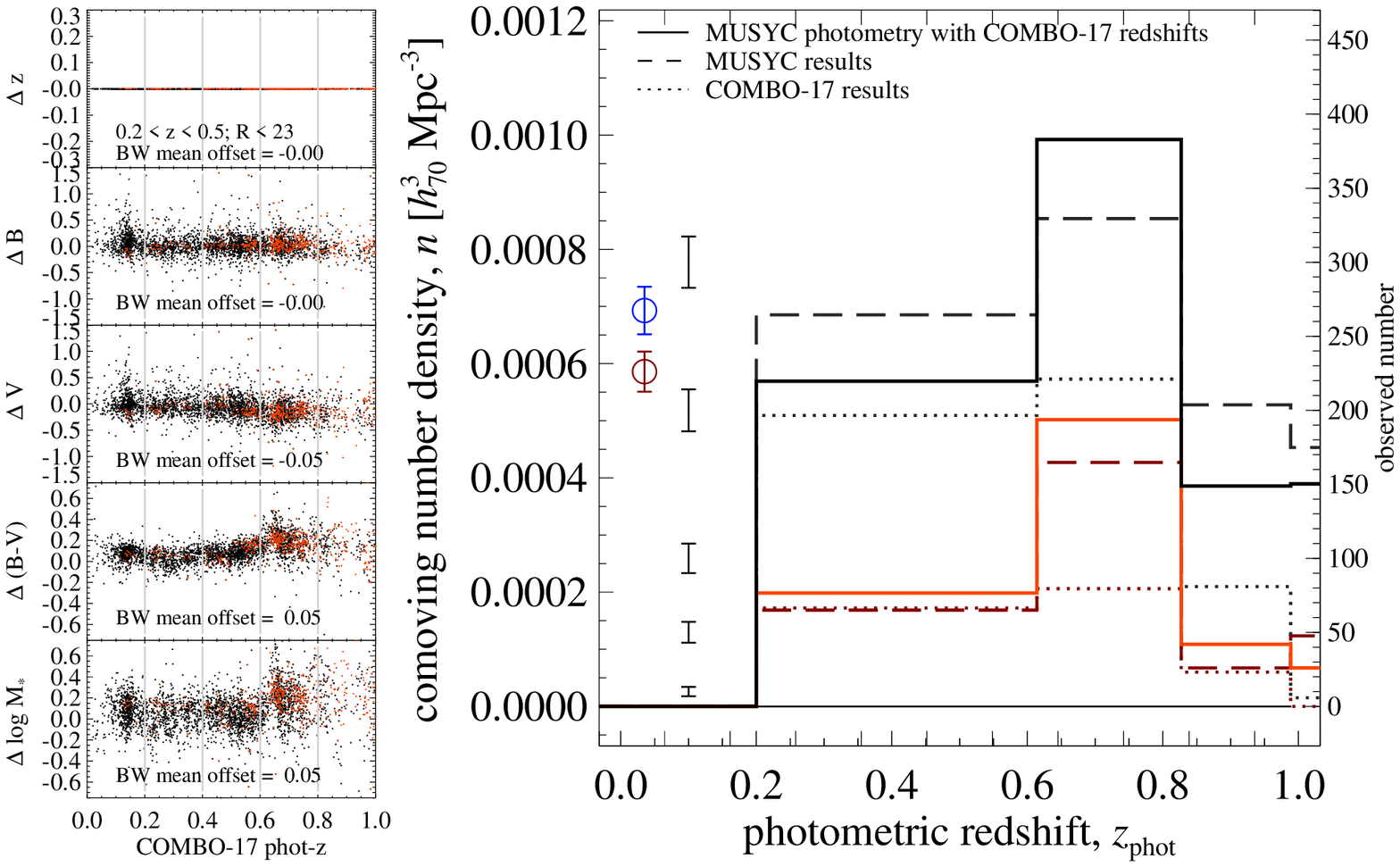}
\caption{Reanalysis of the MUSYC data adopting redshifts from
  COMBO-17---{\em Main panel}: The measured number density of $M_* >
  \masslimit$ M\sun\ galaxies for $z \lesssim 1$ based on the MUSYC
  photometry, but adopting redshifts from the COMBO-17 catalog in
  place of our own ({\em solid histograms}).  These results should be
  compared to the fiducial results from MUSYC ({\em dashed
    histograms}) and from COMBO-17 ({\em dotted histograms}).  The red
  and black histograms are for red and all galaxies, respectively.
  Where the COMBO-17 and MUSYC results differ by 40 \%, the effect of
  using of COMBO-17 redshifts is less than 10 \%; the difference
  between the two surveys' results cannot be explained by differences
  in the photometric redshifts.  {\em Left panels}: the difference
  between the ({\em from top to bottom}) photometric redshifts,
  absolute $B$ and $V$ magnitudes, restframe $(B-V)$ colors, and
  stellar masses used to produce the results shown in the main panel,
  in comparison to those found in the COMBO-17 catalog, plotted as a
  function of the COMBO-17 redshift.  In comparison to COMBO-17, even
  using the same redshifts, the MUSYC data and/or data analysis make
  galaxies appear brighter and redder, and so more massive.
  \label{c17:photz} }
\end{figure}

The essential differences between the COMBO-17 and MUSYC datasets are
COMBO-17's twelve medium bands, allowing much more precise photometric
redshift determinations for $z \lesssim 1$, and the MUSYC $z'JHK$ data,
which open the door to the $z \gtrsim 1$ universe.  The first and most
obvious concern is thus that the difference between the two surveys'
$\zphot \lesssim 1$ results are a product of their different
photometric redshift accuracies.  To test this, we have tried simply
adopting the COMBO-17 photometric redshifts in place of our own
determinations, and repeated our analysis.

The results of this test are shown in Figure \ref{c17:photz}: the main
panel shows our primary result (\viz , the number density of $M_* >
\masslimit$ M\sun\ galaxies as a function of redshift; see Figure
\ref{fig:evo}).  The solid histograms show this trial re-analysis of
the MUSYC data adopting COMBO-17 redshifts; this should be compared to
the fiducial results from MUSYC ({\em dashed histogram}) and COMBO-17
({\em dotted histograms}).  We have derived these `COMBO-17' results
from the public catalog presented by \citet{WolfEtAl}, supplemented
with the stellar mass determinations used by \citet{BorchEtAl}.

At least for $\zphot < 0.8$, the use of COMBO-17 redshifts in place of
our own does not have a great effect on our results: the solid
histograms lie quite close to the dashed ones.  Quantitatively,
adopting COMBO-17 redshifts in place of our own leaves the number of
$M_* > 10^{11}$ M\sun\ galaxies over $0.2 < \zphot < 0.8$ unchanged to
within 1 \%; the effect is slightly greater for the red population,
leading to a red fraction that is 17 \% higher over the same interval.
This implies that, at least for $\zphot \lesssim 1$, our results are
not seriously affected by our lower photometric redshift accuracy
($\sigma_z \approx$ 0.03 for MUSYC, versus 0.02 for COMBO-17, for $R <
24$ and $\zspec < 1$).

To the left of the main panel in Figure \ref{c17:photz}, we show the
difference in ({\em from top to bottom}) the photometric redshifts,
absolute B, and V magnitudes, restframe ($B-V$) colors, and stellar
masses used to produce the solid histograms, and those from the
COMBO-17 catalog.  In all cases, the `$\delta$' is in the sense of
`MUSYC re-analysis' minus `COMBO-17 catalog', and is plotted as a
function of the COMBO-17 catalog redshift.  Within each panel, we
also give the biweight mean offset for each quantity, evaluated for
sources with $R < 23$ (to reduce random scatter due to photometric
uncertainties) and $0.2 < \zphot < 0.5$ (where the COMBO-17 photometry
still samples restframe $V$) in the COMBO-17 catalog.

Looking now briefly at these panels, we see that for $\zphot \gtrsim
0.5$, where the COMBO-17 value for the restframe $V$ magnitude comes
from an extrapolation of the best-fit template spectrum, there is a
progressive offset between the restframe $V$ luminosities inferred by
the MUSYC and COMBO-17 data and analyses---even while using the same
redshifts.  Even for $\zphot < 0.5$, however, we see a systematic
offset of 0.05 mag in ($B-V$) color.  Finally, we note that the
greater number of $0.6 < \zphot < 0.8$ galaxies in comparison to $0.2
< \zphot < 0.6$ that we see in the MUSYC data is also present in the
COMBO-17 photometric redshift distribution; the difference is that
these galaxies do not have $M_* > \masslimit$ M\sun\ in the COMBO-17
catalog.

We therefore conclude that the difference between the COMBO-17 and
MUSYC results is not a product of the two surveys' different
photometric redshift accuracies: the use of COMBO-17 rather than MUSYC
redshifts affects our results only by a few percent.  This leaves the
possibilities that the difference between the two results is due to
the different methods used to infer galaxies' restframe properties, or
to differences in the data themselves.

\subsection{Derivation of Restframe Properties: MUSYC analysis of the COMBO-17 photometry}

\begin{figure} [b] \centering 
\includegraphics[width=12cm]{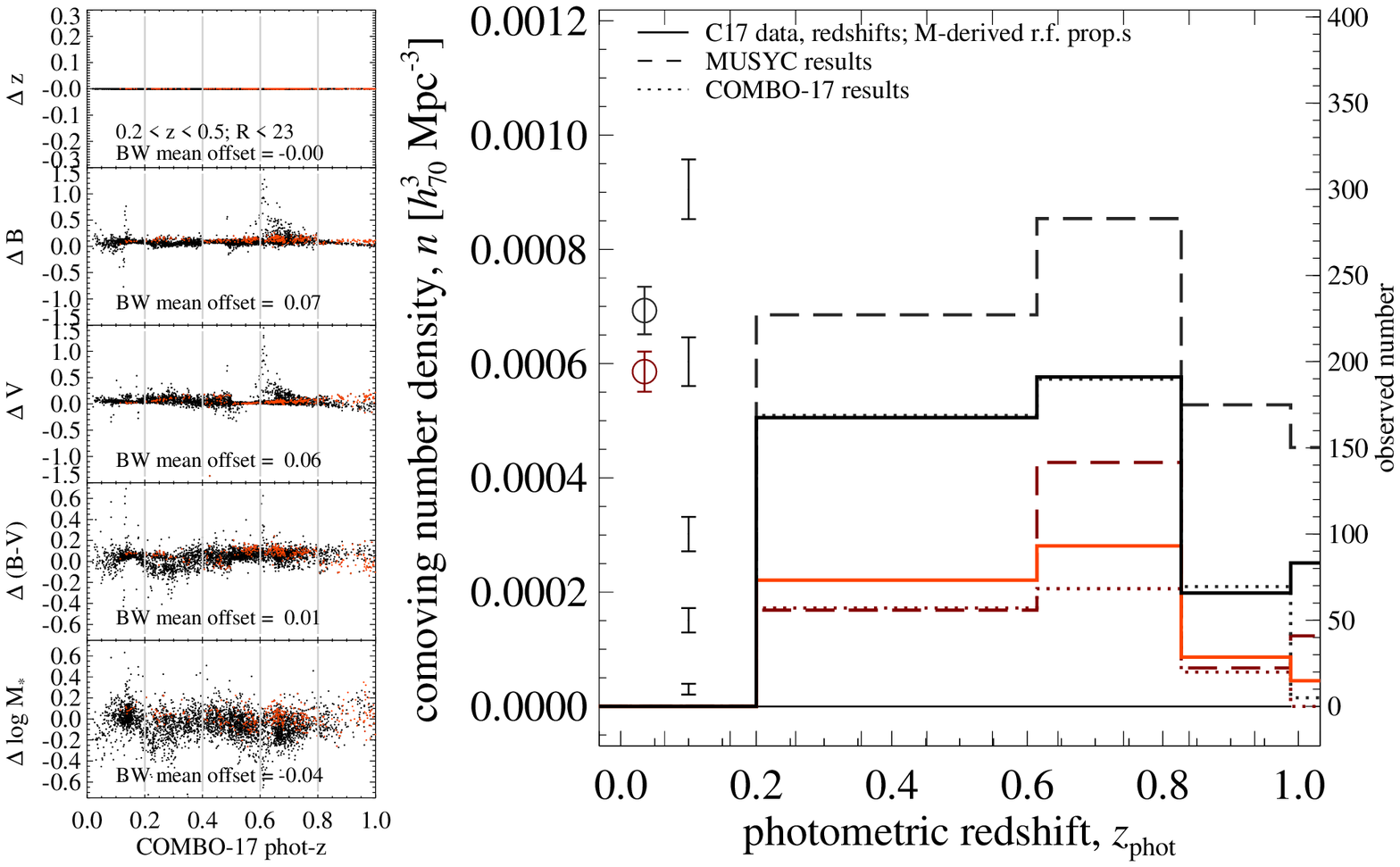}
\caption{MUSYC analysis of the COMBO-17 photometry---The solid
  histograms in the main panel show the measured number density of
  $M_* > \masslimit$ galaxies, applying the MUSYC analysis to the
  COMBO-17 photometry.  In order to isolate the effect on our
  measurements due to the different methods for deriving restframe
  properties, we continue to adopt the COMBO-17 redshifts for this
  test.  All other symbols and their meanings are as in Figure
  \ref{c17:photz}.  Applying the MUSYC analysis to the COMBO-17
  photometry, we reproduce the COMBO-17 result to better than 1 \%;
  the difference between the COMBO-17 and MUSYC results cannot be
  explained solely by our different analytical methods.
  \label{c17:analysis}}
\end{figure}

Are the discrepancies between the COMBO-17 and MUSYC results due to
different systematic effects inherent in our different methods for
deriving restframe photometry and stellar masses?  In order to address
this concern, we have reanalyzed the COMBO-17 data using the
procedures described in the main text.  For this test, in order to
isolate the effect of the different analytical methods, we also adopt
the COMBO-17 redshift determinations; the difference between this test
and the previous one is thus the use of the COMBO-17, rather than the
MUSYC, photometry.

The results of this re-analysis are shown as the solid histograms in
Figure \ref{c17:analysis}; all other symbols and their meanings are as
in Figure \ref{c17:photz}.  The results of the MUSYC re-analysis of
the COMBO-17 data agree very well with the COMBO-17 analysis proper:
the solid black histogram lies very near the dotted black histogram.
Quantitatively, the MUSYC analysis of the COMBO-17 data leads to a red
fraction which is 33 \% higher than for COMBO-17's own analysis; the
total number density of $M_* > \masslimit$ M\sun\ galaxies agree to
better than 1 \%.

Comparing the MUSYC-- and COMBO-17--derived restframe properties ---
again, based on the same photometry and redshifts --- we do see
systematic differences.  We make three specific observations: 1.\
there appear to be discrete redshift regimes where the MUSYC-- and
COMBO-17--derived restframe fluxes compare differently; 2.\ for $0.2
\lesssim \zphot \lesssim 0.35$, we see a `bimodal' offset in the
MUSYC-- and COMBO-17--derived ($B-V$) colors, corresponding to red and
blue galaxies; 3.\ the MUSYC--derived restframe fluxes appear to be
systematically brighter than those derived by COMBO-17, even for the
same photometry.  Remarkably, even despite these differences, our
stellar mass estimates agree to within 0.04 dex, with a scatter of
just 0.11 dex.  Finally, we note that the progressive offset in
extrapolated $V$ luminosities for $\zphot \gtrsim 0.5$ seen in the
previous test is not seen here; that is, applied to the same data,
the two techniques for extrapolating restframe photometry yield
similar results.

We therefore conclude that the difference between the COMBO-17 and
MUSYC results in the ECDFS cannot be explained by differences in the
analytical methods employed by each team: the MUSYC reanalysis of the
COMBO-17 data agrees with the COMBO-17 fiducial results.  Instead, it
seems that the different results are really in the data themselves.


\subsection{Photometric Calibration: MUSYC analysis of the recalibrated COMBO-17 photometry}

\begin{figure}[b] \centering 
\includegraphics[width=12cm]{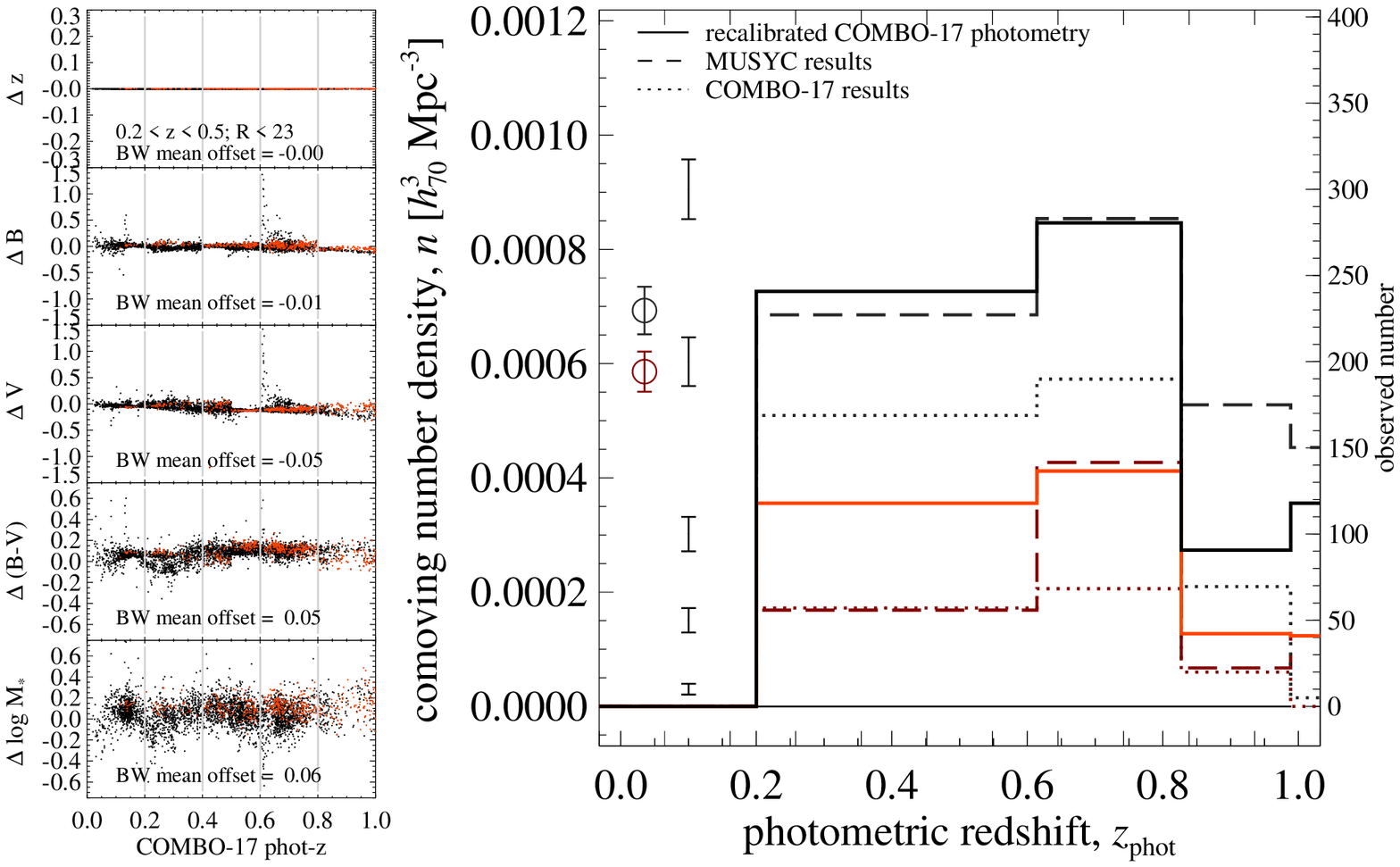}
\caption{Recalibration of the COMBO-17 photometry---the solid
  histograms in the main panel show the measured number density of
  $M_* > \masslimit$ galaxies, applying the MUSYC analysis to the
  COMBO-17 photometry, recalibrated to match the MUSYC photometry.
  Specifically, the UBVRI bands have been scaled up by 0.00, 0.06,
  0.08, 0.08, and 0.14 mag; each of the medium bands has been scaled
  to match the nearest broad band.  We continue to adopt the COMBO-17
  redshifts for this test.  All other symbols and their meanings are
  as in Figure \ref{c17:photz}.  The effect of this rescaling is to
  increase the measured values by a further 30 \%, in comparison to the
  previous test.  Together, the combined effect of this recalibration
  is nearly 50 \%, almost completely explaining the difference between
  the COMBO-17 and MUSYC results.
  \label{c17:seds} }
\end{figure}

A direct, object-by-object comparison of COMBO-17 and MUSYC photometry
reveals significant differences in the photometric calibration of the
two surveys (Paper I).  Specifically, we see differences of 0.00,
0.06, 0.08, 0.08, and 0.14 mag between the MUSYC and COMBO-17
$UBVRI$--band photometry for galaxies, such that the MUSYC photometry
is systematically brighter and redder.  This has subsequently been
confirmed to be due to an error in the photometric calibration of the
COMBO-17 data \citep{combocalib}.  Can this difference in photometric
calibration explain the different results found by COMBO-17 and MUSYC?

To address this issue, we have simply scaled the COMBO-17 photometry
to match the MUSYC photometry, and repeated our analysis.  The
COMBO-17 team has calibrated each of the medium bands relative to the
nearest broad band; accordingly we have scaled each of the medium
bands by the MUSYC--minus--COMBO-17 offset for the nearest broad band.
The results of this test are shown in Figure \ref{c17:seds}.  The
MUSYC re-analysis of the recalibrated COMBO-17 photometry agrees
extremely well with the fiducial MUSYC results: the measured number
density of $M_* > \masslimit$ M\sun\ galaxies at $0.2 < \zphot < 0.8$
agree to better than 1 \%.

There are two separate aspects to this re-calibration: galaxies are
both brighter and redder in the MUSYC catalog than in COMBO-17.
Since COMBO-17 measures total fluxes from their $R$ band image, the
0.08 mag offset between the COMBO-17 and MUSYC zeropoints makes
galaxies appear 7.6 \% brighter (and thus more massive) in the MUSYC
catalog; this effect is responsible for approximately 70 \% of the
change in the measured number density.  At the same time, the
reddening of the SED shape implies a higher mass--to--light ratio
after recalibration; this effect is responsible for the other 30 \% of
the change in the measurements.

\vspace{0.2cm}

We therefore conclude that the primary cause for the disagreement
between the results of the MUSYC and COMBO-17 surveys in the ECDFS is
the different photometric calibrations of the two surveys: reanalyzing
the COMBO-17 data set, recalibrated to match the MUSYC photometry, the
results agree with the MUSYC fiducial analysis.



\begin{thebibliography}{199}


\bibitem[\protect\citeauthoryear{Arnouts et al.}{2001}]
  {ArnoutsEtAl2001} Arnouts S et al., 2001, A\&A 379, 740

\bibitem[\protect\citeauthoryear{Arnouts et al.}{2007}]{Arnouts2007}
  Arnouts S et al., 2007, A\&A 476, 137

\bibitem[\protect\citeauthoryear{Baldry et al.}{2004}]{Baldry2004}
  Baldry I K, Glazebrook K, Brinkmann J, Ivezi\'c \u Z, Lupton R H,
  Nicol R C, Szalay A S, 2004, ApJ 600, 694

\bibitem[\protect\citeauthoryear{Baldry et al.}{2006}]{Baldry2006}
  Baldry I K, Balough M L, Bower R G, Glazebrook K, Nichol R C,
  Bamford S P, Budavari T, 2006, MNRAS 373, 469

\bibitem[\protect\citeauthoryear{Balough et al.}{2003}]{BaloughEtAl}
  Balough M L, Baldry I K, Nichol R, Miller C, Bower R, Glazebrook K,
  2003, ApJ 615, L101

\bibitem[\protect\citeauthoryear{Bell \& de Jong}{2001}]{BellDeJong}
  Bell E F, de Jong R S, 2001, ApJ 550, 212

\bibitem[\protect\citeauthoryear{Bell et al.}{2003}]{Bell2003}
  Bell E F, McIntosh D H, Katz N, Weinberg M D, 2003, ApJS 149, 289

\bibitem[\protect\citeauthoryear{Bell et al.}{2004a}]{Bell2004-Morph}
  Bell E F et al., 2004, ApJ 600, L11

\bibitem[\protect\citeauthoryear{Bell et al.}{2004b}]{Bell2004}
  Bell E F et al., 2004, ApJ 608, 752

\bibitem[\protect\citeauthoryear{Bell et al.}{2007}]{Bell2007}
  Bell E F, Zheng X Z, Papovich C, Borch A, Wolf C, Meisenheimer K,
  2007, ApJ (in press)

\bibitem[\protect\citeauthoryear{Ben\'itez}{2000}]{Benitez} Ben\`itez
  N, 2000, ApJ 536, 571

\bibitem[\protect\citeauthoryear{Bertin \& Arnouts}{1996}]
  {BertinArnouts} Bertin E \& Arnouts S, A\&A 117, 393

\bibitem[\protect\citeauthoryear{Blanc et al.}{2008}]{Blanc2008} Blanc
  G A et al., 2008, ApJ, submitted

\bibitem[\protect\citeauthoryear{Blanton et al.}{2003}]{Blanton2003}
  Blanton M R et al., 2003, ApJ 594, 186

\bibitem[\protect\citeauthoryear{Blanton et al.}{2005a}]
  {Blanton2005Env} Blanton M R, Eisenstein D, Hogg D W, Schlegel B K,
  Brinchmann J, 2005, ApJ 629, 143

\bibitem[\protect\citeauthoryear{Blanton et al.}{2005b}]
  {BlantonEtAl-lowz} Blanton M R, Schlegel D J, Strauss M A, Brinkmann
  J, Finkbeiner D, Fukugita M, Gunn J E, Hogg D W, Ivezi\'c \u Z, Knapp
  G R, Lupton R H, Munn J A, Schneider D P, Tegmark M, Zehavi I, 2005,
  AJ 129, 2562

\bibitem[\protect\citeauthoryear{Blanton et al.}{2005c}]
  {BlantonEtAl-lf} Blanton M R, Lupton R H, Schlegel D J, Strauss M
  A, Brinkmann, Fukugita M, Loveday J, 2005, ApJ 631, 208
  
\bibitem[\protect\citeauthoryear{Blanton \& Roweis}{2007}]
  {Blanton-kcorrect} Blanton M R \& Roweis S, 2007, ApJ 133, 734

\bibitem[\protect\citeauthoryear{Bolzonella, Miralles \& Pello}{2000}]
  {hyperz} Bolzonella M, Miralles J-M \& Pell\'o R, 2000, A\&A 363,
  476

\bibitem[\protect\citeauthoryear{Borch et al.}{2006}]{BorchEtAl} Borch
  A, Meisenheimer K, Bell E F, Rix H-W, Wolf C, Dye S, Kleinheinrich
  M, Kovacs Z, Wisotzki L, 2006, A\&A 453, 869

\bibitem[\protect\citeauthoryear{Brammer et al.}{2008}]{eazy}
  Brammer G B, Van Dokkum P G, Coppi P, 2008, submitted

\bibitem[\protect\citeauthoryear{Brinchmann et al.}{2004}]
  {BrinchmannEtAl} Brinchmann J, Charlot S, White S D M, Tremonti C,
  Kauffmann G, Heckman T, 2004, MNRAS 351, 1151

\bibitem[\protect\citeauthoryear{Brown et al.}{2007}]{BrownEtAl} Brown
  M J I, Dey A, Jannuzi B T, Brand K, Benson A J, Brodwin M, Croton D
  J, Eisenhardt P R, 2007, ApJ 164, 858

\bibitem[\protect\citeauthoryear{Brown et al.}{2008}]{BrownEtAl2008}
  Brown M J I, Zheng Z, White M, Dey A, Jannuzi B T, Benson A J, Brand
  K, Brodwin M, Croton D J

\bibitem[\protect\citeauthoryear{Bruzual \& Charlot}{2003}]
  {BruzualCharlot} Bruzual G, Charlot S, 2003, MNRAS 344, 1000

\bibitem[\protect\citeauthoryear{Bruzual}{2008}]
  {Bruzual} Bruzual G A, 2008, astro-ph/0703052 (v2)

\bibitem[\protect\citeauthoryear{Bundy et al.}{2006}] {Bundy2006}
  Bundy K et al., 2006, ApJ 651, 120

\bibitem[\protect\citeauthoryear{Cattaneo et al}{2006}]{Cattaneo2006}
  Cattaneo A, Dekel A, Devriendt J, Guiderdoni B, Blaizot J, 2006,
  370, 1651

\bibitem[\protect\citeauthoryear{Cattaneo et al}{2008}]{CattaneoEtAl}
  Cattaneo A, Dekel A, Faber S M, Guiderdoni B, 2008, MNRAS
  (submitted), arXiv:0801.1673

\bibitem[\protect\citeauthoryear{Cimatti et al.}{2002}]{K20} Cimatti
  A, Mignoli M, Daddi E, Pozzetti L, Fontana A, Saracco P, Poli F,
  Renzini A, Zamorani G, Broadhurst T, Cristiani S, D'Odorico S,
  Giallongo E, Gilmozzi R, Menci N, 2002, A\& A 392, 395

\bibitem[\protect\citeauthoryear{Cirasuolo et al.}{2007}]
  {CirasuoloEtAl} Cirasuolo M, McLure R J, Dunlop J S, Almaini O,
  Foucaud S, Smail I, Sekiguchi K, Simpson C, Eales S, Dye S, Watson M
  G, Page M J, Hist P, 2007, MNRAS 380, 585

\bibitem[\protect\citeauthoryear{Cole et al.}{2001}]{ColeEtAl}
  Cole S et al., 2001, MNRAS 326, 255

\bibitem[\protect\citeauthoryear{Coleman, Wu \& Weedman}{1980}]{cww}
  Coleman G D, Wu C C, Weedman D W, 1980, ApJS 43, 393

\bibitem[\protect\citeauthoryear{Connolly et al.}{1997}]
  {ConnollyEtAl1998} Connolly A J, Szalay A S, Dickinson M,Subbarao M U,
  Brunner R J, 1997, ApJ 486, L11

\bibitem[\protect\citeauthoryear{Conselice et al.}{2007}]
  {ConseliceEtAl} Conselice C J, Bundy K, Trujillo I, Coil A,
  Eisenhardy P, Ellis R S, Georgakakis A, Huang J, Lotz J, Nandra K,
  Newman J, Papovich C, Weiner B, Willmer C, 2007, MNRAS 381, 962

\bibitem[\protect\citeauthoryear{Cooper et al.}{2007}]{CooperEtAl}
  Cooper M J et al., 2007, MNRAS 376, 1445


\bibitem[\protect\citeauthoryear{Croton et al.}{2006}] {Croton2006}
  Croton D J, Springel V, White S D M, De Lucia G, Frenk C S, Gao L,
  Jenkins A, Kauffmann G, Navarro J F, Yoshida N, 2006, MNRAS 365, 11

\bibitem[\protect\citeauthoryear{Daddi et al.}{2005}]{Daddi2005}
  Daddi E et al., 2005, ApJ 626, 680

\bibitem[\protect\citeauthoryear{Damen et al.}{2008}]{DamenEtAl}
  Damen M et al., 2008, ApJ (submitted)

\bibitem[\protect\citeauthoryear{Dekel \& Birnboim}{2006}]
  {DekelBirnboim} Dekel A, Birnboim Y, 2006, MNRAS 368, 2

\bibitem[\protect\citeauthoryear{De Lucia et al.}{2007}]{DeLucia2007}
  De Lucia G et al., 2007, MNRAS 374, 809

\bibitem[\protect\citeauthoryear{Dickinson et al.}{2002}]{DickinsonEtAl}
  Dickinson M E, Giavalisco M, et al., 2002, astro-ph/0204213 (v1)

\bibitem[\protect\citeauthoryear{Driver et al.}{2006}]{Driver2006}
  Driver S P, Allen P D, Graham A W, Cameron E, Liske J, Ellis S C,
  Cross N J G, De Propris R, Phillipps S, Couch W J, 2006, MNRAS 368,
  414

\bibitem[\protect\citeauthoryear{Drory}{2004}]{Drory2004} Drory N et
  al., 2004, ApJ 608, 742

\bibitem[\protect\citeauthoryear{Ellis et al.}{2005}]{Ellis2005} Ellis
  S C, Driver S P, Allen S D, Liske J, Bland-Hawthorn J, De Propris R,
  2005, MNRAS 363, 1257

\bibitem[\protect\citeauthoryear{Erben et al.}{2005}]{ErbenEtAl} Erben
  T et al., 2005, AN 326, 432

\bibitem[\protect\citeauthoryear{Faber et al.}{2007}]{Faber2007} Faber
  S M et al., 2007, ApJ 665, 265

\bibitem[\protect\citeauthoryear{Fontana et al.}{2004}]{Fontana2004}
  Fontana A et al., 2004, A\&A 424, 23

\bibitem[\protect\citeauthoryear{Fontana et al.}{2006}]{Fontana2006}
  Fontana A et al., 2006, A\&A 459, 745

\bibitem[\protect\citeauthoryear{F\" orster-Schreiber et al.}{2006}]
  {NFS} F\"orster-Schreiber N M, Franx M, Labb\'e I, Rudnick G,
  Van Dokkum P G, Illingworth G D, Kuijken K, Moorwood
  A F M, Rix, HW., R\" ottgering H, Van der Werf P, 2006, AJ 131, 1891

\bibitem[\protect\citeauthoryear{Franx et al.}{2003}]{FranxEtAl} Franx
  M et al., 2003, ApJ 587, L79
  
\bibitem[\protect\citeauthoryear{Franzetti et al.}{2007}]
  {FranzettiEtAl} Franzetti P et al., 2007, A\&A, 165, 711

\bibitem[\protect\citeauthoryear{Gawiser et al.}{2006}] {GawiserEtAl}
  Gawiser E et al., 2006, ApJS 162, 1

\bibitem[\protect\citeauthoryear{Giacconi et al.}{2001}]{Giacconi2001}
  Giacconi R, Rosati P, Tozzi P, Nonino M, Hasinger G, Norman C,
  Bergeron J, Borgani S, Gilli R, Gilmozzi R, Zheng W, 2001, ApJ 551,
  624

\bibitem[\protect\citeauthoryear{Grazian et al.}{2006}]{GrazianEtAl}
  Grazian A, Fontana A, De Santis C, Nonino M, Salimbeni S, Giallongo
  E, Cristiani S, Gallozzi S, Vanzella E, 2006, A\& A 449, 951

\bibitem[\protect\citeauthoryear{Gronwall et al.}{2007}]{GronwallEtAl}
  Gronwall C et al., 2007, ApJ, 667. 79

\bibitem[\protect\citeauthoryear{Gunn \& Stryker}{1983}]{bpgs} Gunn J
  E \& Stryker L L, 1983, ApJSS 52, 121

\bibitem[\protect\citeauthoryear{Hildebrandt et al.}{2006}]
  {HildebrandtEtAl} Hildebrandt H et al., 2006, A\&A 452, 1121


\bibitem[\protect\citeauthoryear{Hogg et al.}{2003}]{Hogg2003} Hogg D
  W, Blanton M R, Eisenstein D J, Gunn J E, Schlegel D J, Zehavi I,
  Bahcall N A, Brinkmann J, Csabai I, Schneider D P, Weinberg D H,
  York D G, 2003, ApJ 585, L5

\bibitem[\protect\citeauthoryear{Hopkins}{2004}]{Hopkins2004} Hopkins
  A M, 2004, ApJ 615, 209

\bibitem[\protect\citeauthoryear{Im et al.}{2002}]{ImEtAl} Im M et
  al., 2002, ApJ 571, 136

\bibitem[\protect\citeauthoryear{Juneau et al.}{2005}]{Juneau2005}
  Juneau S et al., 2005, ApJ 619, L135

\bibitem[\protect\citeauthoryear{Kauffmann et al.}{2003}]
  {Kauffmann2003} Kauffman G, Heckman T M, White S D M, Charlot S,
  Tremonti C, Peng E W, Seibert M, Brinkmann J, Nichol R C, SubbaRao
  M, York D, 2003, MNRAS 341, 54

\bibitem[\protect\citeauthoryear{Kinney et al.}{1996}]{Kinney} Kinney
  A L, Calzetti D, Bohlin R C, McQuade K, Storchi-Bergmann T, Schmitt
  H R, 1996, ApJ 467, 38

\bibitem[\protect\citeauthoryear{S Koposov et al.}{in prep.}]
  {KoposovEtAl} Koposov et al., 2008 (in prep.)

\bibitem[\protect\citeauthoryear{Kriek et al.}{2006}] {Kriek2006}
  Kriek M, et al., 2006, ApJ, 649, L71

\bibitem[\protect\citeauthoryear{Kriek et al.}{2008}] {Kriek2008}
  Kriek M, Van der Wel A, Van Dokkum P G, Franx M, Illingworth G D,
  2008, ApJ (accepted); arXiv:0804:4175v1

\bibitem[\protect\citeauthoryear{Kroupa}{2001}]{Kroupa2001} Kroupa P,
  2001, MNRAS 322, 231

\bibitem[\protect\citeauthoryear{Labb\'e et al.}{2003}] {LabbeEtAl}
  Labb\'e I et al., 2003, AJ 125, 1107

\bibitem[\protect\citeauthoryear{Labb\'e et al.}{2005}] {Labbe2005}
  Labb\'e I et al., 2005, ApJ 624, L81

\bibitem[\protect\citeauthoryear{Labb\'e et al.}{2007}] {Labbe2007}
  Labb\'e I, Franx M, Rudnick G, Forster Schrieber N M, Van Dokkum P
  G, Moorwood A, Rix H-W, R\:ottgering H, Trujillo I, Van der Werf P,
  2007, ApJ 665, 964

\bibitem[\protect\citeauthoryear{Lawrence et al.}{2007}]{LawrenceEtAl}
  Lawrence A et al., 2007, MNRAS 379, 1599

\bibitem[\protect\citeauthoryear{Le Borgne \& Rocca-Volmerange}{2002}]
  {pegase} Le Borgne D \& Rocca-Volmerange B, 2002, A\&A 386, 466

\bibitem[\protect\citeauthoryear{Le F\` evre et al.}{2004}]{VVDS} Le
F\` evre O et al., 2004, A\& A 428, 1043

\bibitem[\protect\citeauthoryear{Lilly et al.}{1996}]{Lilly1996} Lilly
  S, Le F\` evre O, Hammer F, Crampton D, 1996, ApJ 460, L1

\bibitem[\protect\citeauthoryear{Lin et al.}{2007}]{LinEtAl} Lin L et
  al., 2007, ApJ 660, L51

\bibitem[\protect\citeauthoryear{Madau et al.}{1996}]{Madau1996} Madau
  P, Ferguson H C, Dickinson M E, Giavalisco M, Steidel C C, Fruchter
  A, 1996, MNRAS 283, 1388

\bibitem[\protect\citeauthoryear{Maraston}{2005}]{Maraston} Maraston
  C, 2005, MNRAS 362, 799

\bibitem[\protect\citeauthoryear{McCarthy}{2004}]{McCarthy} McCarthy P
  J, 2004, ARAA, 42, 477

\bibitem[\protect\citeauthoryear{McGrath et al.}{2007}]{McGrath2007}
  McGrath E J, Stockton A, Canalizo G, 2007, ApJ 669, 241

\bibitem[\protect\citeauthoryear{Menci et al}{2005}]{Menci2005} Menci
  N, Fontana A, Giallongo E, Salimbeni S, 2005, ApJ 632, 49

\bibitem[\protect\citeauthoryear{Moy et al.}{2003}] {MoyEtAl} Moy E et
  al., 2003, A\&A 403, 493

\bibitem[\protect\citeauthoryear{Nagmine et al.}{2006}]{NagamineEtAl}
  Nagamine K, Ostriker J P, Fukugita M, Cen R, 2006, ApJ 653, 881

\bibitem[\protect\citeauthoryear{Panter et al.}{2007}]{PanterEtAl}
  Panter B, Jimenez R, Heavens A F, Charlot S, 2007, MNRAS 378, 1550

\bibitem[\protect\citeauthoryear{P\'erez-Gonzal\'ez et al.}{2008}]
  {PerezGonzalez} P\'erez-Gonzal\'ez P G et al., 2008, ApJ 675, 234

\bibitem[\protect\citeauthoryear{Popesso et al.}{2008}] {VIMOS}
  Popesso P et al., 2008, arxiv:0802.2930v1

\bibitem[\protect\citeauthoryear{Pozzetti et al.}{2007}]{PozzettiEtAl}
  Pozzetti L et al., 2007, A\&A, 474, 447

\bibitem[\protect\citeauthoryear{Quadri et al.}{2007}]{QuadriEtAl} 
  Quadri R et al., 2007, AJ 134, 3

\bibitem[\protect\citeauthoryear{Ravikumar et al.}{2006}]{IMAGES}
  Ravikumar C D et al., 2006, A\& A 465, 1099

\bibitem[\protect\citeauthoryear{Reddy et al.}{2006}]{Reddy2006} Reddy
  N A, Steidel C C, Fadda D, Yan L, Pettini M, Shapley A E, Erb D K,
  Adelberger K L, 2006, ApJ 644, 792

\bibitem[\protect\citeauthoryear{Rix et al.}{2004}]{RixEtAl} Rix H-W
  et al., 2004, ApJSS 152, 163

\bibitem[\protect\citeauthoryear{Rudnick et al.}{2003}] {RudnickEtAl}
  Rudnick G et al., 2003, ApJ 599, 847

\bibitem[\protect\citeauthoryear{Ruhland et al.}{2008}]
  {Ruhland2008} Ruhland C, Bell E F, H\"au\ss ler B, Taylor E N,
  Barden M, McIntosh D, 2008, submitted

\bibitem[\protect\citeauthoryear{Salpeter}{1955}]{Salpeter}
  Salpeter E E, 1955, ApJ 121, 161

\bibitem[\protect\citeauthoryear{Scarlata et al.}{2007}]{Scarlata2007}
  Scarlata C et al., 2008, ApJSS, 172, 510

\bibitem[\protect\citeauthoryear{Somerville et al.}{2004}]
  {SomervilleEtAl} Somerville R S, Lee K, Ferguson H, Gardner J P,
  Moustakas L A, Giavalisco M, 2004, ApJ 600, L171

\bibitem[\protect\citeauthoryear{Strateva et al.}{2001}]{Strateva2001}
  Strateva I et al., 2001, AJ, 122, 1861

\bibitem[\protect\citeauthoryear{Szokoly et al.}{2004}]{Xray} Szokoly
  G P et al., 2004, ApJSS 155, 271

\bibitem[\protect\citeauthoryear{Taylor et al.}{2008}]{PaperI} Taylor
  E N et al., 2008 (Paper I), ApJSS (submitted)

\bibitem[\protect\citeauthoryear{Treister et al.}{2008}]{TreisterEtAl}
  Treister E, Virani S, Gawiser E, Urry C M, Lira P, Francke H, Blanc
  G A, Cardamone C N, Damen M, Taylor E N, Schawinski K

\bibitem[\protect\citeauthoryear{Tremonti et al.}{2004}]
  {TremontiEtAl} Tremonti C A et al., 2004, ApJ 613, 898

\bibitem[\protect\citeauthoryear{Tresse et al.}{2007}]{TresseEtAl}
  Tresse L et al., A\&A 472, 403

\bibitem[\protect\citeauthoryear{Van der Wel et al.}{2005}]{vdwel05}
  Van der Wel A, Franx M, Van Dokkum P G, Rix H-W, Illingworth G D,
  Rosatti P, 2005, ApJ 631, 145

\bibitem[\protect\citeauthoryear{Van der Wel}{2008}]
  {vdWel2008} Van der Wel A, 2008, ApJ 675, L13

\bibitem[\protect\citeauthoryear{Van Dokkum \& Franx}{2001}]
  {VanDokkumFranx} Van Dokkum P G \& Franx M, 2001, ApJ 553, 90

\bibitem[\protect\citeauthoryear{Van Dokkum et al.}{2006}]
  {VanDokkum2006} Van Dokkum P G et al., 2006, ApJ 638, 59

\bibitem[\protect\citeauthoryear{Vanzella et al.}{2008}]{FORS2}
  Vanzella et al., 2008, A\& A 471, 83

\bibitem[\protect\citeauthoryear{Weiner et al.}{2005}]{WeinerEtAl}
  Weiner B J et al., 2005, ApJ 620, 595

\bibitem[\protect\citeauthoryear{Willmer et al.}{2006}]{WillmerEtAl}
  Willmer C N A et al., 2006, ApJ, 647, 853

\bibitem[\protect\citeauthoryear{Wolf et al.}{2003}] {WolfEtAl}
  Wolf C et al., 2004, A\&A 421, 913

\bibitem[\protect\citeauthoryear{Wolf et al.}{2008}]{combocalib} Wolf
  C, Hildebrandt H, Taylor E N, Meisenheimer K, 2008, A\&A
  (submitted), arXiv:0809.2066

\bibitem[\protect\citeauthoryear{Wuyts et al.}{2008}]{WuytsEtAl} Wuyts
  S et al, 2008, ApJ (in press), arXiv:0804.0615v2

\bibitem[\protect\citeauthoryear{Wyder et al.}{2007}]{WyderEtAl} Wyder
  T K, 2007, ApJSS 173, 293

\bibitem[\protect\citeauthoryear{York et al.}{2000}]{YorkEtAl} York D
  G et al., 2000, ApJ 120, 1579




\end{thebibliography}
\end{document}